\documentclass[a4paper, 10pt, journal]{IEEEtran}
\usepackage[table]{xcolor}
\usepackage{url}
\usepackage{graphicx}
\usepackage{amsfonts}
\usepackage{color}
\usepackage{stfloats}
\usepackage{color,soul}
\usepackage{amsmath}
\usepackage{multirow}
\usepackage{hhline}
\usepackage{url}
\usepackage{float}
\usepackage{enumitem}
\usepackage{cite}
\usepackage{multicol}
\usepackage{blindtext}
\usepackage{rotating}
\usepackage{array}
\usepackage{tikz}
\usepackage{tabularx}
\usepackage{booktabs}
\definecolor{Gray}{gray}{0.9}
\usepackage{float}
\usepackage{tabularx}
\usepackage{longtable}
\usepackage{pdflscape}
\usepackage{afterpage}
\usepackage{capt-of}
\usepackage[bottom]{footmisc}
\usepackage{threeparttable}
\usepackage{color, colortbl}
\usepackage{xurl}

\begin{document}
\title{Applications of Explainable AI for 6G: Technical Aspects, Use Cases, and Research Challenges}
    
\author
{          
 ~Shen Wang, ~\IEEEmembership{Member,~IEEE}, 
 M. Atif Qureshi, 
 Luis Miralles-Pechuán, 
 Thien Huynh-The,~\IEEEmembership{Member,~IEEE}, \\
 Thippa Reddy Gadekallu,~\IEEEmembership{Senior Member,~IEEE},
 and Madhusanka~Liyanage,~\IEEEmembership{Senior Member,~IEEE}

\thanks{Shen Wang is with the School of Computer Science, University College Dublin, Ireland, email: shen.wang@ucd.ie}
\thanks{M. Atif Qureshi is with ADAPT Centre, eXplainable Analytics Group, Faculty of Business, Technological University Dublin, Ireland, email: muhammadatif.qureshi@tudublin.ie}
\thanks{Luis Miralles-Pechuán is in the School of Computing, Technological University Dublin, Central Quad, Grangegorman, Dublin, Ireland, email: luis.miralles@tudublin.ie}
\thanks{Thien Huynh-The is with the Department of Computer and Communication Engineering, Ho Chi Minh City University of Technology and Education, Vietnam, email: thienht@hcmute.edu.vn}
\thanks{Thippa Reddy Gadekallu is with the School of Information Technology and Engineering, Vellore Institute of Technology, India, as well as with the Department of Electrical and Computer Engineering, Lebanese American University, Byblos, Lebanon. email: thippareddy.g@vit.ac.in}
\thanks{Madhusanka Liyanage is with the School of Computer Science, University College Dublin, Ireland, email: madhusanka@ucd.ie }
}

\maketitle


\begin{abstract}
When 5G began its commercialisation journey around 2020, the discussion on the vision of 6G also surfaced. Researchers expect 6G to have higher bandwidth, coverage, reliability, energy efficiency, lower latency, and an integrated ``human-centric" network system powered by artificial intelligence (AI). Such a 6G network will lead to an excessive number of automated decisions made in real-time. These decisions can range widely, from network resource allocation to collision avoidance for self-driving cars. However, the risk of losing control over decision-making may increase due to high-speed, data-intensive AI decision-making beyond designers' and users' comprehension. The promising explainable AI (XAI) methods can mitigate such risks by enhancing the transparency of the black box AI decision-making process. This paper surveys the application of XAI towards the upcoming 6G age in every aspect, including 6G technologies (e.g., intelligent radio, zero-touch network management) and 6G use cases (e.g., industry 5.0). Moreover, we summarised the lessons learned from the recent attempts and outlined important research challenges in applying XAI for 6G in the near future. 

\end{abstract}


\section{Introduction} \label{sec:intro} 
\begin{table}[h!]
\centering
\caption{Summary of Important Acronyms}
\label{table:acronyms}
\begin{tabular}{|p{1.5cm} p{6.5cm}|}
\hline
\rowcolor{gray!30}
\textbf{Acronym} & \textbf{Definition}\\
\hline
\hline
5/6G & Fifth/Sixth Generation \\
AI & Artificial Intelligence \\
A/C/D/RNN & Artificial/Convolutional/Deep/Recurrent Neural Network \\
A/M/V/XR & Augmented/Mixed/Virtual/eXtended Reality \\
B5G & Beyond 5G \\
CAV & Connected Autonomous Vehicle  \\
CV & Computer Vision \\
DARPA & Defense Advanced Research Projects Agency \\
DRL & Deep Reinforcement Learning \\
E2E & End to End \\
ETSI & European Telecommunications Standards Institute \\
eMBB & Enhanced Mobile Broadband \\
GDPR & General Data Protection Regulation \\
IBN & Intent-Based  Networking \\
ICT & Information and Communications Technology \\
IoE/T & Internet of Everything/Things \\
KNN & K-Nearest Neighbors \\
LIME & Local Interpretable Model-Agnostic Explanations \\
LRP & Layerwise Relevance Propagation \\
LSTM & Long Short-Term Memory \\
mMTC & Massive Machine-Type Communications \\
ML & Machine Learning \\
MSCA & Marie Skłodowska-Curie Actions \\
NFV & Network Function Virtualisation \\
NLP & Natrual Language Processing \\
PIPL & Personal Information Protection Law \\
PIRL & Programmatically Interpretable Reinforcement Learning \\
QoE/S/T & Quality of Experience/Service/Trust \\
RL & Reinforcement Learning \\
SCS & System Causability Scale \\
SDN & Software Defined Network \\
SHAP & SHapley Additive exPlanations \\
SVM & Support Vector Machines \\
UAV & Unmanned Aerial Vehicles \\
URLLC & Ultra-Reliable Low Latency Communications \\
XAI & eXplainable AI \\
ZSM & Zero-Touch Service and Network Management \\
\hline
\end{tabular}
\end{table}
The mobile network has been drastically revolutionised in the last few decades. The first generation mobile network (1G) started in the 1980s, and this system supported people calling from a moving place rather than a fixed location. The second-generation (2G) changed the signal transmitted from analog to digital. It enabled services such as Short Messaging Service (SMS) so that both callers and receivers did not have to be ``online" at the same time. The third-generation (3G) mobile network increased the data rate to the level of Mbps, which accelerated access to essential Internet services such as web browsing. The fourth-generation (4G) provides a much higher data rate of up to 1Gbps by fully integrating with all-IP packet switching networks so that mobile users can easily access data-intensive services such as sharing video through TikTok wherever connected. The ongoing fifth-generation mobile network (5G) technology supports services such as Enhanced Mobile Broadband (eMBB), Ultra-Reliable Low Latency Communications (URLLC), and Massive Machine-Type Communications (mMTC). 5G realises the Internet of Things (IoT) by significantly increasing the density (100x) of various connected devices with much higher data rates (10x) and (10x) less latency than 4G.

5G network is being commercialised and deployed across the world. Many organisations have started planning beyond 5G (B5G) to develop the next generation of wireless cellular networks (6G). Although B5G and 6G are used interchangeably in the literature\cite{David2018privacy,benzaid2020ai,zhu2019millimeter}, this paper uses 6G only for simplicity. 6G will further extend the connection coverage by achieving space-air-ground-sea integrated networks\cite{kato2019optimizing} to facilitate the Internet of Everything (IoE). The increased data rate in 6G will also support more data-intensive applications such as full-sensory digital reality. The super reliable and low latency that 6G provides can be well-suited for mission-critical scenarios such as autonomous driving and smart health care.  5G has a highly softwarised network infrastructure thanks to the software-defined network (SDN) and the network function virtualisation (NFV). Building on top of this 5G feature, fully automated network management will be feasible with the power of Artificial Intelligence (AI) in the 6G era to increase the efficiency of network maintenance. More 6G features can be found in \cite{letaief2019roadmap, dang2020should, Saad2020privacy,zhang2019vision6G}. Additionally, as first pointed out in \cite{dang2020should}, we agree that the design of 6G will be ``human-centric" rather than ``machine-centric", as all the previous network generations mainly focused on improving the network performance technically while ignoring their impact on the human or society. 6G emphasises more on implementing a fully automotive network powered by AI (such as an intent-based network or intelligent radio) to satisfy humans' needs without violating personal privacy (e.g., intelligent health and wearable). As a result, for a given time interval, there will be an excessively higher number of AI decisions automatically made due to the high-performance 6G network compared with 5G. Note that the number of incorrect AI decisions is also increasing, which leads to a high risk of the overall AI-based 6G systems. Therefore, such a black-box intelligent 6G system requires promising technologies such as eXplainable AI (XAI) to enhance trust between humans and the network. The role of AI and XAI towards 6G will be discussed further in the following paragraphs.


\subsection{Role of AI for 6G} 
AI will play a critical role in realising 6G networks and their applications. There are several ways in which AI can be used in 6G, including the conventional use of AI for prescriptive, predictive, diagnostic, and descriptive analytics. \textit{Prescriptive analytics} can be used for making decisions or predictions related to edge AI such as cache placement, AI model migration, dynamically scaling network slices and adapting its service function chains, and optimal automatic allocation of resources (e.g., spectrum, cloud, and backhaul). AI-based \textit{predictive analytics} help to predict the future from the acquired data in real-time for events like resource availability, preference, user behaviour, user locations, and traffic patterns, then proactively change the network. Proactive actions can fine-tune the resource allocation, deployment of proactive security solutions, pre-migration of edge services, and edge AI models. \textit{Diagnostic analytics} concerns with detecting faults in the network, thus detecting network anomalies, service impairments, network faults, and the root causes of these network faults, which ultimately helps enhance network security and reliability. Due to the high scalability of the 6G network in terms of users, devices and services, AI-enabled automatic services are essential for 6G. \textit{Descriptive analytics} heavily rely on historical data to enhance the service provider’s and network operator’s situational awareness. The applications include user perspectives, channel conditions, traffic profiles, network performance, and so on. Furthermore, handling, generating, and processing large volumes of data in real-time and in a collaborative way is yet another complicated task that requires scalable AI. AI will also play a vital role in controlling and orchestrating the 6G network. For instance, novel 6G network controlling and orchestrating concepts of Intent-Based Networking (IBN) as well as Zero Touch Service and Network Management (ZSM) are primarily dependent on AI technologies\cite{liyanage2022survey}. Novel concepts such as Open RAN or O-RAN will define the development of RAN (Radio Access Network) for future 6G networks. AI is heavily utilized to realize the critical features in O-RAN, such as RAN Intelligent Controller (RIC) frameworks\cite{singh2020evolution}.

\begin{figure}[htb]
    \centering
    \includegraphics[width=0.49\textwidth]{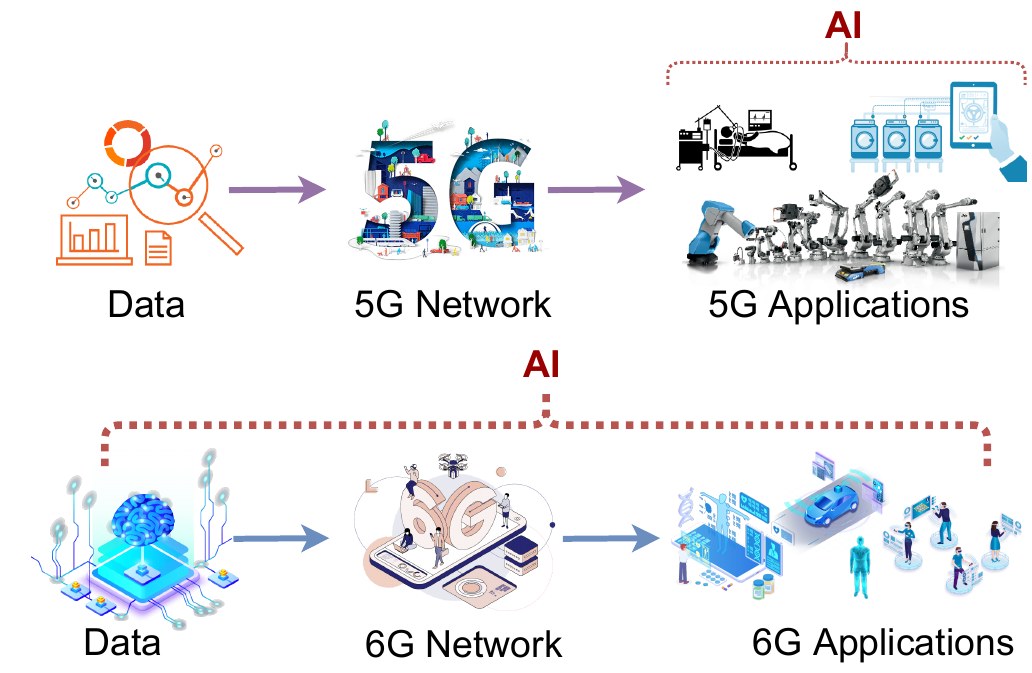} 
    \caption{An illustration of the role of AI in 5G and 6G Networks, whereas AI for 5G is application-driven. In contrast, AI will be used to improve the design of every aspect of 6G (e.g., reliable data sensing, efficient network management, and applications such as connected autonomous vehicles).}
    \label{fig:ai_5g6g}
\end{figure}

Figure \ref{fig:ai_5g6g} illustrates the role of AI in 5G and 6G networks. AI will be lightly used in some of the 5G applications. However, AI will be integrated into the E2E (End-to-End) processes in 6G networks.

\begin{figure*}[htb]
    \centering
    \includegraphics[width=0.99\textwidth]{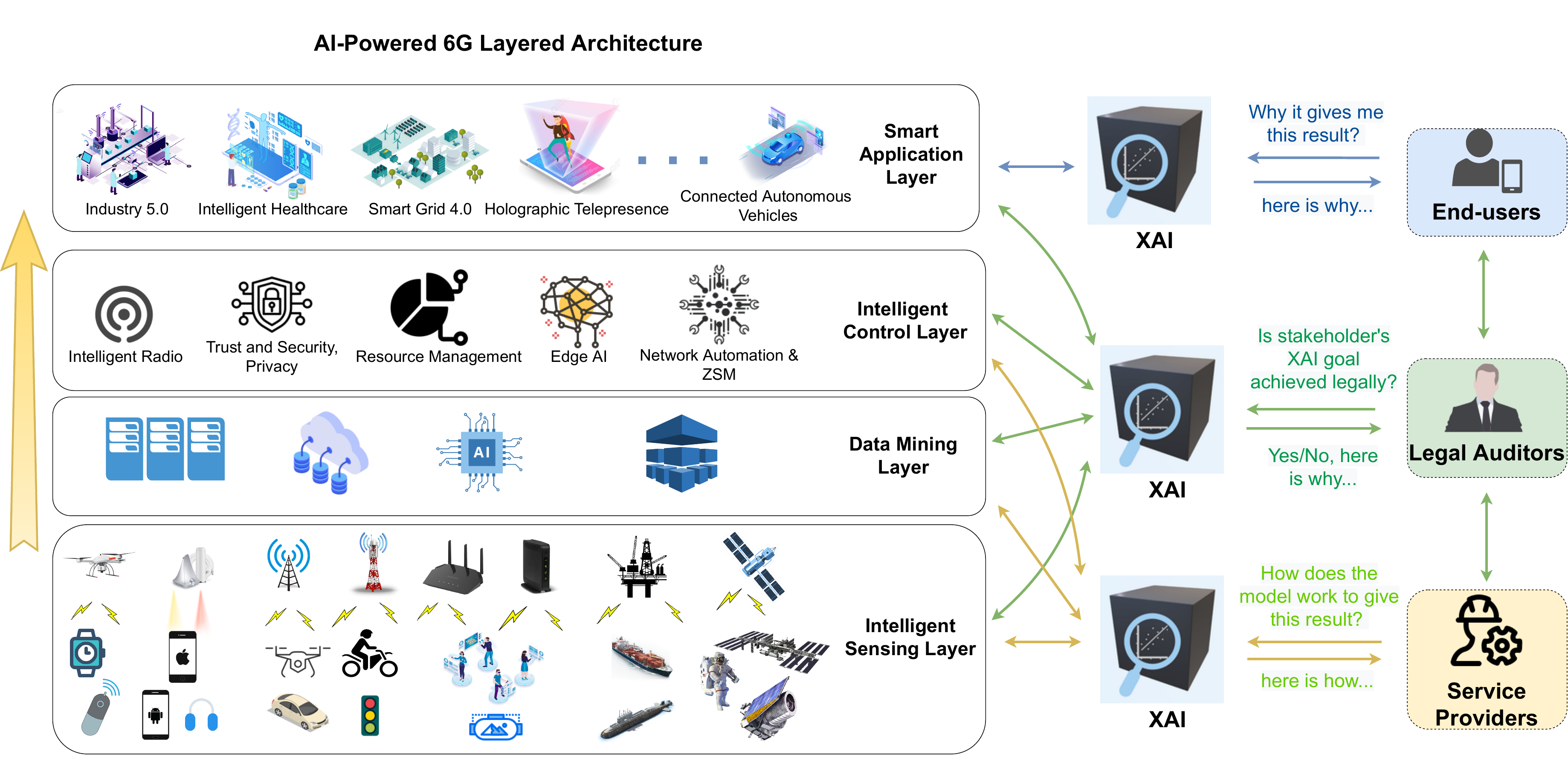} 
    \caption{An illustration of the benefits (i.e., question-and-answer interactions) of introducing XAI to three typical stakeholders (i.e., end-users, legal auditors, and service providers) across all four layers \cite{yang2020ai6Gnetwork} of AI-powered 6G network. Note that 6G technical aspects discussed in Section \ref{sec:tech_spec} are illustrated in the intelligent control layer, while some typical 6G use cases discussed in Section \ref{sec:use_cases} are illustrated at the smart application layer. XAI is built on top of AI so can be deployed on any of these four layers according to specific scenarios and stakeholders.}
    \label{fig:xai_6g}
\end{figure*}

\subsection{Role of XAI for AI-powered 6G} 


XAI is a promising set of technologies that increases the AI black-box models' transparency to explain why certain decisions were made. Especially the high-stakes ones that are made for 6G stakeholders, such as service providers, end-users, and legal auditors. XAI is the key to implementing the ``human-centric" AI-powered 6G network. 

Fig. \ref{fig:xai_6g} shows that AI is penetrated in all four layers of the AI-powered 6G network architecture that is proposed in \cite{yang2020ai6Gnetwork}. From the bottom-up, the first layer is the \textit{intelligent sensing layer}. It is designed to collect data using various sensors (e.g., phones, watches, drones, or vehicles) in multiple scenarios (e.g., space, sea, road, sky, or factory). AI technology can facilitate massive data collection to be a real-time, robust, and scalable process. For instance, this could be done by smartly utilising the scarce spectrum resources and automatically reporting unreliable data events such as broken sensors. XAI can ensure that the whole process works as expected by providing additional information regarding the AI black-box model. For example, legal auditors may use XAI to check for any privacy violations in the whole data collection process (e.g., using LIME to check if the most important features leading to a given decision contain private information such as race, gender, or nationality.), which empowers the downstream AI system. 

The second layer of the AI-powered 6G architecture is the \textit{data mining layer}. Due to the broad and diverse coverage of 6G networks, a massive amount of data will be collected from the intelligent sensing layer with a stringent latency requirement. Therefore, the objective of the data mining layer is to perform automatic feature engineering tasks, such as dimension reduction techniques, so that only the most relevant part of the data will be kept for the follow-up processing in the layer of \textit{intelligent control}. This third layer will utilise the filtered data for making decisions such as resource allocations and network management to ensure a certain level of system performance that meets the application requirement. For both the data mining and intelligent control layers, XAI is particularly helpful for service providers to diagnose the root cause of incorrect decisions by AI systems. As the top layer of the 6G architecture, the \textit{smart application layer} interacts with the end-users who are not the technical experts in various scenarios. For example, in the autonomous driving use case, when the AI system suggests turning right, XAI will provide more user-friendly information explaining that the right turn will save five minutes of journey time but can have more curvy roads ahead. Instead of executing decisions straight away, XAI will enhance the trust between stakeholders and the AI-powered 6G networks for prescriptive, predictive, diagnostic, and descriptive analytics.

\subsection{Motivation}

Recent research shows the great potential of XAI for computer visions such as medical imaging \cite{kaur2021trustworthy}. However, the challenges when deploying such systems on a large-scale (e.g., upcoming 6G systems) still remain unclear. A huge loss is likely to happen when XAI is not functioning for some 6G use cases as expected. Such malfunctions may have various causes (e.g., biased spectrum resource allocation\cite{guo2020explainable}, inappropriate data collection\cite{grant2020show}, AI model attack\cite{qiu2019review}, etc.) as 6G will be deeply coupled with AI in a ``full-stack" manner. Although there are many surveys on XAI and 6G separately, however, there is a lack of comprehensive survey that jointly explores the potential of XAI for implementing AI-enabled human-centric 6G networks. XAI has attracted the attention of many researchers since the Defense Advanced Research Projects Agency (DARPA) launched its XAI program in 2017 \cite{gunning2019darpa}. Das and Rad \cite{das2020opportunities} compared and analysed commonly used XAI techniques in terms of their algorithmic mechanisms, taxonomies, and successful applications. Their paper proposed several promising future directions and challenges for XAI. However, the great potential of XAI in realising the ``human-centric" 6G network is missing in existing XAI surveys. Saad et al. \cite{Saad2020privacy} have broadly described the vision of 6G, which is far beyond utilising more spectrum by including more technological trends and driving applications. Morocho-Cayamcela et al. \cite{morocho2019machine} focus on the applications of AI at each main aspect in implementing B5G/6G, ranging from wireless communications to e-health. It also mentioned the trade-off between interpretability and AI algorithms' accuracy but did not extend the discussion on XAI to enhance trust in using 6G cellular systems. Porambage et al. \cite{porambage2021roadmap} review the recent progress of 6G in security and privacy areas. These areas will likely have many high-stakes decisions by AI systems. However, their contribution lacks discussion of the importance and challenges of XAI for managing the risk of such high-stakes decisions.



\begin{table*}[htbp]
\caption{Summary of important surveys on XAI and 6G}
\label{tab:surveys}
\renewcommand{\arraystretch}{1}
  \begin{tabular}{| p{0.5cm}|p{0.33cm}|p{0.33cm}|p{0.33cm}|p{0.33cm}|p{0.33cm}|p{12cm}|}
  \hline
      \rowcolor{Gray}
    	\multicolumn{1}{|c|}{\textbf{Ref.}} 
         & \multicolumn{1}{|c|}{\textbf{{\rotatebox[origin=c]{90}{AI}}}}
         & \multicolumn{1}{|c|}{\textbf{{\rotatebox[origin=c]{90}{XAI}}}}
         & \multicolumn{1}{|c|}{\textbf{{\rotatebox[origin=c]{90}{B5G/6G Technical Aspects}}}}
         & \multicolumn{1}{|c|}{\textbf{{\rotatebox[origin=c]{90}{B5G/6G Use cases}}}}
         & \multicolumn{1}{|c|}{\textbf{{\rotatebox[origin=c]{90}{Future Challenges}}}}
         & \multicolumn{1}{|c|}{\textbf{Remarks}}\\ [12ex]
    \hline
    \hline
        \multicolumn{1}{|c|}{\cite{guo2020explainable}} & \cellcolor{yellow!30} M & \cellcolor{green!30} H & \cellcolor{green!30} H & \cellcolor{red!30} L & \cellcolor{yellow!30} M & A concise survey on the potentials of using XAI for 6G wireless enabling technologies at the physical and MAC layer.\\[2ex]
    \hline
        \multicolumn{1}{|c|}{\cite{das2020opportunities}} & \cellcolor{yellow!30} M  & \cellcolor{green!30} H & \cellcolor{red!30} L & \cellcolor{red!30} L & \cellcolor{yellow!30} M & A comprehensive survey on the XAI covering commonly used algorithms and applications.\\[2ex]
    \hline
        \multicolumn{1}{|c|}{\cite{morocho2019machine}} &  \cellcolor{green!30} H & \cellcolor{yellow!30} M & \cellcolor{green!30} H & \cellcolor{yellow!30} M & \cellcolor{yellow!30} M & A comprehensive survey mainly focus on the application of machine learning for the future 6G/B5G network.\\ [2ex]
    \hline
        \multicolumn{1}{|c|}{\cite{porambage2021roadmap}} &  \cellcolor{yellow!30} M & \cellcolor{red!30} L & \cellcolor{green!30} H & \cellcolor{green!30} H & \cellcolor{yellow!30} M & A comprehensive survey on the security aspect of 6G using AI solutions \\[2ex]
    \hline
        \multicolumn{1}{|c|}{\cite{saad2019vision}}  & \cellcolor{red!30} L & \cellcolor{red!30} L & \cellcolor{green!30} H & \cellcolor{green!30} H & \cellcolor{green!30} H & A comprehensive survey of 6G visions, technical challenges, use cases, and open research directions.\\[2ex]
    \hline
        \multicolumn{1}{|c|}{\cite{li2020trustworthy}} & \cellcolor{yellow!30} M & \cellcolor{red!30} L  & \cellcolor{yellow!30} M & \cellcolor{yellow!30} M & \cellcolor{yellow!30} M & A concise paper on measuring the explainability of deep learning applications in the 6G network automation.\\[2ex]
    \hline
        \multicolumn{1}{|c|}{\textbf{This paper}} & \cellcolor{green!30} \textbf{H} & \cellcolor{green!30} \textbf{H} & \cellcolor{green!30} \textbf{H} & \cellcolor{green!30} \textbf{H} & \cellcolor{green!30} \textbf{H} & \textbf{A comprehensive survey of using XAI for trustworthy and transparent 6G including use cases, requirements/vision, technical aspects, projects, research work, standardization approaches and future research directions.}\\[2ex]
    \hline

  \end{tabular}
  
\begin{flushleft}
    
\begin{tikzpicture}
\node (rect) at (6,2) [draw,thick,minimum width=0.5cm,minimum height=0.35cm, fill= red!30, label=0:Low Coverage (The paper never or rarely discusses the certain topic. )] {L};
\node (rect) at (6,2.5) [draw,thick,minimum width=0.5cm,minimum height=0.35cm, fill= yellow!30, label=0:Medium Coverage (The paper discusses the certain topic but not in a great technical detail or missed some important subtopics.)] {M};
\node (rect) at (6,3) [draw,thick,minimum width=0.5cm,minimum height=0.35cm, fill= green!30, label=0:High Coverage (The paper discusses the certain topic in a great technical detail with all important sub-topics covered. )] {H};
\end{tikzpicture}

\end{flushleft}
  
\end{table*}

Guo \cite{guo2020explainable} carefully discussed the potential of XAI in the key enabling technologies, such as radio resource management, for 6G at the physical layer and the MAC layer. Besides, it also proposed some initial plans for measuring the level of explainability, which was later formalised as the quality of trust (QoT) in \cite{li2020trustworthy} to the users of 6G networks, especially for deep-learning based 6G autonomy. Their paper lacked broader discussions on 6G, especially about the new use cases and the technical aspects that need XAI to uncover the myth of the decision-making process. As mentioned in earlier subsections, there is a high necessity of introducing XAI into AI-powered 6G. Therefore, a comprehensive survey of the state-of-the-art XAI and its potential in building the future 6G networks with a holistic view will be helpful to guide the researchers and practitioners. 

In Table \ref{tab:surveys}, we concisely summarise the comparison of important related survey papers. The gap in existing surveys is highlighted, which is the lack of comprehensive analysis of XAI for developing a trustworthy, responsible, and transparent AI-powered 6G network.

\subsection{Our Contributions}
The main contributions of this paper are summarised as follows:
\begin{itemize}
\item \textbf{Bridging the gap between XAI and 6G}. Most of the existing surveys in XAI \cite{das2020opportunities, adadi2018peeking, arrieta2020explainable} are about pure AI applications such as natural language processing (NLP) and computer vision (CV). The discussions of XAI to 6G, which is the enabling infrastructure of future AI applications, are unfortunately missing. Similarly, many recent surveys in 6G \cite{Saad2020privacy, morocho2019machine, porambage2021roadmap} attempt to cover all possible enabling technologies and applications extensively, without a particular focus on interactions between human and 6G networks, where XAI can play an important role. This survey paper bridges this gap by comprehensively overviewing both XAI and 6G and their connections. 
    \item \textbf{A comprehensive survey of XAI to all key aspects of 6G}.  Compared to the previous surveys that briefly mentioned XAI impact on 6G\cite{guo2020explainable, li2020trustworthy}, this paper extends the scope of 6G in which XAI can help. Specifically, this paper goes beyond the smart radio resource management at the physical and MAC layers. Every key 6G technical aspect (e.g., network automation, security, and privacy) and 6G use cases (e.g., industry 5.0, Smart Grid 4.0, Metaverse, and Holographic communication) are examined to investigate how XAI can help in enhancing the transparency and trustworthiness of all 6G stakeholders. The relevant 6G and XAI standards, legal framework, and research projects are also reviewed. Moreover, this paper discusses several implementation challenges and possible solutions in applying XAI to 6G.
\end{itemize}

\subsection{Paper Outline}
\begin{figure}[htb]
    \centering
    \includegraphics[width=0.5\textwidth]{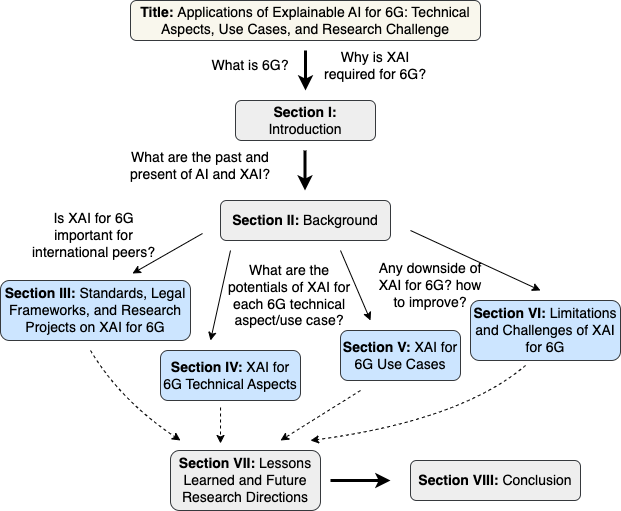} 
    \caption{Structure and relationships between the sections of the paper.}
    \label{fig:paper_structure}
\end{figure}

As shown in Fig. \ref{fig:paper_structure}, the organisation of this paper is described as follows: the introduction section highlights the motivation and the overall contribution of this paper, which is followed by the second section that briefly introduces AI and XAI in terms of their history, technology evolution, popular algorithms and their applications, and the trend of being applied to 6G areas. In Section \ref{sec:tech_spec}, for each of the six main technical aspects of 6G, namely: intelligent radio, trust and security, privacy, resource management, edge network, and network automation, we introduce its motivations, technical requirements, and challenges, and discuss how XAI can improve the level of its trustworthiness. Similarly, Section \ref{sec:use_cases} discusses each of the six typical 6G use cases, with a particular emphasis on how XAI can help in advancing some of their technical limitations in the 6G age. To demonstrate the importance of our work in the convergence of XAI and 6G, in Section \ref{sec:standards}, we list several standards, legal frameworks, and ongoing important research projects world-widely about 6G and XAI. Same as many other new technologies, XAI also has its limitations which are discussed in Section \ref{sec:xai_6g_lim}, along with its corresponding challenges in the future. Section \ref{sec:future} summarises the learned lessons and future research directions for Section \ref{sec:bg}, \ref{sec:tech_spec}, \ref{sec:use_cases}, and \ref{sec:xai_6g_lim}. Finally, we conclude the paper in the last section.

\section{Background} \label{sec:bg}
This section overviews the background of XAI that forms the prerequisite for understanding the potential of XAI for 6G. We overview the motivation of XAI, concepts relevant to XAI, the taxonomy of XAI and algorithms and applications, XAI stakeholders in 6G, and the 6G deployment of XAI with a case study for smart transport.

\subsection{Motivations of XAI} 

The main problem with the existing AI algorithms, especially the most accurate ones, is the black-box models. Their high internal complexity justifies the recent interest in XAI to develop new methods to illustrate how ML models work. XAI will also make users adapt and trust ML and incorporate them in their work \cite{gunning2019darpa, arrieta2020explainable}. 

The relationship between XAI and AI is described as follows. AI can be an independent technology, but XAI cannot exist without AI as it is designed for explaining the decisions made by AI. Normally, AI does not always need XAI as not all decisions need to be inspected for debugging systems or for legal purposes. However, when explanations are highly required (e.g., high-stake decisions in autonomous driving), XAI can be integrated with all major stages of AI models including data collection (e.g., feature engineering), model training (e.g., a self-explanatory model such as decision tree), and model deployment (e.g., post-hoc XAI model such as LIME). XAI can provide various types of explanations such as text, visuals, rules, linear model weights, feature importance, etc. The XAI developers will work closely with XAI stakeholders to determine the specific explanation type that makes the most sense for given AI decisions.

\subsection{Concepts relevant to XAI}\label{subsec:explain_and_inter}
We discuss some popular concepts relevant to XAI below:
\begin{itemize}
    \item \textit{Interpretability}: Interpretability is basically how easy the AI/ML model \textit{per se} is to be understood. This property purely depends on how the model is inherently designed. For example, using a simple rule-based model like a decision tree that a human can easily understand would be an example of having good interpretability. 
    \item \textit{Explainability}: Explainability is the ability to provide extra details to clarify the internal functioning of an AI/ML model for a given audience. It is a characteristic to provide information in the form of statements, that clarify, give context, or justify a particular prediction to a given audience. An example of explainability is to create a complex model but capable of providing statements of why it has taken a particular decision. Note that the target audience is key for explainability since the type of explanations and how easy something is to understand largely depend on the person who is getting the information. What for a group of doctors may seem obvious, to a layperson can be completely incomprehensible.

\item \textit{Interpretability v.s. Explainability}: Explainability requires an extra layer of skill to provide customised knowledge that a specific user can understand, whereas interpretability focuses on the essence of the model internally. Some researchers claim that Interpretable ML (better interpretability) rather than XAI (better explainability) is the preferable option \cite{rudin2021interpretable,rudin2019stop}. They argue that using models to explain black-box models may lead to errors. However, XAI and interpretable ML do not exclude each other, and sometimes one approach would be more suitable than the other. As a suite of algorithms, XAI clarifies and simplifies the internal logic of black boxes and is a great instrument to know whether or not they can be trusted \cite{ribeiro2016should}. Note that ``in-model" explainability (refer to the later subsection C) is basically in the use of interpretable ML models either alone or in conjunction with black box models.
    \item \textit{Accountability}: Accountability is essential in data protection; thus, it covers many disciplines such as finance and accounting. Therefore, institutions have become more accountable for restoring trust in financial institutions due to high-level scandals in the 90s. Accountability refers to making companies and individuals responsible for their actions. For example, if an accountant does not detect an apparent anomaly in a company, the accountant can be held and face legal consequences. Furthermore, companies had to be examined by external auditors periodically. For XAI in particular, accountability is related to providing explanations to a given audience to justify certain actions for which someone is responsible. If an algorithm discriminates against a person, that person deserves an explanation to serve justice. To provide such justification, the people in charge of the system need to be accountable, implying companies and organisations have to make an effort to align their technology with the principles of the different regulations such as GDPR \cite{vedder2017accountability}.
    \item \textit{Trustworthiness}: We cannot expect professionals to have blind faith in models’ predictions of the models if we do not provide clear and solid explanations of why decisions are made. Researchers have been focused on making models as accurate as possible for years. But they did not realise the importance of explainability until they realised that professionals were reluctant to use them unless clear explanations were provided\cite{kentour2021analysis}. If we want professionals to trust ML models in their daily work to make decisions, they need people to feel confident towards them \cite{ribeiro2016should}. Trustworthiness is the ability ML models need to have to make individuals feel confident in their decisions and rely on them. This is mainly achieved by giving evidence and logic on why a prediction has been taken \cite{mueller2019explanation}.
    \item \textit{Causality}: Many of the latest ML methods like Deep Learning rely on more and more data to be trained to find correlations and patterns with all the possible causes. However, we have to distinguish between correlation and causation. Models based on correlation lack the ability to generalize and adapt to new scenarios \cite{arrieta2020explainable}. For example, a model trained with a certain camera struggles with pictures from a different camera. We need models that can detect cause-effect relations and adapt to environmental changes. Examples of cause-effect relations are when a thing breaks when it is hit or falls from an edge of the table. Developing models able to understand the underlying logic between different events will allow the creation of better and more satisfactory explanations \cite{scholkopf2022causality}.
    \item \textit{Confidence}: Some ML models have a confidence score that is generally a value between 0 and 1, indicating the probability that a decision is correct. Laypeople’s confidence in a model depends on the overall performance of the model and also on the confidence score of individual predictions \cite{rechkemmer2022confidence,biran2017explanation}. If we say that the model has an accuracy of 98\% and that the predicted decision is correct with a 0.95 confidence score it will make users more prone to trust the model. But even giving users low confidence scores can be very helpful because they will know that if the model is not sure and they do not have to take that decision so seriously, which otherwise will be misleading. Giving users certainty on how accurate predictions are, increases their trust in them.
    \item \textit{Fairness and ethical decision}: Algorithms are used in decisions that affect people’s lives, therefore we need to guarantee that the decisions taken by ML models do not discriminate in terms of protected attributes such as gender, ethnicity, race, or sexual orientation \cite{arrieta2020explainable}. In certain fields like banking or criminal justice, historical records can perpetuate some bias in the future. According to previous publications \cite{corbett2018measure}, to maintain equity across different groups three norms can be applied: a) anti-classification: the protected attributes are not considered to make decisions; b) classification parity: errors (false positives and negatives) are constant across the different groups; and c) calibration: the percentage of the decisions of the models such as granting a loan should be the same across the different groups.

\end{itemize}

\subsection{Taxonomy of XAI algorithms and applications}

\renewcommand{\aboverulesep}{0pt}
\renewcommand{\belowrulesep}{0pt}
\begin{table*}[htbp]
    \centering
    \caption{Typical XAI algorithms with their types and AI algorithms to explain.}
    \rowcolors{2}{gray!15}{white}
    \begin{tabular}{p{2.5cm} | p{2.5cm} p{2.5cm} p{2cm} p{1.5cm} | p{2.5cm}}
        \rowcolor{gray!50}
        \toprule
        \multirow{2}{*}{XAI algorithm} & \multicolumn{4}{c|}{XAI Taxonomy}  & AI/ML algorithms to explain\\ \cline{2-6}
        \rowcolor{gray!50}
         & Model Specific / Agnostic & Local / Global & Surrogate / Visual /Textual & Pre / In / Post model & Supervised / Unsupervised / DL / RL   \\
         \midrule
          Facets &  Agnostic & Both & Visual  & Pre  & Unsupervised \\
          
         LIME &  Agnostic & Local  & Surrogate  & Post  & Supervised\\
         
         SHAP & Agnostic  & Local but also global  & Surrogate  &  Post & Supervised\\
         
         Counterfactual & Both  & Local but also global  & Surrogate  & Post & Supervised\\
         
         LRP & Specific  & Local  & Visual  & Post & DL\\
         
         Knowledge Graphs & Agnostic  & Global and Local  & Textual  & Post & DL\\
         
         Bayesian N. & Specific  & Local  & Visual / Textual  & In & Supervised\\
         
         PIRL & Specific  & Global  & Surrogate  & In & RL\\
         
         Hierarchical Policies & Specific  & Local  & Surrogate  & In & RL\\
         
         Struc Causal Model & Specific  & Local  & Both  & Post & RL\\ 
         
         Bayesian RL & Specific and agnostic  & Local  & Visual / Textual  & In/Post & RL\\ 
         
      \bottomrule
    \end{tabular}
    \label{tab:xai_ai}
\end{table*}

Based on different criteria, the XAI methods can be classified into several arrangements\cite{stiglic2020interpretability, arya2019one, vilone2020explainable} which can be overlapping or otherwise. In this section, we discuss the taxonomy of XAI inspired by \cite{singh2020explainable} as well as the popular XAI methods.

\paragraph{Model-Agnostic vs Model-Specific}
Model-agnostic methods are the ones that do not consider the internal components of the model (i.e., model weight and structural parameters). Therefore, they can be applied to any black-box approach. In contrast, model-specific methods are defined using parameters of the individual model, such as interpreting weights of linear regression or using inferred rules from a decision tree that would be specific to the trained model \cite{arras2019evaluating}. There are some advantages of model-agnostic methods \cite{ribeiro2016model} such as high flexibility for developers to choose any ML model for generating interpretation which is different from the actual black-box model that generates decisions.

\paragraph{Local vs Global}
Based on the scope of explanations, provided methods can be classified into two classes: local and global methods. Local interpretable methods use a single outcome, or particular prediction or classification results of the model \cite{ribeiro2016should} to generate explanations. In contrast, the global interpretable methods use the entire inferential ability of the model or overall model behaviour \cite{samek2019towards} to generate explanations. In the local interpretable methods, only specific features and characteristics are essential. For the global methods, feature importance can be used to explain the general behaviour of the model.

\paragraph{Surrogate vs Visual Aid}
A popular way to explain a black-box model is to apply an interpretable approximate model that replaces the black-box model for explaining decisions. This interpretable approximate model is called the \textit{surrogate model}, which is trained to approximate the predictions of a black box and is later used to draw explanations interpreting the decisions from the black-box model. An example of a black-box model can be a deep neural network (DNN), whereas any interpretable model can be a surrogate such as decision trees or linear models. Besides surrogate models, visual explanations aid to generate explanations in a more presentable way showing the inner working of many model-agnostic (e.g., which pixels or area of pixel help differentiate between a cat and a dog, or which words help decide if the document is spam or not). The visual aids can be graphs, scatter plots, heatmaps, etc.

\paragraph{Pre-model, In-model vs Post-model Strategies}
XAI can be applied throughout the entire developmental pipeline of the model. The goal of \textit{pre-modelling} explainability is to describe the dataset to gain better insights into the dataset used to build a model. The general objectives of the pre-model are to perform data summarisation, dataset description, explainable feature engineering, and exploratory data analysis. Google Facets\footnote{https://pair-code.github.io/facets/} is an example of pre-model explanations that enable the learning of patterns from large amounts of data. 
In contrast, the goal of \textit{in-model} explainability is to develop inherently explainable models instead of generating black-box models. Methodologically, there are different strategies or ways to construct in-model explanations. The most obvious is to adopt an inherently explainable model, such as linear models, decision trees, and rule sets. However, some efforts are needed to generate explanations using these methods, like picking important features. Other approaches are proposed beyond inherently explainable models, such as hybrid models, joint prediction and explanation, and explainability through architectural adjustments. In the hybrid approach, complex black-box methods are coupled with inherently explainable models to devise a high-performance and explainable model, such as combining a deeply hidden layer of neural network with a KNN model \cite{papernot2018deep}. Also, the model can be trained to provide a prediction and the corresponding explanation jointly \cite{hind2019ted}. The idea here is to produce a training dataset, where the decision is supplemented with the user’s rationale for the decision. Lastly, explanations through architecture adjustments focus on deep network architecture to enhance explainability, such as pushing higher layer filters to represent an object part, as opposed to a mixture of patterns \cite {zhang2018interpretable}. These in-model XAI approaches have two main shortcomings: first, they assume the availability of explanations in the training dataset, which is often not the case. Second, explanations generated by these methods are not necessarily reflective of how model predictions were made, but rather what humans would like to see as an explanation. 
The \textit{post-model} explainability method extracts explanations that are inherently not explainable to describe a pre-developed model. These popular post-hoc XAI methods generally operate over four key characteristics: the target, what is to be explained concerning the model; the drivers, what is causing the decision to be explained; the explanation family, and how an explanation is going to be presented to a user; and the estimator, the computational process generating the explanation \cite{arrieta2020explainable}. 

The following popular XAI techniques are summarised in Table \ref{tab:xai_ai} in terms of their corresponding XAI type under different taxonomies with the types of AI algorithms it can explain.

Among model-agnostic XAI strategies, \textit{LIME} \cite{ribeiro2016should} is one of the popular post-hoc algorithms that explain an instance prediction: target, input features as drivers, importance scores as explanation family, and computed through local perturbations of the model input as the estimator. LIME implements a linear interpretable model, a surrogate model,  in the local area as a local approximation to explain a prediction. Because of the local approximation, LIME works on all black-box models and data types (i.e., text, tabular data, images, and graphs). 

\textit{SHAP} \cite{lundberg2017unified} fundamentally differs from LIME in generating feature importance using Shapley values \cite{shapley1953} rather than using a family of interpretable models to estimate the behavior of a more complex model. Shapley's value comes from cooperative game theory which is used for estimating marginal contribution. Therefore, SHAP generally performs better than LIME and it can also be used to explain the global behaviour of a model, rather than a single given instance. In general, SHAP is model agnostic. There are also model-specific versions of SHAP to speed up the performance. For example, TreeSHAP \cite{lundberg2018consistent} and Deep SHAP \cite{lundberg2017unified} are designed for decision trees and DNN respectively. 

\textit{Counterfactual} is another algorithm that is available for both model-agnostics \cite{sharma2019certifai} and model-specific \cite{wachter2017counterfactual} variants. Counterfactual builds on explaining the prediction of the predictor algorithm by finding the slightest change in the input feature values causing the change in the original prediction. For instance, if changing the BMI of the person has resulted in flipping the original prediction from illness to being healthy, then using the BMI value is an indicative explanation for correlating with the original prediction, thereby counterfactual human-friendly explanations. The explanation is straightforward but with multiple possible explanations. Thus, it becomes tricky to know which is suitable, the simplest or the most complex (i.e., the combination of several features).

\textit{Layerwise Relevance Propagation (LRP)} \cite{montavon2019layer} is an algorithm designed to explain a DNN with an assumption that a classifier can be decomposed into different layers, making it a model-specific method. LRP is designed with the intuition that certain layers of inputs are relevant for the prediction. Furthermore, to gain insights into which neurons are significant, activation scores of each neuron are considered through back-pass and eventually learn about the input data. It is generally applied to image data so as to highlight meaningful pixels that enable a certain prediction.

\textit{Bayesian Networks (BN)} are a well-known class of probabilistic models \cite{ben2008bayesian}. BNs are based on directed acyclic graphs that compute random variables (nodes) and their relationships (edges) to predict the probability of certain events related to those variables \cite{ben2008bayesian}. BNs use bayesian inference (causality) to estimate such likelihood. Each node of the graph has a probability distribution P$(X_i|$Parent$(X_i)$) which represents the conditional probability with respect to the parent of that node. The main advantage of BNs is that the graph and the relationships between the variables are interpretable. The predicting reasoning is computed by following the direction of the edges in the graph and the distributions can be visualized. Therefore, BNs are a handy tool for understanding probability distributions, knowledge discovery, and detecting anomalies.

\textit{Knowledge Graphs (KGs)} represent domains using entities (countries, people, places...) and relationships (part of, has a, located in...). By doing so, they allow combining knowledge from multiple sources and adding context to the different entities. KGs are able to represent companies, organisations, or areas of expertise. KGs have been used for a very long time but it has been in the last decades with the importance of acquiring semi-structured knowledge, especially on the web, when they have become popular. The term KG was coined by Google in 2012 \cite{paulheim2017knowledge} and it allows us to respond to users' queries based on concepts and their relationships which is more effective than just matching keywords with text \cite{paulheim2017knowledge}. This is called the semantic web and some examples of public KGs are YAGO, Wikidata, or DBpedia \cite{paulheim2017knowledge,tiddi2022knowledge}. KGs need to grow in size to cover more information but this makes them more likely to contain errors or inaccurate information. Therefore, size and correctness need to be balanced out. KG can be constructed in an automatic fashion using ML techniques and they can be explored later on to extract useful information. KGs provide structure knowledge as ontologies that facilitate interpreting data in the form of horn clauses (logical formulas to express rules) \cite{russ2011knowledge} from neural networks. KG can be integrated with ML techniques and they also have been proven to be very useful tools for XAI \cite{tiddi2022knowledge}. For example, KGs have been integrated with DL to give users explanations about a news recommendation system \cite{paulheim2017knowledge}. Another example of KG and ML is Dedalo, which is a framework that uses inductive logic programming to extract information from linked data that combines datasets from different areas to give explanations about the created clusters by an algorithm \cite{tiddi2014dedalo}.



Explainable Reinforcement Learning (XRL) is a promising but challenging research branch of XAI \cite{heuillet2021explainability} as the RL model often contains an enormous number of states and actions with a complex reward system. Nevertheless, XRL could accelerate the process of RL design by facilitating developers debugging RL systems. A. Heuillet et al \cite{heuillet2021explainability} presents a rigorous classification of XRL methods based on types of explanations (text or images), on the level of explanation (local if it was for predictions or global if the whole model was explained), and also which algorithm was explained. Specifically, we present the following three recent XRL algorithms:

\begin{itemize}
    \item Programmatically Interpretable Reinforcement Learning (PIRL) is an example of a Global-Intrinsic method \cite{verma2018programmatically}. The idea of PIRL is to use a programming language much closer to the human way of thinking to emulate the behaviour of the DRL model. PIRL uses a framework called Neurally Directed Program Search (NDPS) to learn the behaviour of the DRL model by imitation learning. Thus there are two steps, the construction of the DRL and the extraction of knowledge of the DRL model to create a sequence of instructions. The PIRL instructions are not as accurate as those of the neural network but they can be quite close and much more understandable. The PIRL model has been successfully applied to The Open Racing Car Simulator (TORCS) \cite{wymann2000torcs}.

\item Hierarchical and interpretable skill acquisition in multi-task RL \cite{shu2017hierarchical} is an example of the intrinsic-local category. This approach consists of presenting a policy with high-level actions as a sequence of simpler actions because they are more familiar to humans.  This approach was used to play the game Minecraft, and it implements a hierarchical policy based on two layers using the actor-critic algorithm. It was used to explain a multi-task RL model playing Minecraft. The model also uses a stochastic temporal grammar model to capture the relationships between the actions to create a hierarchical policy. Humans just say to park the car, instead of defining all the actions related to the steering wheel, clutch, accelerator, and brake. In the same way, before you move an object, you need to find it.  This method presents high-level instructions such as “Stack blue”. Where “Stack blue” is composed of “Find/Get/Put/ Blue”, and, at the same time “Find Blue” is composed of multiple less low-level “Go left, Move forward, Turn right...”.

\item Structural Causal Models (SCM) are a very clear way to represent causal-effect relationships of events. In this case, P. Madumal et al. \cite{madumal2020explainable} proposed a framework, that falls in the post-hoc local category, based on SCM to explain the behaviour of model-free RL agents. They use a directed acyclic graph (DAG) in which the nodes represent the states and the edges of the actions. By traversing the graph it can be observed which actions take place from one state to the other. The process has three main steps: creating the DAG; using multivariate regression models to approximate the relationships using the minimum number of variables; and generating the explanations by analyzing the variables of the DAG to respond to the questions: “Why action A?” and “Why not action B?”. The latter question is answered by simulating the counterfactual in the DAG. In their research, they created a model based on casual structures to evaluate six domains using six different RL algorithms to play Starcraft II. They also conducted a study in which a group of 120 people evaluated the quality of explanations, and they found that the people in the study preferred their explanations based on casual models to other baselines.

\item Bayesian Reinforcement Learning (BRL) is an alternative to Neural Networks applied to RL and based on Bayesian Networks (BN) \cite{ghavamzadeh2015bayesian,vlassis2012bayesian}. BRL performs the learning process by adding information using the bayesian inference to calculate the bayesian distribution \cite{ghavamzadeh2015bayesian}. The probability of a particular action of the agent is calculated by applying rules based on previous probabilities. BRL deals with the trade-off between exploration and exploitation using the Bayesian posterior which recalculates the probabilities once the previous values are updated \cite{ghavamzadeh2015bayesian,vlassis2012bayesian}. BRLs improve the regularization of the models because since we can include prior knowledge the outlier points are less likely to cause overfitting in the model and make it less likely to adapt to new scenarios. Additionally, incorporating previous knowledge accelerates converging to optimal solutions. Lastly, BRL is able to handle uncertainty in a better way since values can model the distribution explicitly \cite{ghavamzadeh2015bayesian}. The downsides of BRL are that they are more difficult to model because they need the right configuration for representing previous information and also they require more computation for processing the sequential decision-making results \cite{ghavamzadeh2015bayesian}.
\end{itemize}

\subsection{Types of explanations by XAI for 6G}
The most commonly seen type of explanation generated by XAI is the feature importance, which ranks the input data features according to their corresponding contributions to the final output. For example, \cite{ayoub2022using} uses LIME and SHAP to identify the failure in microwave networks from the set of features of link characteristics, G.828 metrics, and power values. The rules of the decision tree-based AI models are also a typical explanation as long as they are not too deep for the stakeholders to understand. For instance, a 6-layer DT is used for enhancing trust management for network intrusion detection systems. Saliency maps are well-known explanations in computer vision tasks which comprise the highlighted parts of images that lead to the AI output. We believe it will have great potential in analysing time-series network data such as anomaly detection by highlighting only the key interval that contributes to the prediction results the most. Counterfactual explanations can also be used in future 6G network root cause analysis by telling the engineer which and by how much the features need to be changed to fix an unhealthy service.







\subsection{XAI stakeholders in 6G}
Nearly every sector has a demand for automated algorithmic decision-making, and this demand is evolving into supplementing decisions with explanations generated by the XAI model. With the upcoming 6G making internet bandwidth faster and available to almost every other device, the demand for AI will be enhanced by XAI within the ecosystem. However, the question remains, who needs XAI, and at what level of explanation sounds reasonable? Also, it is vital to note that different stakeholders have different expectations from the explanations \cite{ribera2019can}, and based on the user requirement of XAI \cite{wick1989reconstructive}, stakeholders' demands can be classified broadly into three categories. 

\renewcommand{\aboverulesep}{0pt}
\renewcommand{\belowrulesep}{0pt}
\begin{table}[h]
    \centering
    \caption{XAI requirements for key 6G stakeholders.}
\label{tab:stakeholders}
\rowcolors{2}{gray!15}{white}
\begin{tabular}{p{1.4cm} p{2.1cm} p{1.5cm} p{2cm}}
\rowcolor{gray!50}
\toprule
Stakeholders & Examples & XAI Goals & Preferred Explanation \\
\midrule
Service Providers & System designers, data scientists, software developers & To improve 6G AI performance & Logical explanation that can clearly expose the decision-making process \\
End-Users & Consumers, business partners, policymakers & To trust the 6G AI systems & Easy to understand like visual explanation \\
Legal Auditors & Solicitors, auditors & To ensure XAI goals are well justified and met & Clear enough explanation to verify if XAI goals can be met\\
\bottomrule
\end{tabular}
\end{table}

\begin{itemize}
    \item The demand that will be useful for \textit{service providers} to help them identify problems or bugs within the system that produce a decision and improve the performance by troubleshooting the decision-making process. Service providers can be system designers, data scientists, AI/XAI researchers, software developers/testers, etc. 
    \item The demand of the \textit{end-users}, who would be interested in understanding the decision for usage and application \cite{qureshi2019eve} purpose. For the end-user, the interface of explanations is essential, which should explain the decision in the form of a story that the end-user can easily understand \cite{qureshi2017lit}. End-users can be businesses, non-technical people, consumers of technology, and policymakers.
    \item The demand of the \textit{legal auditors}, who would be interested in auditing legal compliance of automated decision-making algorithm. Here, the legal auditors would look for confirmation that ensures compliance, such as no racial discrimination or gender bias while approving loan applications. These stakeholders can be auditors and other legal professionals.
\end{itemize}
We summarised the XAI requirements for different 6G stakeholders in Table \ref{tab:stakeholders}.
\subsection{Deploy XAI on 6G: A Case Study for CAV Collision Avoidance}
\begin{figure}[htb]
    \centering
    \includegraphics[width=0.5\textwidth]{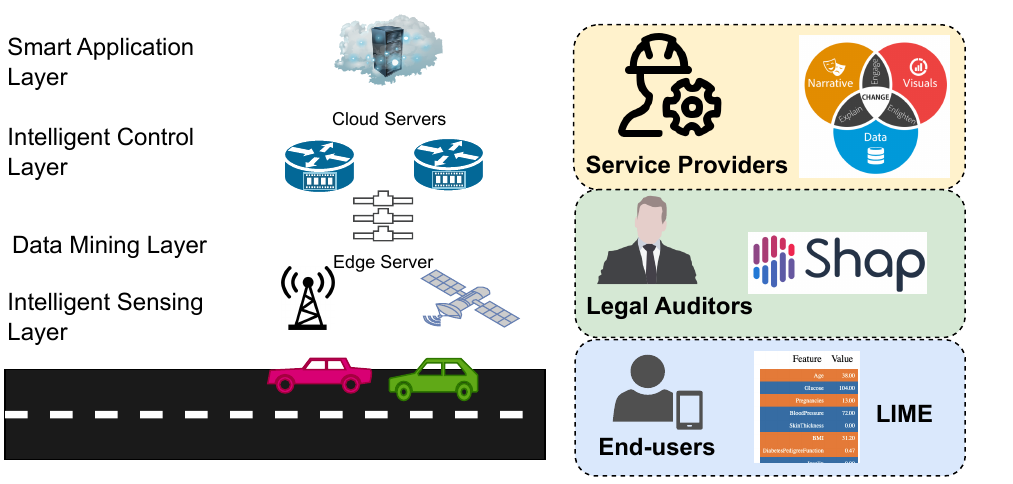} 
    \caption{An illustration of Deploy XAI on 6G: A Case Study for CAV Collision Avoidance. Specifically, the driver will need LIME for generating feature importance explanations quickly to decide whether the suggested acceleration should be taken. Legal auditors need SHAP to generate more reliable explanations to make sure that the AI system behaves as it promised. Lastly, the service provider will need to utilise all kinds of explanations to debug in every stage of the AI pipeline from data collection to decision inference.}
    \label{fig:deploy_xai_6g}
\end{figure}

Recall that Fig. \ref{fig:ai_5g6g} shows AI will be a fully integrated solution of all aspects of 6G rather than a standalone solution solely at the application layer of existing 5G. Accordingly, deploying XAI on 6G networks often requires a cross-layer consideration, as shown in Fig. \ref{fig:xai_6g}, even if it is still designed for a particular application to the end user. In general, to deploy XAI technology on the 6G system, the first step is to identify stakeholders and typical scenarios. This step tries to answer the question of \textit{who / when needs explanations to what AI scenarios}. Normally, as depicted in Fig. \ref{fig:xai_6g}, the end-users expect the XAI solution can be deployed at the higher layer of the 6G AI network. At the same time, service provider demands AI explanation more towards the lower layer to make their AI solutions more robust. In the second step, we deal with a technical system design issue to achieve a trade-off. Specifically, this system trade-off refers to how much model accuracy, system performance, and energy consumption would be compromised to achieve a certain level of interpretability. Normally, the interpretability of AI models increases as the complexity and accuracy decrease. Therefore, higher demand on the interpretability and model performance lead to low energy efficiency due to costly computation and communication resources for running AI and XAI methods. In the layer stack of AI-powered 6G network shown in Fig. \ref{fig:xai_6g}, the end-users usually require higher interpretability as they are not AI 6G experts. Service providers can sacrifice interpretability a bit to guarantee a satisfactory performance level of their 6G system. As the explainability is more or less subjective, the stakeholders' option matters the most to the success of the AI-powered 6G system. Thus, the last step is to make our XAI-6G an ever-evolving system driven by the stakeholders' feedback on the provided explanations. 
As shown in Figure \ref{fig:deploy_xai_6g}, we discuss a case study of deploying XAI in the context of 6G smart transport with fully automated and connected vehicles. We first identify a specific scenario, for example, to avoid an unexpected car crash. We also need to identify the corresponding XAI stakeholders of this scenario, such as passengers (end-users), legal auditors, and 6G smart transport developers (service providers). The goal of embedding the XAI system for this 6G use case is to prevent more car crashes by investigating possible reasons. Drivers need a visual XAI method such as LIME to generate feature importance explanations to quickly decide whether the suggested acceleration should be taken. Legal auditors should use global XAI methods for instance SHAP to generate more reliable explanations to make sure that the AI system behaves as it promised. The developers will need various types of XAI explanations (e.g., visual, numerical, narrative, etc.)at the data mining, smart control, and smart sensing layers to ensure that the AI model or feature engineering needs further improvement. As the AI decisions involved in these lower layer happens more frequently and are diverse due to the technical requirements of 6G, local XAI methods such as LIME can be suitable due to their computation efficiency. Another thing worth noting is that at the 6G AI data sensing layer, the XAI deployment will likely be in a distributed manner as the relevant AI decisions will be triggered at the edge (not the cloud) to respond quickly (e.g., cooperative driving on the highway). Eventually, the 6G smart transport XAI system will improve iteration based on the above-mentioned stakeholders' feedback. The service provider and XAI 6G developers will review this feedback periodically to decide how to upgrade the existing system.

\section{Standards, Legal Frameworks, and Research Projects on XAI for 6G} \label{sec:standards}

This section presents the important standardization activities, legal frameworks, and research projects related to the 6G XAI.

\subsection{Existing XAI Standards for 6G}

IEEE has initiated several standardization activities related to XAI. In 2020, the IEEE Computer Society/Artificial Intelligence Standards Committee (C/AISC)  approved a project to specify an architectural framework for XAI\cite{IEEEWG2}. This framework facilitates the adoption of XAI by defining standards and application guidelines for the following areas:

\begin{itemize}
\item Specify the definition of XAI.
\item Defines the taxonomy of XAI.
\item Specify the application scenarios of XAI.
\item Defines the performance evaluation methods for XAI systems.
\end{itemize}

In 2021, the IEEE Intelligence Society - Standards Committee (CIS/SC) approved a new project to develop the standard for XAI\cite{IEEEWG1}. This group focuses on defining standards on mandatory and optional requirements that need to be satisfied by AI algorithms or systems to be recognized as explainable. 

In addition, the National Institute of Standards and Technology (NIST) has published a report on a set of principles that can be used to judge the explainability of AI decisions\cite{phillips2020four}. This report defines the four principles of XAI:

\begin{itemize}
\item \textit{Explanation:} Ability to provide reasons for the outcomes of the system.
\item\textit{Meaningful:} The provided explanation should be understandable and meaningful to the users. 
\item \textit{Explanation Accuracy:} The provided explanation should accurately describe the process of generating the outcome.
\item \textit{Knowledge Limits:} The system should understand the cases which are not designed or approved to operate or are unable to operate reliably.
\end{itemize}

However, none of these XAI standardization activities is focused on 6G or communication networks. Thus, it is yet to initiate the more focused XAI standardization activities for B5G and 6G domains.

\subsection{Legal framework for explainability}

As explained in Section \ref{sec:bg}, XAI is important for auditors to evolve a legal framework to protect consumer rights under technology usage. Currently, there is no unified law that protects consumer rights for XAI technology. Nevertheless, different regions have started reacting to the evolution of AI and XAI. In the future, with the advancement towards 6G and XAI, we can expect international regulations which are mutually approved. For now, we list the adoption of legal frameworks emanating from different regions of the world concerning user privacy and rights to ensure fairness.

\begin{itemize}
    \item \textit{EU/EEA:} The GDPR\cite{Web-GDPR} is a regulation in EU law on data protection and privacy in the European Union (EU) and the European Economic Area (EEA) and came into effect on 25 May 2018. The GDPR law sets obligations for businesses and grants rights to citizens. Under GDPR, businesses require data protection compliance to ensure data protection concerning users and privacy. Failure to comply can cost up to 20 million euros or 4\% of their global revenue. Under GDPR compliance, users have the ``right to explanation" in algorithmic decision-making \cite{qureshi2019eve}, primarily AI systems. In addition, the regulation protects the fair usage of data collection, processing, and application, while maintaining an up-to-date and accurate reflection of data. Finally, it allows users to demand a copy of their personal data from the business. This regulation comes closest to realising and facilitating XAI goals of transparency and explanation.
    \item \textit{USA}: 
    The US has taken a different approach when it comes to data protection. Instead of having a general data protection regulation, the US has chosen to implement sector-specific privacy and data protection policies that work with state laws to protect American citizens' interests. Some of the key sectors are healthcare under HIPAA\cite{Web-HIPAA}, finance sector and consumer rights under GLBA\cite{Web-GLBA}, federal agencies under FISMA\cite{Web-FISMA}, and protection of Controlled Unclassified Information in an in non-federal information systems and organisations under NIST 800-171\cite{Web-NIST-800-171}. Overall, the US is concerned with data integrity as a commercial asset. In contrast, GDPR gives more to an individual instead of looking at it from the interest of businesses. However, this diversity of legal framework will benefit the adoption of XAI in full spectrum in the 6G world. With this diversity, businesses will communicate with each other through devices and with individuals who would be end-users.
    \item \textit{Rest of the world}: 
    Ethics, consent, user privacy, law, and transparency are now part of global values. Nearly all countries are bringing forward policies and regulations to ensure their understanding of it to ensure governance, including data-driven decision making. China developed a new personal data law (PIPL)\cite{Web-(PIPL} which came  into effect in Nov 2021, drawing its inspiration from GDPR. PIPL tightens how technology giants use data and move private data overseas, with violations resulting in fines up to 5\% of the annual revenue of the previous year or CNY50 million. Similarly, PIPL's articles mention automated decision-making related to finance, health, credit status, and more under fair usage and transparency concerning user rights, similar to GDPR. In Russia, a law concerning personal data was adopted in 2006 under Russian Federal Law No. 152-FZ. Even though the law protects individual rights to a certain degree, it is not as comprehensive as GDPR. Perhaps a revision or expansion of the law can be expected under changing technology and algorithmic decision-making to uphold the spirit of privacy and user protection. Brazil's Lei Geral de Proteçao de Dados (LGPD)\cite{Web-LGPD}, Australia's Privacy Amendment (Notifiable Data Breaches) \cite{Web-AU-NDB}, Japan's Act on Protection of Personal Information\cite{Web-JP-PPI} are steps in similar directions as to those discussed in 2016 for GDPR, which came into effect in 2018.
    
    
    
\end{itemize}

\subsection{Ongoing reputable research projects for 6G using XAI}

\subsubsection{European Union (EU) Funded Projects}

Due to the popularity of XAI topics, several funding organizations have offered funding for XAI-related topics. European Union (EU) is one of such leading funding organizations that has funded several projects in XAI.

Horizon H2020 (H2020) is one of the biggest funding programs supported by the EU. H2020 is a seven-year funding program that operated from 2014–2020 and offered an estimated €80 billion of funding\cite{H2020}. Under direct H2020 funding, several XAI-related projects got funded as listed below. 
\begin{itemize}
\item \textit{AI4EU:} A European AI On Demand Platform and Ecosystem (2019-2021) \cite{AI4EU}.
\item \textit{FeatureCloud} (2019-2024) \cite{FeatureCloud}.
\item \textit{XMANAI:} Explainable Manufacturing Artificial Intelligence (2020-2024) \cite{XMANAI}.
\item \textit{DEEPCUBE:} Explainable AI Pipelines for Big Copernicus Data (2021-2023)\cite{DEEPCUBE}.
\item \textit{SPATIAL:} Security and Privacy Accountable Technology Innovations, Algorithms, and Machine Learning (2021-2024) \cite{SPATIAL}.
\item \textit{STAR:} Safe and Trusted Human Centric Artificial Intelligence in Future Manufacturing Lines (2021-2023) \cite{STAR}.
\end{itemize}

The AI4EU \cite{AI4EU} project is building Europe's first  AI on-demand platform, which will be used to disseminate AI resources developed by other EU-funded projects. The AI4EU project focuses on XAI and the other interconnected AI domains, such as Collaborative AI, Physical AI, Integrative AI, and Verifiable AI. The FeatureCloud project\cite{FeatureCloud} is focusing on designing secure and trusted medical health systems to reduce the impact of cybercrimes and fuel cross-border collaborative data-mining efforts. To realize this objective, the FeatureCloud project integrates XAI with blockchain and federated learning techniques. The XMANAI project\cite{XMANAI} is focusing on the use of XAI for manufacturing to increase trust in AI-based manufacturing processes.

Moreover, the practical utilization of XAI is demonstrated by XMANAI  projects in four industrial plants. The DEEPCUBE project\cite{DEEPCUBE} is focusing on utilizing XAI Pipelines for extensive data analysis. Primarily, it analyses the Copernicus data, which is collected by European Union's  Copernicus Space Programme \cite{Copernicus}. The SPATIAL project\cite{SPATIAL} is focusing on the development of accountable, resilient, and trustworthy AI-based security and privacy solutions for future networks and ICT systems. Thus, the SPATIAL project focuses on using XAI to ensure the security and privacy of 5G and 6G networks. Several B5G and 6G use cases such as healthcare and IoT services are considered in this project. The STAR \cite{STAR} project is studying the use of XAI techniques to increase the transparency of AI-based manufacturing processes and also to improve the user trust level in AI systems.

In addition, H2020 has an element called Marie Skłodowska-Curie actions (MSCAs) \cite{MSCA} which offers grants for all stages of researchers' careers. Under the H2020 MSCA funding, there are two projects for training Early-stage researchers (ESRs) in the domain of XAI applications. 
\begin{itemize}
\item \textit{NL4XAI}: Interactive Natural Language Technology for Explainable Artificial Intelligence\cite{NL4XAI}
\item \textit{GECKO}: Building Greener and more sustainable societies by filling the Knowledge gap in social science and engineering to enable responsible artificial intelligence co-creation \cite{GECKO}
\end{itemize}

The NL4XAI \cite{NL4XAI} project focuses on developing self-explanatory XAI systems by utilizing natural language generation and processing, argumentation technology, and interactive technology. The GECKO\cite{GECKO} project is exploring the development of interpretable XAI models to mitigate unintentionally harmful and poorly designed AI models.


\subsubsection{United State Government Funded Projects}
Defense Advanced Research Projects Agency (DARPA) Information Innovation Office (I2O) in the United States has started a funding program called Explainable Artificial Intelligence (XAI) \cite{DARPAXAI}. Under this program, DARPA has funded several projects focusing on different aspects of XAI:
\begin{itemize}
\item \textit{Driving-X}: Study the use of XAI for self-driving vehicles.
\item \textit{Rollouts}:  Use XAI to establish comfortable human-robot interaction.
\item \textit{StarCraft}: Design a self-explaining AI model to play video games. 
\item \textit{Learning and Communicating Explainable Representations for Analytics and Autonomy}: Design an XAI framework for multi-model analytics and autonomy by recognition, reasoning, and planning domains.
\item \textit{COGLE}: Design a system to provide explanations of the learned performance capabilities of an autonomous system.
\item \textit{Explainable AI for Assisting Data Scientists}: Study the effectiveness of the XAI system in debugging common ML models.
\item \textit{DARE}: Use XAI to improve the accuracy of deep learning models to enable multiple modes of explanation.
\item \textit{EQUAS}: Develop a new Explainable QUestion Answering System (EQUAS) based on pedagogical and argumentation theories.
\item  \textit{Model Explanation by Optimal Selection of Teaching Examples} Analyze the Explanation-by-Examples system to improve the user understanding of black-box ML models.

\end{itemize}

\subsubsection{Other Projects}

In 2017, The Ministry of Science and ICT (MSICT) in South Korea funded the XAI Center\cite{XAICenter} which focuses on the research and development of XAI technologies. The XAI center has supported several research activities which are mainly focused on the medical and financial sectors. In 2019, Europe-based Christ-Era organization funded 12 XAI projects under ``Explainable Machine Learning-based Artificial Intelligence"\footnote{\url{https://www.chistera.eu/projects-call-2019}} funding call. Some of these projects focus on B5G and 6G applications such as digital medicine, robotics, and predictive maintenance.   

Although there are many global-level research activities being initiated, many of these activities have 6G, 6G technologies, and 6G applications as minor focus. There are still mainly focusing on B5G developments. Table \ref{tab:projects} highlights the relevance of these research projects to different aspects of 6G networks.

\begin{table*}
  \centering
        \caption{Important research projects on XAI and their relevance to 6G.}
        \label{tab:projects}
        \renewcommand{\arraystretch}{1.2}
  \begin{tabular}{|p{3cm}||c||c||c|c|c|c|c|c||c|c|c|c|c|c|}
  \hline 
  \rowcolor{gray!25}
  &  & &\multicolumn{6}{c||}{\textbf{Relevant 6G Technical Aspect}} &\multicolumn{6}{c|}{\textbf{Relevant 6G Applications}} \\
  \hline
   \rowcolor{gray!25}
      	&  {\rotatebox[origin=c]{90}{~\textbf{Funded by}~}}
&  {\rotatebox[origin=c]{90}{~\textbf{Relevance to 6G}~}}
        
         &{\rotatebox[origin=c]{90}{~Intelligent Radio~}}
         &{\rotatebox[origin=c]{90}{~Trust and Security~}}
         &{\rotatebox[origin=c]{90}{~Privacy~}}
         &{\rotatebox[origin=c]{90}{~Resource Management~}}
         &{\rotatebox[origin=c]{90}{~Edge Network/ EDGE AI~}}
         &{\rotatebox[origin=c]{90}{~ Network Automation, ZSM~}}
         &{\rotatebox[origin=c]{90}{~ Intelligent  Health  and  Wearable~}}
         &{\rotatebox[origin=c]{90}{~ Industry 5.0, Collaborative Robots, Digital Twin~}}
         &{\rotatebox[origin=c]{90}{~Connected Autonomous Vehicles, UAVs~}}
         &{\rotatebox[origin=c]{90}{~Smart Grid 2.0~}}
         &{\rotatebox[origin=c]{90}{~Multi-sensory XR Applications, Holographic Telepresence~}}
         &{\rotatebox[origin=c]{90}{~Smart Governance~}}

         \\

  \hline
%

SPATIAL\cite{SPATIAL}    & EU H2020 & \cellcolor{red!15}H  & \checkmark   & \checkmark  & \checkmark &  & \checkmark  & \checkmark   &   & \checkmark &  &  &  &   \\
\hline

AI4EU\cite{AI4EU}    & EU H2020 & \cellcolor{yellow!15}M  & \checkmark  & \checkmark  & \checkmark & \checkmark &  &    &  \checkmark  & \checkmark & \checkmark & \checkmark & \checkmark &  \checkmark \\

\hline

FeatureCloud\cite{FeatureCloud} & EU H2020  & \cellcolor{yellow!15}M  &   & \checkmark  & \checkmark & \checkmark  &   &    & \checkmark  & &  &  &  &  \\

\hline

XMANAI\cite{XMANAI}    & EU H2020 & \cellcolor{yellow!15}M   & \checkmark  & \checkmark  & \checkmark & \checkmark & \checkmark  &   &   & \checkmark & \checkmark & \checkmark & \checkmark &  \\

\hline
DEEPCUBE\cite{DEEPCUBE}   & EU H2020 & \cellcolor{green!15}L  &   & \checkmark  & \checkmark & \checkmark &   &    &  \checkmark &  & \checkmark &  &  & \\

\hline
STAR\cite{STAR}   & EU H2020 & \cellcolor{yellow!15}M   & \checkmark  & \checkmark  & \checkmark & \checkmark & \checkmark  &   &   & \checkmark & \checkmark & \checkmark & \checkmark &  \\
\hline
NL4XAI\cite{NL4XAI}   & EU MSCA & \cellcolor{green!15}L  &   & \checkmark  & \checkmark &  &   &    &  \checkmark &  &  &  & \checkmark &  \\
\hline

GECKO\cite{GECKO}    & EU MSCA  & \cellcolor{green!15}L  &   & \checkmark  & \checkmark &  \checkmark &   &    &   &  &  & \checkmark &  &  \\
\hline

Driving-X   & US DARPA  & \cellcolor{red!15}H  &   & \checkmark  & \checkmark &  &  &    &   &  & \checkmark &  &  & \\
\hline

Rollouts   & US DARPA & \cellcolor{red!15}H  &   & \checkmark  & \checkmark &  &  & \checkmark   &   & \checkmark &  &  &  & \\
\hline

StarCraft   & US DARPA  & \cellcolor{red!15}H  &   & \checkmark  & \checkmark & \checkmark  &  &    &   &  &  &  &  \checkmark &  \\
\hline

COGLE   & US DARPA & \cellcolor{green!15}L  &   & \checkmark  & \checkmark &  &   & \checkmark   &   & \checkmark &  \checkmark  &  &  & \\
\hline

DARE   & US DARPA & \cellcolor{green!15}L   &   &   &  & \checkmark &   &   &   &  &  &  &  &  \\
\hline

XAI Center\cite{XAICenter}  & South Korea MSICT & \cellcolor{red!15}H  & \checkmark  & \checkmark  & \checkmark & \checkmark & \checkmark  & \checkmark   & \checkmark  & \checkmark & \checkmark &  \checkmark & \checkmark &  \checkmark \\
\hline


  \end{tabular}
  \begin{flushleft}
\begin{center}
    
\begin{tikzpicture}

\node (rect) at (0,2) [draw,thick,minimum width=0.6cm,minimum height=0.4cm, fill= green!15, label=0:Low Relevance] {L};
\node (rect) at (2.8,2) [draw,thick,minimum width=0.6cm,minimum height=0.4cm, fill= yellow!15, label=0:Medium Relevance] {M};
\node (rect) at (6,2) [draw,thick,minimum width=0.6cm,minimum height=0.4cm, fill= red!15, label=0:High Relevance] {H};
\end{tikzpicture}
\end{center}

\end{flushleft}
  
  \end{table*}

\section{XAI for 6G Technical Aspects} \label{sec:tech_spec}

In this section, we discuss the following primary technical aspects of 6G networks: intelligent radio, security, privacy, resource management, edge networks, and network automation. For each technical aspect, we first introduce the background and motivation of its importance in 6G. Then, besides technical requirements, the prospective challenges of the development of regular AI/ML algorithms in wireless networks are analysed. Finally, we explain how XAI can build trust between humans and AI-enabled machines based on the capability of supporting human-understandable explanations. 

Notably, for all the subsections ``how XAI can help" in this and the subsequent Section \ref{sec:use_cases} is discussed, considering that not many XAI solutions are built specifically for 6G networks in the existing literature, we have additionally incorporated our analysis on the current AI solutions for the mobile networks and then accordingly share our opinions on how XAI can improve (as well as the potential issues) these solutions (i.e., often for only one layer, as shown in Fig. \ref{fig:xai_6g}) in the near future. We hope this section offers guidance for applying XAI to the future fully AI-powered 6G networks across all layers shown in Fig. \ref{fig:xai_6g}.

\subsection{Intelligent Radio}

\subsubsection{Introduction}
The intelligent radio at the intersection of AI and cognitive radio has recently attracted significant attention in solving spectrum problems, including access, monitoring, and management. The rise of modern communication systems with 5G, B5G, and towards 6G has extended radio services to various industrial domains, exposing several challenging issues and complicated problems in wireless communications. 
It is feasible to use AI algorithms with automatic learning models to effectively handle channel modelling, intelligent spectrum access, physical layer design, and other network management issues in wireless communications~\cite{yang2020ai6Gnetwork}. The emergence of ML, especially DL in the era of big data, has enabled revealing the essential and unexplored radio characteristics and boosted the progress of wireless and networking technologies with new architectures and novel analysis of pyramid structures.

In the last decade, it has witnessed the evolution of the traditional black-box base station towards a virtualized next-generation node base (gNB) with the capacity of a functional split, that promotes a new paradigm of Open Radio Access Network (O-RAN) specialized by disaggregated, virtualized and software-based components, connected over open, programmable, and standardized interfaces with full interoperability across different vendors. In 6G, the AI/ML workflow for intelligent radio is identified by O-RAN with Non-Realtime RAN Intelligent Controller (Non-RT RIC))~\cite{oran2021}, which should consist of several steps: data collection and processing, training, validation and publishing, development, AI/ML execution and interface, and model maintenance. xApp, as a microservice that is responsible to supervise and manage radio resources through standardized interfaces and service models, is usually designed to control O-RAN slicing policies with real-time responses, in which different AI/ML solutions can be exploited using different key performance measurements, depending on target tasks in the physical layer~\cite{polese2022coloran}.

\subsubsection{Requirements and challenges}

Extremely high data rates and low latency of massive machine-type wireless communications are realised as the key requirements of 6G. It can be achieved with an advanced-designed physical layer, wherein several fundamental signal processing and analysing tasks (e.g., source coding, modulation, orthogonal frequency-division multiplexing (OFDM) modulation, and multi-input multi-output (MIMO) precoding) are powered by AI algorithms~\cite{zhang2019vision6G}. These tasks are typically deployed by the appropriate modules, which follow a forward procedure at the transmitter and an inverse procedure at the receiver. Previously, numerous intelligent radio signal processing approaches were studied with traditional ML algorithms, where expert knowledge of concerning domains is needed to fine-tune radio features and learning models. Being superior to conventional ML, DL has been recently applied to the intelligent radio area to improve performance significantly thanks to its great ability to deal with large-noisy-confusing raw datasets of radio signals~\cite{pham2021radioSurvey}. For example, the accuracy of automatic modulation classification in 5G was improved with CNN architectures while keeping a reasonable complexity~\cite{huynhthe2020amc}. Although DL can extract underlying features from raw radio signals at multi-scale representations to learn complex discrimination patterns, it usually represents black-box models with insufficient interpretability and weak explainability~\cite{huynh2021automatic}. Consequently, understanding why and how an AI model can predict outcomes is the key to helping network engineers and communication system designers improve the system's performance and manage operation risks sustainably.

Although it is currently difficult to identify the changes in 6G architecture, if compared with 5G, concretely, at least the following must be considered in the physical layer: AI/ML algorithms should be deployed as native parts for intelligent beamforming, cognitive intelligent-based autonomous radio resource management, intelligent channel coding and modulation, channel estimation, intelligent multiple access and spectrum sharing. Moreover, for future RANs with flexibility, massive interconnectivity, and spectral efficiency, xApps with AI/ML-enabled non-RT RIC should be carried out at the intelligent control layer to optimize radio resource utilization and minimize traffic congestion~\cite{polese2022understanding}. A fully-user-centric network architecture with ML-driven layers can leverage distributed AI with RAN decisions made by end terminals automatically without the need for centralized controllers~\cite{pham2022distributed}, thus enhancing learning efficiency and reducing computing cost if compared with centralized AI.

\begin{figure}[!htb]
    \centering
    \includegraphics[width=0.5\textwidth]{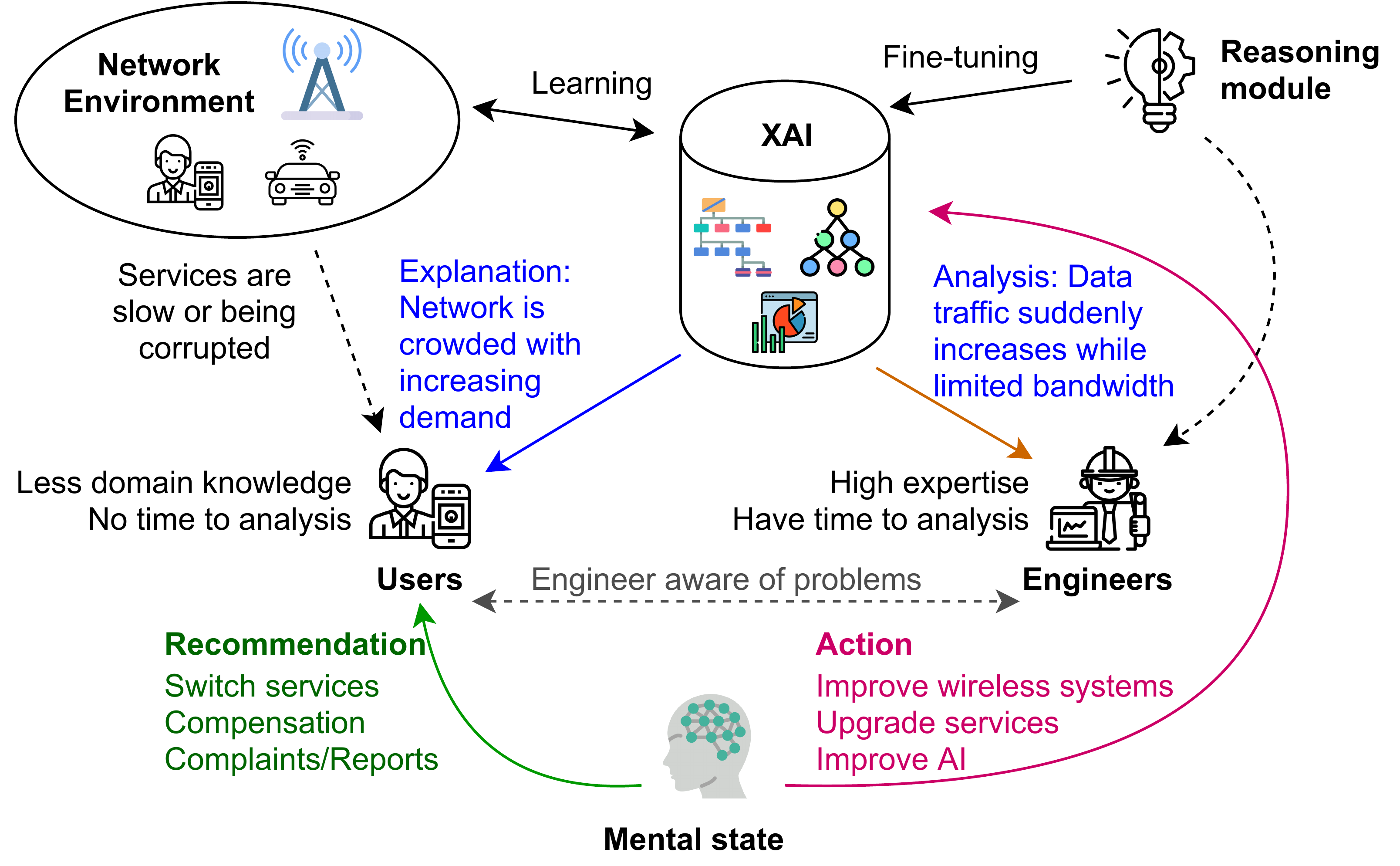} 
    \vspace{1mm}
    \caption{An example of XAI for intelligent radio: XAI can explain abnormal phenomena in wireless networks to end-users by ordinary explanation and to system engineers by specialised analysis.}
    \label{fig:xai_radio}
\end{figure}

\subsubsection{How XAI can help }

Despite being superior to traditional ML algorithms (e.g., decision trees, random forest, KNN, ANN, SVM) in terms of accuracy when dealing with large-messy-confusing practical datasets, DL with various architectures has a black-box nature that exposes a lack of explainability. For instance, Tunze \textit{et al.}~\cite{tunze2020TVT} proposed an advanced automatic modulation classification method with CNN architectures to generally improve accuracy and reduce complexity, however, the method failed to explain why some modulations present better performance than others with the same channel condition and how to predict when the DL model will crash under different practical channel conditions. 

In this context, XAI such as LIME, SHAP, and LRP can help 6G stakeholders (e.g., service providers) understand the relations between input data's quality and learning efficiency, to handle imbalance and overfitting problems, and determine whether the bias existed in the training/testing dataset.
In an effort to provide inherent explainability for DL-based modulation, the concept bottleneck model in~\cite{wong2021explainable} which comprised a regression network to infer a number of potential concepts and a classification network to predict the target modulation based on the set of concepts, enables XAI stakeholders to interrogate the classification decisions and address the out-of-training-set problem (i.e., some classes are not seen during training).
In~\cite{guo2020explainable}, Guo pointed out the weak transparency of DL compared with traditional ML for radio resource allocation in the physical layer and MAC layer, and then recommended some trustworthy AI techniques to improve explainability. 
For instance, a case study-based deep feature visualization XAI technique allows the manipulation of key features to optimise a deep model along with different network traffic and channel conditions.
Besides, hypothesis testing and didactic statements with human-machine interfaces are beneficial for elucidating the model learning and decision-making processes, where they play the role of human-like reasoning modules in an XAI integration framework as shown in Fig.~\ref{fig:xai_radio}).
For channel modeling with DL, Lee \textit{et al.}~\cite{ lee2021channel} proposed a model-agnostic metamodeling method that can interpret any data-driven channel model into a more understandable form with many transparent mathematical expressed via a symbolic representation technique.

To achieve intelligent radio in 6G, several new RAN architectures like O-RAN, cloud RAN, virtual RAN, and massive RAN have cooperatively operated and connected together in a dense radio environment. XAI will improve the connectivity between mobile users and base stations relying on a set of trustworthy wireless evaluations with dynamic cell selection, intelligent beamforming, channel estimation, adaptive coding and decoding, and automatic modulation recognition, thus optimizing the utilization efficiency of limited radio spectrum while maintaining high data transmission rate. Indeed, xAI can help to identify the outliers of radio signal data for training AI models; determine inefficient performing layers and modules in deep neural network architectures for modulation classification and channel estimation; point out which data,  model configuration, and training option that induce the failure and performance degradation of AI models for intelligent beamforming; and provide additional evaluation metrics to consolidate the AI decision of cell selection in 6G networks. XAI is also realized as the key to the next evolution of xApps to dApps~\cite{doro2022dapps} which enables real-time inference and control in O-RAN by cooperatively learning AI models in a distributed manner using locally collected data at RAN nodes. Moreover, XAI can become a sustainable solution to mitigate the difference in hardware configuration and software performance caused by network vendors and operators having different services level agreements, especially, in open network architectures like O-RAN featured by the incorporation of multi-vendor elements, the interoperability and compatibility should be in place. Despite having great potential to revolutionize intelligent radio in 6G with the interpretability and explainability of black-box models, XAI may increase the system's complexity (e.g., where XAI is deployed as an attached module to assist the primary AI model in offering intuitive explanations).

\subsection{Trust and Security}
\subsubsection{Introduction}

Since complex 6G networks may contain several heterogeneous dense sub-networks via intelligent connections with cloud-based infrastructures, they will expose some trust and security problems at multiple network connection levels. 
In this context, 6G communication systems should automatically detect proactive threats with intelligent risk mitigation and self-sustaining operation. To this end, AI-based trust and security become promising solutions to accurately identify and quickly respond to potential threats in an automatic manner~\cite{Mucchi2021security}. Besides some new threats, 6G networks must cope with existing security issues\cite{liyanage2017secure} in previous-generation networks, e.g., SDN, multi-access edge computing (MEC), and NFV. A distributed network that relies on the expansion of device-to-device communication with mesh networks and multi-connectivity is vulnerable and sensitive to the attacks of malicious parties. A hierarchical security protocol can be suitable for wide-area network security and sub-network communication security. Multiple radio units can be attacked over user and control plan microservices at the edge in coexistence scenarios of centralised radio access networks (RAN) and distributed core functions. In the perspective of AI-enabled 6G to achieve full automation, ML systems may become the target of several data-based and model-based security threats~\cite{Porambage2021security}, such as data injection, data manipulation, model evasion, and model modification.

\subsubsection{Requirements and challenges}
To guarantee high-quality services, the latency criteria of security mechanisms should be taken into consideration in enhanced ultra-reliable, low-latency communications (eURLLC)~\cite{Porambage2021openSecurity}. Some specialized 6G services with data plane, such as game streaming and remote surgery/telesurgery, require a reliable security solution that can not only effectively defend against cyberattacks but also perform AI/ML-based security analysis promptly with ultra-low latency, to guarantee a certain level of user experience. Moreover, high reliability will demand several extraordinary security solutions to maintain the availability of service operations effectively. Extreme data rate transmission can disclose some traffic processing security issues, e.g., traffic analysis, AI/ML-related processing flow, and pervasive encryption. These existing issues can be partly handled with distributed security solutions, where raw data and information should be processed locally and on the fly, in decentralised systems or even in partitioned parts of a network. Distributed ledger technology (with some distinctive features such as transparency, redundancy, and security) and ultra-massive machine type communication (umMTC) can be applied to satisfy security requirements. However, implementing and integrating AI/ML algorithms for resource-constrained devices are still challenging besides multi-threat analysis on big datasets. Some other issues that may arise from AI-driven security solutions are the responsibility for mistakes made by AI, the scalability and feasibility of AI models in diversified storage and computing infrastructures, and the vulnerability of AI models in distributed systems~\cite{Tang2020security}. For example, uploading local parameter sets from edge devices to a federated centre and broadcasting an updated global model to devices can become the target of poisoning attacks.


\subsubsection{How XAI can help}
Regarding AI security technologies, transparency in verifying how securely AI systems operate against adversarial machine learning (AML – an attack intentionally fools the AI model by entering deceptive data to make network system unstable, malfunctioning or unavailable) should be ensured to protect subscribers and mobile communication systems from AML. Besides being created in a reliable system, it is necessary to check whether the AI models operating in user equipment (UE), radio access network (RAN) and core have been maliciously modified or altered by a malicious attack. Open radio access networks (ORANs) with an intelligent controller can execute self-healing or recovery procedures if any malicious or abnormal event is detected in AI models.
Many recent methods have exploited some advanced ML and DL algorithms to effectively deal with different cybersecurity problems in ORAN and virtualized radio access networks (vRANs)~\cite{Porambage2021security}: poisonous attacks by tampering the Internet of Everything (IoE) data for training with malicious samples, evasion attacks by injecting disorders and outliers to testing data to circumvent the learned model, API-based attacks to pilfer prediction outcomes, infrastructure physical attacks and communication tampering by interfering communication-computing connections and shutting AI systems down~\cite{Siriwardhana2021security}. 
However, most existing ML/DL-powered security mechanisms are unable to explain their final decisions (e.g., how a system achieves threat detection more precisely than attack classification) and response actions (e.g., how an action accordingly responds on time to protect networks from cyberattacks)~\cite{vigano2020explainable}, and consequently cause uncertainty to 6G stakeholders.

XAI presents great potential for improving cybersecurity in 6G IoT networks (i.e., effectively preventing wireless connections, sensory data, intelligent controllers, and applications from different common attacks like poisoning attacks, evasion attacks, physical layer attacks, and model inversion attacks), ensuring the extreme reliability of IoT-based latency-sensitive services (e.g., industrial automation, emergency response, and remote surgery), and enhancing the trustworthiness of AI-aided security solutions (i.e., AI models have the capability of interpretability and explainability).
In~\cite{zolanvari2021trust}, Zolanvari \textit{et al.} introduced transparency relying upon statistical theory (TRUST), a model-agnostic XAI concept that acts as a surrogate explainer to offer multi-level interpretability without sacrificing performance or imposing any restrictions, for cybersecurity in Industrial IoT. 
By determining the AI's behavior representatives and reasoning on the highest class probability, TRUST provisions transparency of the final decisions made by the AI model. 
For attack classification, TRUST provides explanations comprehensively for new random samples while presenting high accuracy of over $98\%$ and outperforms LIME in terms of speed and explainability. 
In~\cite{chai2022explainable}, an explainable multi-modal hierarchical model (MMHAM) for phishing website detection was proposed to overcome the limited interpretability of deep models. 
MMHAM leverages a novel shared dictionary learning approach and a hierarchical attention mechanism to align deep representations of fraud cues and facilitate systematic interpretability at different levels, respectively.
Although MMHAM can help XAI stakeholders to detect phishing websites and proactively react to preventive actions based on explainable insights, it is unable to interpret complicated objects like phishing patterns.
In~\cite{gulmezoglu2021xai}, LIME and saliency map XAI methods were applied to interpret AI models developed to detect and classify fingerprinting attacks, in which the most dominant features extracted from raw data are discovered as two post-hoc XAI methods to explain leakage sources in cyber threat intelligence systems.
Relying on comprehensive benchmarks with the remove and retrain metric, the two XAI techniques are proficient and compatible with different AI models, including random forest and neural networks.
This work also concludes that LIME consumes much more computations than the saliency map in calculating the weights of CNN models.
Undoubtedly, XAI will provide much more meaningful and helpful information to stakeholders  and administrators about trust and security, such as identifying the most critical and vulnerable features to cyberattacks, ranking the most important features of different AI models to combat against diversified attacks, skipping some hidden layers in deep networks to accelerate attacking detection speed, formulating the network configuration and AI training setup to achieve the highest cyberattacks classification, and drawing the relationship between predictive variables and dependent variables under various attack scenarios and system conditions.

\subsection{Privacy}

\begin{figure}[!t]
    \centering
    \includegraphics[width=0.475\textwidth]{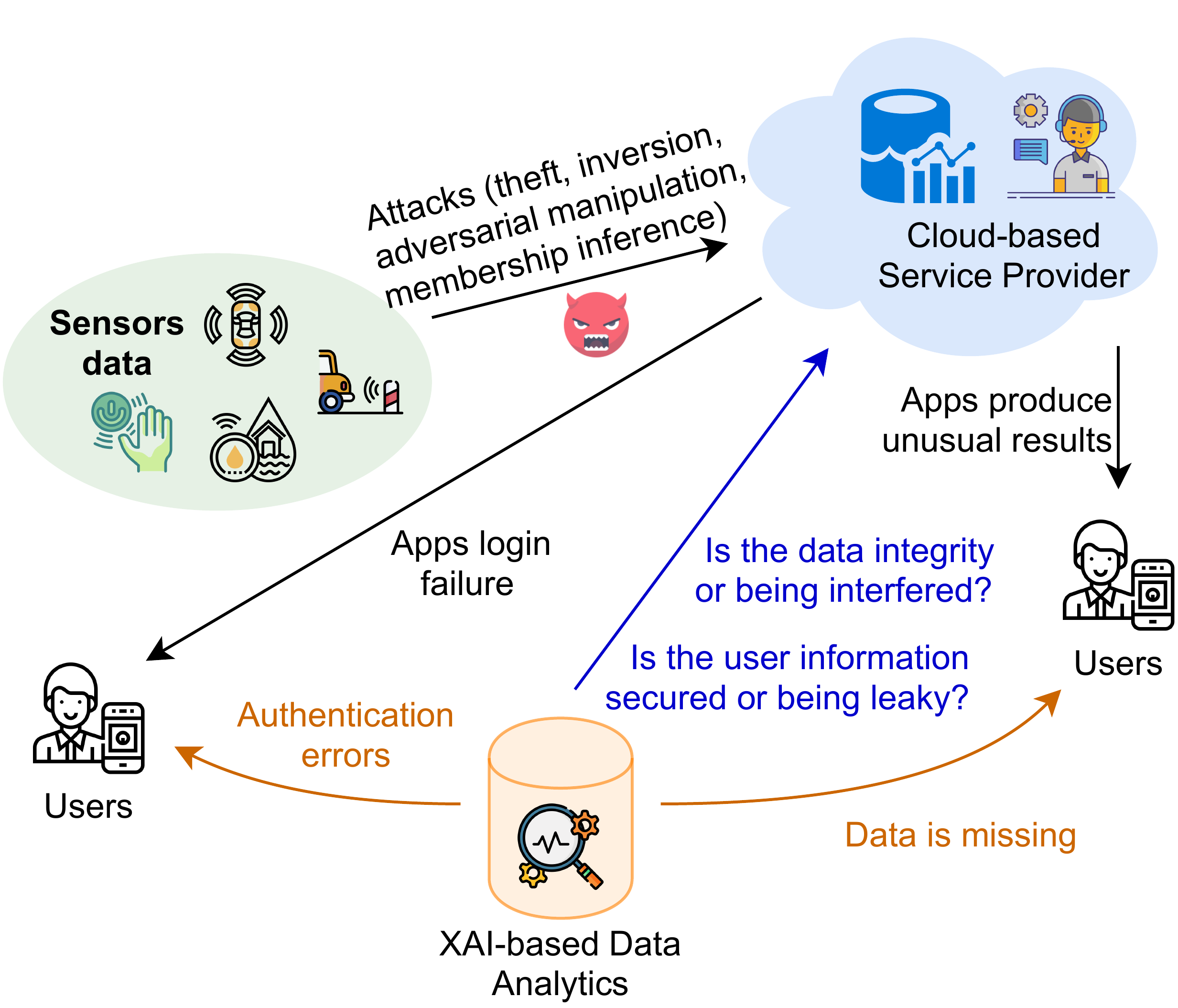} 
    \vspace{1mm}
    \caption{An example of XAI for data privacy: XAI-based data analytics can provide specialised analysis and professional explanations to cloud-based service providers regarding undesirable issues of data privacy from end-users.}
    \label{fig:xai_privacy}
\end{figure}

\subsubsection{Introduction}
Data leakage can violate users' privacy, which can be prevented through comprehensive privacy-preserving algorithms. When the number of end devices progressively increases with high data variety, transmitting data over wireless networks, storing data in storage infrastructures, and processing data in computing infrastructures are burdensome with the inserted privacy protection mechanisms~\cite{Sun2020privacy}. The possible number of wireless connections in 6G would be greater than 5G up to $1,000$ times. Therefore, ensuring high data privacy without service performance degradation is challenging (i.e., obtaining a good trade-off between the enhancement of data privacy and the preservation of service performance in terms of accuracy and processing speed). Moreover, the massive amount of data serving the learning process of statistical AL/ML models will expose a significant challenge for user privacy, which has attracted much more attention from various industrial and academic communities.

\subsubsection{Requirements and challenges}
With the increasing number of smart devices that allow collecting massive sensitive data effortlessly, the data privacy concerns in 6G would be significant due to the lack of data transparency~\cite{David2018privacy}. Running intelligent applications on mobile devices and at the edge of the network can become the vulnerable target of privacy attacks. Many problematic privacy concerns will be more severe in the era of big data with 5V features, including volume, velocity, variety, veracity, and value. Adding privacy protection mechanisms will increase communication and computational costs and may not ensure the high quality of 6G-based applications and services unless addressed. Therefore, privacy protection mechanisms should be designed to be cost-efficient besides detecting potential privacy threats automatically and effectively~\cite{Saad2020privacy}. Due to the diversity and variety of applications and services, ensuring privacy concerns are addressed is a significant challenge. First, easier data acquisition and accessibility can cause regulatory difficulties with data storage, permission, and utilisation. Second is the development and integration of AI/ML algorithms into advanced privacy protection mechanisms which may provoke overloading on resource-constrained devices. Finally, it is noteworthy to balance between the high accuracy of services and the robust protection of user privacy, especially from the perspective of data ownership, access authorisation, and other regulations.

\subsubsection{How XAI can help}
Several AI algorithms, including ML and DL, have been applied for privacy protection, in which privacy threats and attacks should be automatically detected and classified to consequently select the most appropriate privacy-defense mechanism.
In~\cite{fu2019keeping}, a regular ML framework with feature extraction and multi-class classification was developed to detect and assess the privacy risks of mobile phone applications. Despite achieving high classification accuracy of over $90\%$, some classifiers like SVM and Naïve Bayes presented poor interpretable capabilities, thus being unable either to explain AI’s decisions or to provide assessment insights. 
In~\cite{wu2022ensemble}, a privacy-preserving data mining framework was proposed to identify malicious adversaries that attempt to collect sensitive user data on edge computing platforms.
An ensemble learning algorithm with random decision trees was studied to increase the accuracy of data leakage detection, but reasoning insights that help service providers plan effective data protection and data recovery solutions based on the level of data fragmentation were disregarded.
It is observed that the trustworthiness of AI-aided privacy protection methods is questionable.
In this context, some XAI algorithms like LIME and SHAP can contribute as the post-doc models to interpret AI models and expound their decisions by giving a set of explanations representing the impact of each user feature or each data attribute to the final prediction for every single input sample.
Furthermore, XAI helps to rank different encryption methods to hide sensitive data, identify which data needs to be shuffled or masked to disassociate its original attributes, and measure the correlations between privacy metrics and data leakage probabilities.
Remarkably, some XAI algorithms and methods are beneficial to explainability enhancement of black-box based privacy models, for example, Deep Learning Important Features (DeepLIFT) finds neurons and weights that significantly affect the final decisions made by DL-aided malicious adversaries detection models, Layer-wise Relevance Propagation (LRP) with a set of purposely-designed output-oriented propagation rules measures the positive and negative impacts of each layer in DL-based threat detection and classification models, and ProfWeight~\cite{dhurandhar2018} transfers knowledge from a trained data privacy protection model to other data recovery models and from a data leakage detection model to other secure data storage and integrity models.

Compared with traditional private data release methods that add noise to the original data to improve privacy, several ML and DL algorithms have recently been exploited to generate synthetic fake datasets with little accuracy degradation~\cite{liu2021when}. 
Xu \textit{et al.}~\cite{xu2019GAN} proposed GANobfuscator, a differentially generative adversarial network (GAN), to mitigate sensitive information leakage. 
GANobfuscator can obtain differential privacy by adding specialised noise to the original dataset and adopting gradient pruning techniques during the learning stage.
Based on the comprehensive evaluations with different datasets and network architectures, the artificial data generated by GANobfuscator can guarantee the privacy of the original one without information loss.
In~\cite{yoon2020anonymization}, the synthetic data of electronic health records were generated by a conditional GAN framework to protect sensitive patient information from privacy attacks by adversaries. 
This GAN framework can prevent patient reidentifiability from statistical measurements of the similarity between the whole dataset and the data of individual patients.
Although GAN-based methods are beneficial to privacy protection when generating artificial data to dupe adversaries, they restrain the interpretability and explainability (e.g., reasoning the classification output given by the discriminator to consequently enrich corrective feedback to the generator in a GAN framework). 
In~\cite{nagisetty2020xai}, Nagisetty \textit{et al.} developed an XAI-guided gradient descent method, denoted XAI-GAN, which can assess the discriminator’s decision to explain the feedback passing to the generator in a standard GAN.
Especially, XAI-GAN was built as an in-model agnostic explainer to work with different XAI algorithms, such as LIME, DeepSHAP, and saliency maps.  
Based on the extensive experimental results on different datasets, some key points were concluded: (i) XAI-GAN is better than standard GANs in terms of generated sample quality, (ii) XAI-GAN outperforms standard GANs in terms of classification accuracy when training with the same amount of data, and (iii) XAI-GAN handles the trade-off between data efficiency and training time more effectively than standard GANs.

\subsection{Resource Management}
\subsubsection{Introduction}
Resource management is challenging due to the inherent scarcity of radio resources in wireless communications. It then becomes more difficult in advanced and complicated networks like 6G, wherein the number of smart devices increases rapidly and are involved in different IoT networks (such as cellular IoT, cognitive IoT, and mobile IoT). The overall system performance of a wireless network certainly depends on how it monitors and manages hyper-dimensional resources (e.g., time slots, frequency bands, modulation types, and orthogonal codes) effectively. Besides, the incorporation of wireless channel variations and traffic load attributes in the network design phase impacts connectivity among users having diverse quality-of-service (QoS) requirements. In the context of which new IoT applications demand high data rates, low latency, efficient spectral utilisation, and the expansion of personalised IoT services, the problems of resource monitoring and management are crucial\cite{ahmad2018towards}. In the last decades, AI/ML algorithms and DL architectures have been widely used to tackle several challenging resource allocation and management issues in 6G-related areas ~\cite{Hussain2020resourceManagement}. Concretely, they have contributed to many aspects including massive channel access, power allocation, and interference management, user association and hand-off management, energy management, ultra-reliable and low-latency communication, and heterogeneous QoS. Traditional mechanisms cannot optimise the non-convex problem of resource allocation and are time-consuming to manage resources in a crowded and complicated network. This drawback motivated the discovery of several data-driven ML-based resource allocation and management solutions, in which the high learning capability of AI/ML models is beneficial to the dynamic nature of 6G-enabled IoT networks~\cite{Yu2021resourceManagement}.

\subsubsection{Requirements and challenges}

Cellular IoT networks in 6G are specialised with extremely high data rates and solid connectivity between heterogeneous devices/users and access points/base stations. The diverse requirements of various IoT services and applications can be met by carefully selecting various network parameters (e.g., channel state information and traffic characteristics) and communication parameters (e.g., angle of arrival and modulation types). These parameters are now remarkably identified by ML and DL in terms of high estimation accuracy and good ability to deal with big raw data. Some smart devices demand autonomous access to the available spectrum and adaptive tuning of transmission power to mitigate interference and save energy~\cite{Zhang2020resourceManagement}. Some relative system parameters, such as the position and velocity of high-mobility users, propagation conditions, and interference patterns should be considered when designing an effective ML-based resource allocation solution. Notably, several traditional optimisation schemes cannot deal with diversified contexts for integration and cannot respond to varying environments. In numerous IoT applications, ubiquitous and heterogeneous devices have diversified QoS requirements and randomly varying access to network resources, demanding upgrading traditional ML algorithms with RL to fully adapt to the diversity of network requirements and the variation of network conditions~\cite{Lin2020resourceManagement}. Additionally, the high computational complexity of heuristics-based resource allocation schemes should be considered to be implemented on resource-constrained devices.
Compared with 5G, the explosion of new smart terminals in 6G along with a variety of applications in vertical industry markets is pushing mobile network operators (MNOs) to deal with more complex scenarios and deliver more diverse services. So far, the dynamic and diverse demands of 6G users through real-time micro-management of multiple resources including communication, computing, and storage should be taken into consideration completely. To this end, XAI can be deployed and integrated into customized network slicing procedures at MEC servers, wherein the E2E slice consists of a number of interconnected NFVs from RAN, transport, and core networks, to reliably handle incoming differentiated resource requests from a group of users and reasonably address the demand changes in resource requirements of an individual user.

\subsubsection{How XAI can help}
In the last decades, many ML/DL-based solutions have been introduced for different resource management tasks in 5G RAN, transport, and core networks, through NFV to decouple software and hardware by visualizing network functions to run on virtual machines, including scheduling and duty cycling, resource allocation, power allocation and interference management, resource discovery, cell selection, and mobility estimation. However, most of them are lacking the interpretability and explainability to build trust in network operators and end-users.
In 6G, resource management methods should be more reliable and trustworthy with the help of XAI besides ensuring high performance.
Specifically, for the network operators, XAI will improve the efficiency of computation offloading, resource allocation, and resource management in a variety of communication networks (including 5G-mmWave communication, O-RAN, long-distance and high-mobility communications (LDHMC), extremely low-power communications (ELPC), and vehicle-to-everything communication)~\cite{guan2021customized} and ensure the primary trained AI models are robust and well-performed with diverse wireless scenarios having various environment agents (e.g., wireless impairments, mobility, and available computing resource).
For the end-users, the detailed parameters or explanations of XAI will optimize traffic allocation and minimize the energy consumption of intelligent resource-constrained edge devices, thus enhancing the trust of end-users while offering a high QoS and QoE.
Nascita \textit{et al.}~\cite{nascita2021xai} introduced MIMETIC-ENHANCED, a multimodal DL-based mobile traffic classification framework for RAN and transport networks (e.g., V2X - vehicle-to-everything network), in which a general XAI module was deployed to be familiar with different methods, such as addition feature attribution and Deep SHAP.
The XAI module in MIMETIC-ENHANCED can typically identify which set of inputs presents the highest confidence probability associated with the model’s output.
Especially, Deep SHAP allows the classification framework to produce the local explanations based on quantifying the importance value of each input and the global explanations by aggregating the importance values of inputs belonging to each modality. 
MIMETIC-ENHANCED was experimentally evaluated in terms of trustworthiness (how much XAI stakeholders can trust an estimated confidence entity) and interpretability (the intrinsic reasoning that allows XAI to operate properly).
For internet traffic classification, Callegari et al.~\cite{callegari2021explainable} built a collection of fuzzy rule-based classifiers by means of a multi-objective evolutionary learning scheme, in which each classifier plays the role of the XAI classification model and its trade-off between the classification accuracy and the explainability level is optimized individually. 
Accordingly, based on the input data attributes and the complexity of the problem, XAI stakeholders can select the most appropriate model to achieve a comfortable performance while aptly providing understandable explanations over linguistic IF-THEN rules.

With the tremendous potential of mobile edge computing (MEC) for 6G mobile services and applications, partially or entirely offloading computations have been advanced with ML and DL algorithms to minimize the energy consumption of user equipment. In~\cite{ali2019deep}, an energy-efficient offloading scheme was proposed by exploiting the advancement of deep networks to improve the accuracy of multi-component binary classification under various network and user attributes, such as the amount of transmission data, delay, network condition, computational load, etc. In~\cite{baek2021heterogeneous}, Baek and Kaddoum deployed a deep recurrent Q-network (DRQN) to cope with a joint request offloading and resource allocation control for heterogeneous services in multifog networks. Compared with deep Q-network and deep convolutional Q-network, DRQN can handle the partial observability problem more effectively thanks to its nodes in recurrent layers to hold internal states and aggregate observations. Many advanced DL-based offloading and resource allocation methods are with low transparency and poor explainability which may lead to some risky operation failures and difficulties to fix or update deep models. In this context, some XAI candidates (e.g., LIME, LRP, SHAP, and saliency map) can be applied to interpret deep networks (e.g., RL, RNN, and CNN) with the input of time series data~\cite{schlegel2019towards}, which help XAI stakeholders identify which are the most important attributes for different users to optimize offloading computation or minimize energy consumption. 
By providing insightful analysis and extra relevant information, XAI may tutor stakeholders and system resource managers to comprehend black-box models, such as identifying the least correlative layers in a deep neural network for removal without performance degradation of resource allocation and utilization, determining a set of configurable parameters (e.g., learning rate and regulation factor) in the training stage to achieve a better learning convergence, governing the relationship between accuracy and complexity of resource allocation models to obtain the overall system resource optimization, and pointing out the underlying correlations between global features and local attributes of different architectures to select the most appropriate deep networks.

\subsection{Edge AI}
\begin{figure}[!htb]
    \centering
    \includegraphics[width=0.500\textwidth]{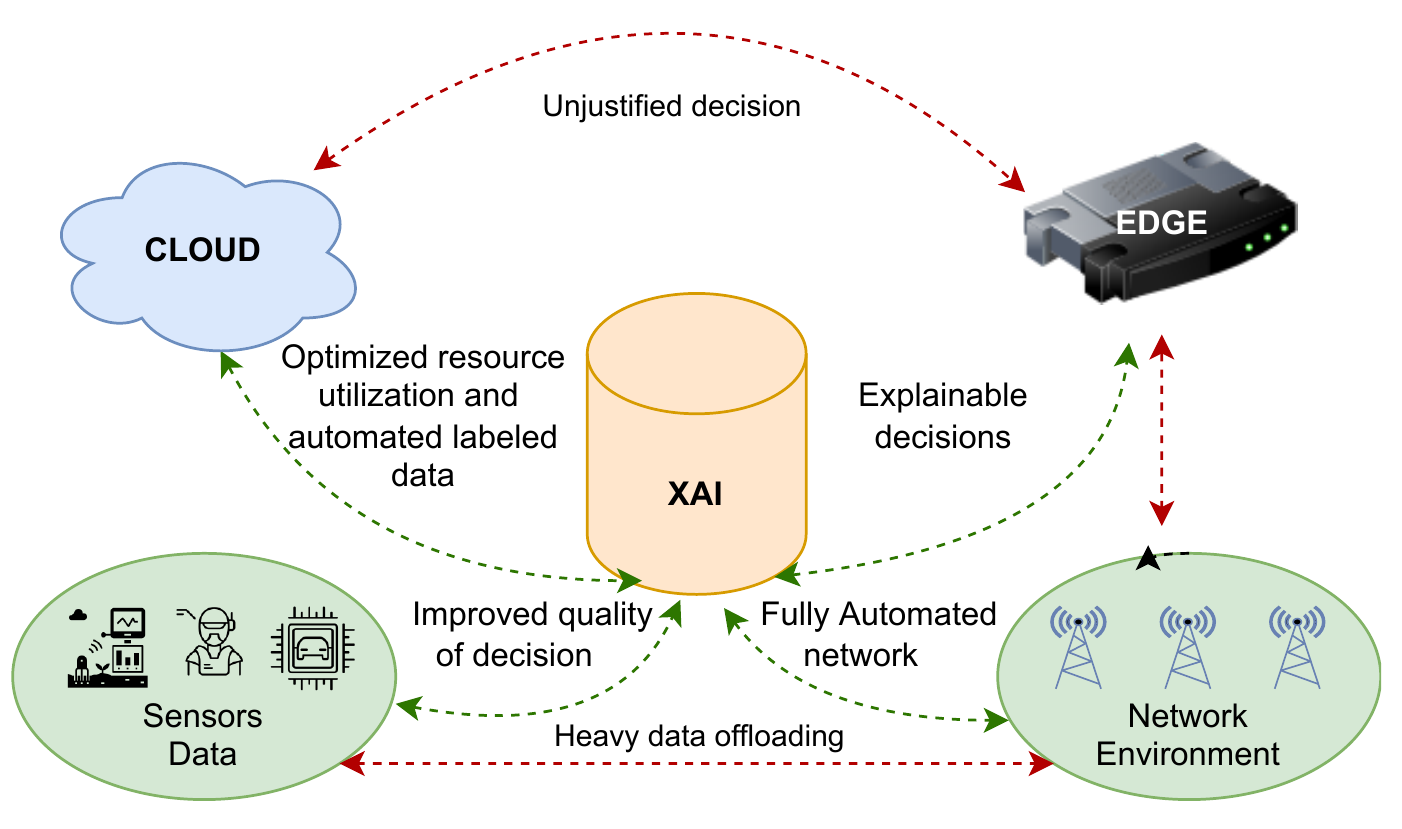} 
    \caption{An illustration of XAI for edge-assisted 6G: more justified decisions improve the resource allocation between cloud, edge, and sensors.}
    \label{fig:xai}
\end{figure}

\subsubsection{Introduction}

Edge AI is one of the essential components missing in the existing 5G communication networks. Edge AI is a framework that focuses on the integration of mobile edge computing, communication networks, and AI seamlessly \cite{wang2019edge} and is considered to be one of the most critical enabling technologies for the futuristic 6G cellular networks \cite{xiao2020toward}. Recently, many researchers have been working to make the 6G cellular network a fully autonomous and highly intelligent system. Edge AI plays a vital role in realising human-like intelligence in all the aspects of 6G network systems to improve the quality of experience of the users \cite{zhao2020edge}. AI-enabled decentralised mobile edge servers are deployed at a massive scale for performing processing and decision-making at near proximity of service requests and data generation. This makes edge AI a vital component providing accelerated and ubiquitous integration of AI into future 6G networks \cite{tomkos2020toward}. 

\subsubsection{Requirements and challenges}
Several resources are needed to execute the AI models in 6G, such as coordination of data, training of the model, caching, and computing \cite{kaur2021machine}. One essential requirement for edge AI in 6G networks is the automatic creation of labels from the massive amount of raw data generated by the wireless devices in the 6G cellular networks. Distributed AI, which performs the computation jointly at the cloud data centres with the distributed edge servers, is one of the key requirements for edge AI. Another essential requirement of edge AI in 6G is personalised AI, through which the decision-making of AI algorithms can be improved by understanding the preferences of the human users \cite{zhao2021survey}. Security is also a crucial requirement for edge infrastructure. Some security threats for edge infrastructure include resource or service manipulation, denial of service attacks, man-in-the-middle attacks, and privacy leakage. AI/ML algorithms can play a significant role in monitoring and predicting security attacks \cite{porambage2019sec,li2021security}. However, edge AI can vastly improve automation and lower the dependency on human intelligence of 6G cellular networks. In some mission-critical situations, humans have to be involved in decision-making. However, humans may not understand the reason for the predictions of edge AI-based 6G applications, making it very difficult to make confident decisions. One of the key goals of edge AI is self-evolution and self-adaptation so that human efforts can be reduced when processing data and making decisions. Furthermore, the development of the model by dynamically adapting it to unknown events based on the environment and the features of the data is another key goal of edge AI. However, due to the black-box nature of some AI algorithms, it will be challenging to evaluate/audit the effectiveness of AI models for the challenges mentioned earlier. To address these challenges, XAI can be used at the edge of AI in the intelligent control layer so that the 6G will be self-evolutionary and self-adaptive making it fully autonomous and the decisions taken by the AI models can be audited easily by humans.

\subsubsection{How XAI can help}


Lack of explainability can be a severe setback for edge-intelligence-enabled communication networks for some of the applications, like vehicle-to-vehicle communications, that require real-time decisions for preventing crashes, providing driver assistance, and enhancing traffic management. Due to this setback for edge-intelligence-enabled communication, humans find it difficult to pinpoint the actual origin of problems in catastrophic scenarios. Some of the factors like human intuition, channel measurements, and theoretical analysis have played a crucial role in the designing of wireless standards and cellular networks. This approach made the domain experts go with either computer simulations or theoretical analysis for validating the building blocks of communication systems. AI models are expected to give justifications or explainability for cellular networks \cite{shafin2020artificial}. 

The base stations integrated with edge intelligence will be granted precise, robust, and high-speed AI algorithms through AI-enabled 6G that will ensure safety-critical mass autonomy for the sub-network settings. The cellular networks are virtually split based on the types of services they offer through network slicing. Future AI-enabled 6G network slicing will be allocated based on several human-centric requirements such as ethics, and safety at the sub-network level. XAI can help in explaining the behaviors of mass control systems in terms of the overall policy as well as for individual instances, integrated with system performance that can lead to trustworthy supervision of 6G-based services \cite{li2020trustworthy}.

 XAI can play a massive role in guaranteeing the performance of edge AI-enabled 6G networks where several network components are integrated for different requirements through justifiable results. Verification, validation, and auditing of the decisions at the edge for addressing the challenges mentioned above will become simple due to the justification of the decisions from XAI algorithms. Also, humans may have to be involved in some stage of decision-making in mission-critical edges AI-enabled 6G applications like drone-assisted telesurgery systems, smart grid, and border surveillance. In those situations, the job of humans becomes simple as they can easily understand the reasons for the decisions/predictions of the AI algorithms at the edge as depicted in Fig. \ref{fig:xai}, thus enabling the humans to take better decisions.
 
 The predominant 6G applications, especially involving XR may use AI incorporating Switched Service Networks that can automatically endure high Key Performance Indicators thereby seamlessly managing resources, functions, and network control. The use of AI would enable the multi-sensory XR applications to automatically provide energy services to the users in order to send and develop 3D radio environment maps. The AI-based 6G functions would be complemented by “collective network intelligence” wherein the network intelligence in association with AI algorithms would run on the edge devices ensuring distributed autonomy and integration of related services \cite{chaccour2021edge}.  The adoption of XAI would provide justification for the relevance of the energy services being disseminated to the users in the development of 3D radio environment maps. It would ensure that the most appropriate services are disseminated, being mapped accurately with the specific user requirements. In \cite{hossain2020explainable}, XAI is proposed to be used by the authors to make the healthcare professionals understand the findings of the DL algorithms to control COVID-19-like pandemics using edge-enabled 5G and beyond networks. The proposed XAI-based solution will ensure that the findings of the ML-based algorithms are justifiable, which will help the ethical acceptance of the deep neural network model to combat the pandemic situations by healthcare professionals.
 
 In this context, some of the XAI frameworks such as LRP, SHAP, and LIME can be applied on edge AI empowered 6G to interpret the decisions of AI algorithms that can help the stakeholders do not have stringent performance requirements in terms of storage, latency, etc for autonomous maintenance of the futuristic cellular networks. However, to fully realize the potential of integrating XAI with edge AI empowered 6G, several issues such as reduced performance of the AI algorithms (to make them more explainable, the complexity of the algorithms may have to be compromised, which in turn may reduce the performance of the AI algorithms), lack of metrics to measure the performance of the XAI algorithms, have to be addressed.

\subsection{Network Automation and ZSM}

\subsubsection{Introduction}

New business models will be unlocked using technologies like SDN, MEC, network slicing, and NFV in 5G and beyond cellular networks\cite{liyanage2017enhancing, liyanage2015leveraging}. The increase in flexibility, cost efficiency, and performance, along with inter-domain cooperation and agility, results in a massive increase in complexity in the management and operation of 5G and beyond networks. Hence, the solutions provided by conventional methods may be inefficient in network and service management. Thus, it is inevitable for management operations through closed-loop automation. Management automation through self-managing capabilities will improve the efficiency and flexibility of delivering services and reduce operating expenses \cite{benzaid2020zsm}. Zero Touch Network and Service Management (ZSM) was established by ETSI to achieve self-managing capabilities. ZSM reference architecture aims to specify an E2E service and network management services that are fully automated, without the intervention of humans \cite{ortiz2020inspire,liyanage2022survey}.

\subsubsection{Requirements and challenges}

AI/ML and Big Data analytics are key enablers for 100\% automation in cellular networks. AI algorithms can learn from the vast amount of data generated in the 6G cellular network. They can play a vital role in the self-management of the network (self-configuration, self-protecting, self-optimisation, and self-healing), resulting in reduced human errors, accelerated time-to-value, and lower operational costs. AI/ML algorithms can be applied to raw data, filter the important events from the large volume of events, identification of problems in the network, and then send the most vital information to the upper layers \cite{AIOPS}.  However, the successful integration of AI/ML techniques in ZSM for full automation depends on the interpretability and transparency of the AI/ML models to ensure transparency, reliability, trustworthiness, and accountability in AI-enabled ZSM \cite{benzaid2020ai}. The type of ML algorithms used and the input data used to train them, which enable the ML algorithms to arrive at the decisions, need to be understood to provide reliable decisions on any automated tasks in ZSM related to delivery, deployment, configuration, assurance, and optimization. The end-user finds it challenging to explain the approach followed by the ML algorithms to arrive at the end results due to the increased complexity of the used ML algorithms. This situation is especially faced by users, mainly when a series of updated models are applied to analyse the large volumes of data generated in 5G and beyond networks. To address these issues, XAI has to be embedded with the intelligent control layer of 6G architecture, so that the 6G cellular networks are fully automated and also the decisions taken by the 6G network based on AI algorithms can be better understood by the network operators.

\subsubsection{How XAI can help}



Emerging networks are becoming more dynamic and complex with each passing day. Hence, the task of data management in cellular networks is more complex in the present day and age. The traditional approach of data management in cellular networks involves the extraction of information/data generated from measurements/logs, and individual events, and then assembling or correlating them together, later creating a summary event that can be used by humans to make decisions. Due to the huge quantity of data generated, simple correlation techniques for data management don't work in the current scenario. Hence, AI/ML algorithms have a huge role to play in managing present-day cellular networks. However, as more sophisticated ML algorithms are being applied, the end user finds it difficult to explain the approach used by the ML algorithms to arrive at the end prediction results, specifically in the scenario where several updated models are applied over a period of time. With its ability to provide justification and interpretability, XAI has a vast potential to address the challenges mentioned above in integrating AI with ZSM. To facilitate XAI for ZSM, the authors in \cite{dutta2021challenge} have proposed to add a dictionary that acts as a repository to capture the AI models used in the system. The authors proposed the usage of factor graphs or directed acyclic graphs to represent the taxonomy of AI models. To add the input/output variables and attributes of specific AI/ML algorithms used, the authors also proposed using the algorithm instance repository. In this way, the resulting events can be tagged or labelled. The analysts can use this metadata to do reverse engineering and come up with an explanation of the results. 

For example, holographic telepresence a typical 6G application in video conferencing, is capable of projecting full-motion 3D images in real-time. In the case of healthcare, holographic telepresence has immense importance in trauma care and surgeries enabling patients and clinicians to communicate across geographically distributed locations. The need for such technology has been experienced at its peak during the COVID-19 pandemic which required high-paced care and rendering of specialized services to the patients.  The dissemination of such applications is dependent on large bandwidth network communication and related services which would transmit clear holographic transmission using smart devices. The holographic projections are pre-programmed and the necessary projection requires huge manual interventions for efficient resource allocation and communication among the stakeholders \cite{kaiser20216g}. The implementation of ZSM will automate resource allocation and ensures seamless data transfer pertaining to holographic communications. The involvement of ZSM includes a collection of full motion 3D images from the source, which is analysed thereby triggering the necessary action to be sent to the executor or the consumer of the analytics function. Thus the intent of the service consumer gets fulfilled autonomously rendering the required services and resources in 6G. The use of XAI ensures justification for the identification of appropriate managed resources and rendering the most suitable services seamlessly to the stakeholders in a holographic telepresence environment.

XAI algorithms such as PIRL can be used by stakeholders such as legal auditors and can trace back the automated decisions taken by ZSM to automate several network management services in 5G and beyond networks. Specifically, network administrators can better understand the details concerning network maintenance, implementation of upgrades, and monitoring of attached network devices. In this way, performance management, configuration management, and fault management can be executed in a seamless manner by the network operators. However, as the decisions taken by ZSM may be mission-critical that may affect the network bandwidth and resource allocation, performance degradation of ML algorithms due to integration of XAI is a challenge that has to be addressed.

\subsection{Other technical aspects}
Providing end-to-end virtual networks that cater to the diverse and customized needs of heterogeneous applications is one of the key technical requirements in 6G, that can be realized by network slicing (NS). Managing resources and functions in NS is a challenging and crucial task, where efficient decisions are required at all levels of the network in real-time. Hence, AI can play a vital role in automating the decision-making for these key tasks in NS \cite{bega2020network}. As these  decisions involve complex management of resources that have financial and service quality, making the human-in-the-loop understand the rationale behind these decisions is of paramount importance. Integrating XAI algorithms in 6G improves the credibility and accountability of resource allocation in NS \cite{barnard2021resource}.

\section{XAI for 6G Use Cases} \label{sec:use_cases}

 Most of the visionary 6G applications need the support of AI and intelligent automation to realize their full capability. This section overviews comprehensively the potentials of XAI for such typical 6G use cases \cite{morocho2019machine, giordani2020toward, yang2020ai6Gnetwork, zhang2019vision6G, saad2019vision} including intelligent health  and wearable, industry 5.0, collaborative robots, digital twin, CAVs and UAVs, smart grid 2.0, holographic telepresence, metaverse and smart governance in 6G. Specifically, for each use case, it first introduces the motivation, which explains why this use case is important and required urgently. Then, it analyzes the enabling technologies required which normally includes the 6G communications technologies and AI algorithms. We also introduce some important existing work in the literature under each use case. Last but not least, we explain why XAI can enhance trust between humans and machines in each specific 6G use case as well as potential issues due to XAI. The XAI requirements of these 6G use cases are summarised in Table \ref{tab:use_cases}.

\subsection{Intelligent Health and Wearable, Body Area Networks}

\subsubsection{Motivation}
The advancement of 6G is expected to drive an innovative development of eHealth systems and improve the performance of medical and healthcare services with advanced AI/ML algorithms~\cite{Alwis2021healthcare}. The upcoming 6G communication can provide eHealth applications and services with ultra-high data rates and ultra-low latency for a huge number of connected devices. In this context, an eHealth system can take responsibility for real-time monitoring and tracking, recording health information, and storing eHealth records in cloud-based computing infrastructures. Furthermore, it can exchange medical records and health reports and provide a remote diagnosis by connecting different health service providers in a network~\cite{Bisio2019healthcare}. Nowadays, 6G eHealth solutions can be expanded to various scenarios, such as hospitals, sports, homes, and pharmacies, in which the QoS for all eHealth applications and services should be ensured in indoor and outdoor environments. More importantly, 6G-enabled Internet of Medical Things (IoMT) networks promisingly provide precise medical services by applying AI/ML algorithms to process healthcare and medical data besides very high-quality connectivity.

\subsubsection{Requirements}

As the healthcare data acquired by various multimodal sources has a large volume, high velocity, and diversified variety, exploiting ML algorithms and DL architectures to develop data-driven solutions has been attracting much more attention from healthcare and medical communities via academic research and industrial products. An AI-aided eHealth system should process different data types (e.g., sequential versus high-dimensional data and structured versus unstructured data) and fulfil high-quality services with very high accuracy (e.g., image-based cancer detection and recognition) and very low latency (e.g., online surgery via video streaming)~\cite{Arunim2021healthcare}. In several healthcare and wellness services, raw sensory data collected from wearable devices usually have noise and outliers. Therefore, AI-based solutions should learn complicated patterns from messy datasets effectively. Compared with data in other sectors, healthcare and medical data are more sensitive to security and privacy, hence ML-based solutions can be studied to automatically detect privacy threats and protect data from cyberattacks in distributed cloud-based systems. Recently, federated learning (FL) was introduced to overcome such kinds of data security and privacy by sharing the information of trained local models instead of the raw data from edge users~\cite{rieke2020healthcare}.
Due to the diversity of healthcare data, including sensory data, electronic medical reports, and medical images, collected by different healthcare centres with heterogeneous formats, 6G should specifically regulate data acquisition, storage, and data interoperability in the data mining and analytics layer of 6G-enabled IoMT networks, besides data privacy to prevent the sensitive patient information from cyberattacks. Moreover, from the perspective of AI-aided intelligent healthcare services, a hybrid architecture, i.e., a combination of centralized architecture and decentralized architecture, should be deliberated for distributed learning with a centralized-training decentralized execution strategy.

\subsubsection{Existing AI solutions}
The last few years have witnessed ubiquitous utilization of ML algorithms and DL architectures for a variety of healthcare and medical applications~\cite{Qayyum2021healthcare}, such as physical activity recognition with time-series sensory data and diabetic retinopathy recognition with multimodal images. The authors in~\cite{Huynh2021healthcare} proposed an intermediate fusion framework for human activity recognition using sensory data of wearable devices, in which the deep local features extracted by a deep convolutional network were combined with descriptive statistic features to improve the recognition rate. Subsequently, the fused feature vectors were passed into an SVM classifier to predict activities. A comprehensive diabetic retinopathy recognition method~\cite{Hua2021healthcare} leveraged DL technology with CNN architectures to learn the amalgamation between fundus images and wide-field swept-source optical coherence tomography angiography. In this method, a twofold feature augmentation mechanism was advanced to enrich the generalization capacity of the feature level and prevent CNN from the vanishing gradient problem. In another work~\cite{Cao2020healthcare}, a two-stage learning model with CNN architecture was presented in a lung nodule detection method to overcome the heterogeneity of lung nodules and the complex pattern of the noisy tomography image dataset. The proposed deep model not only improved the accuracy of early lung cancer detection but also facilitated small-scale datasets with a random mask-based data augmentation scheme. Recently, ML and DL have been applied for many other healthcare and medical applications using different data types, such as sleep analysis with electroencephalography signals collected by wearable in-ear devices~\cite{Nakamura2020healthcare} and retinopathy risk progression monitoring with electronic medical records~\cite{Hua2019healthcare}. 

\subsubsection{How XAI can help}

\begin{figure*}[!htb]
    \centering
    \includegraphics[width=0.75\textwidth]{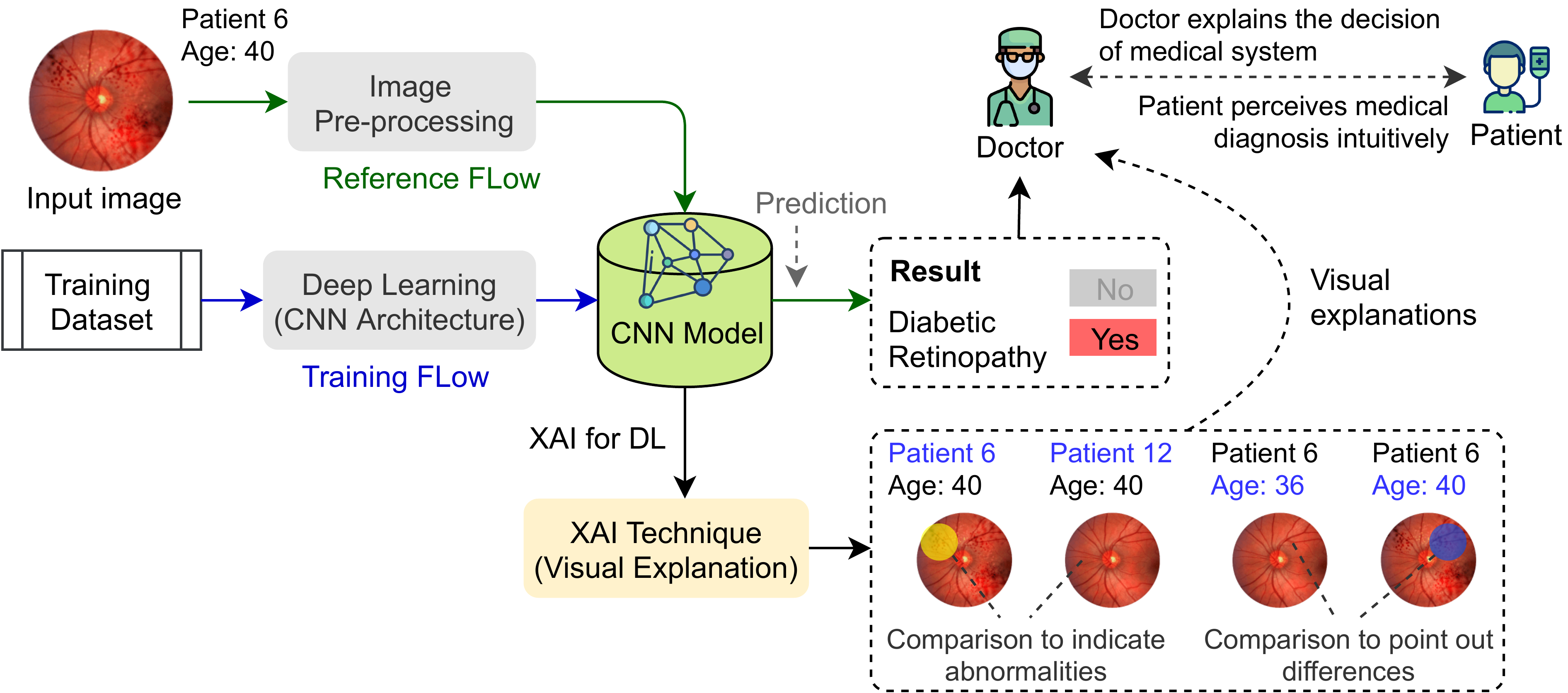} 
    \vspace{1mm}
    \caption{An example of XAI with visual explanation for CNN-based diabetic retinopathy recognition.}
    \label{fig:xai_healthcare}
\end{figure*}
XAI becomes a promising solution for AI models use in the healthcare sector with many benefits such as transparency improvement of AI-based illness diagnosis~\cite{li2020self}, AI-based drug trial result tracking~\cite{liu2020attention}, and model augmentation for health monitoring~\cite{hou2022deep}. Moreover, XAI provides clear explanations on the importance of health parameters such as patient count, patient ages, gender, pre-conditions,  and environments to healthcare stakeholders such as drug developers, doctors, health center, etc~\cite{pawar2020explainable}.
For personalized healthcare in clinical practice, XAI can offer the most appropriate feature set of an intelligent model based on relevant explanations. In~\cite{moreno2021automated}, an XAI processing pipeline, namely feature selection and classification for improving XAI (SCI-XAI), was developed to automatically select feature engineering as well as classification techniques to achieve the best performance of different fundamental tasks, such as detection and classification. SCI-XAI is benchmarked by a fidelity-to-interpretable ratio that measures how much of the model’s interpretability is sacrificed for performance. 
In~\cite{Schoonderwoerd2021human}, Schoonderwoerd et al. proposed a human-centered XAI approach, denoted DoReMi, to generate explanations for clinical decision support systems. 
Remarkably, this approach can provide explanations with multiple levels of interpretability and understandability for different XAI stakeholders, including service providers and end-users (e.g., clinicians and patients).

In~\cite{kragujevac2021model}, four model-agnostic XAI techniques (including LIME, SHAP, Anchors, inTrees) were applied to explain the results derived from an XGBoost classifier for ultrasound image analysis in the application of high-risk asymptomatic carotid plaques prediction. The local explanations generated by different XAI techniques are compared and then synthesized to offer global explanations that can explain the operation specification of the entire model. Four XAI techniques were evaluated using different explainability metrics (including clarity, parsimony, completeness, and soundness) in local explanation generation and global explanation synthesis.
As a result, model designers conceive the advantages and disadvantages of each model in associating and supporting XGBoost, and further measure the correlations between individual local explanations and the global explanation.
With the detailed and clear information derived from XAI, doctors and medical specialists can identify the status of carotid plaques (e.g., shape, location, period, and level) to obtain a more accurate diagnosis and then guide the most appropriate treatment.
DL with CNN architectures have been widely used for medical image analysis and demonstrated high performance in terms of accuracy for various fundamental tasks, including detection, classification, and recognition~\cite{yan2019xAIhealthcare}. With the block-box nature, deep CNN models with a low level of accountability and transparency fail to explain and articulate how their decisions can be reached. Consequently, understanding the operation mechanisms of black-box models is nearly impossible for stakeholders, including end-users and service providers. In this context, developing inherently interpretable models is urgently necessary to render traceable explanations of AI outcomes. 
In~\cite{gulum2021improved}, the combination of saliency maps and gradient class activation mapping (Grad-CAM)~\cite{selvaraju2017gradcam} was explored to improve DL explanations of prostate lesion localization while keeping high performance in terms of accuracy of lesion classification. The saliency maps and Grad-CAM techniques can emphasize not only the individual pixels but also the feature maps that yield the greatest change in class score, which in turn improve the clarity of visual-based explanations and reduce the amount of noise accordingly in localization and segmentation tasks.

For computer-aided skin lesion analysis and diagnosis with human-friendly explanations, Lucieri et al.~\cite{lucieri2022exaid} proposed ExAID, an explainable AI for dermatology framework, which can deal with multimodal inputs. The ExAID framework leveraged an activation vector technique to translate the outcomes of deep networks to human-understandable concepts and explored a localization maps technique to highlight these concepts. Subsequently, the relevant concepts are assembled to construct fine textual explanations which can be combined with concept-wise location information to offer more coherent multimodal explanations.
Besides the ability to access data and come to conclusions, XAI can provide doctors and specialists with the decision routine information to understand how those conclusions are reached. Meanwhile, few conclusions require hints of human interpretation~\cite{Tjoa2020xAIhealthcare}. As a result, with XAI, doctors are able to explain why a certain patient has a high risk of health problems when he should be admitted to the hospital for supervision, and what treatment would be most suitable. An example of applying XAI with a visual explanation technique to recognize image-based diabetic retinopathy is illustrated in Fig.~\ref{fig:xai_healthcare}.
Although current XAI techniques can fulfil the interpretability of black-box models with meaningful explanations to XAI stakeholders, they reveal some serious concerns about the security and privacy of medical data and patient results~\cite{kwon2019retainvis}. In summary, for the services providers, XAI will improve the quality and privacy of healthcare and medical data that is collected in the 6G intelligent sensing layer across heterogeneous IoMT devices and medical systems (e.g., smartphones, smartwatches, electronic medical report systems, and medical imaging systems), enhance the compatibility as well as interoperability of hardware and software produced by different original equipment manufacturers, and improve the performance of healthcare and medical diagnosis with more coherent and confident treatment plans relying on the combination of expert knowledge and explainable result. For the end-users, XAI will provide more detailed numerical analytics and semantic explanations on each medical diagnosis decision to enhance the trust of patients and the confidence of doctors and medical technicians.

\subsection{Industry 5.0, Collaborative Robots, Digital Twin}



\subsubsection{Motivation}
Different from Industry 4.0, which spearheads the explosion of IoT, cognitive computing, big data, and AI over technical interconnectivity and decentralization, Industry 5.0 commits the human touch of business and intelligent systems back into development and production. The primary mission of Industry 5.0 is to create a significant revolution in industrial processes, manufacturing, and business, where problem-solving and creativity-making are the superior objectives instead of replacing repetitive jobs of people with automated robots~\cite{Grau2021Industry}. In this context, the combination of increasingly powerful machines and better-trained experts motivates effective, safe, and sustainable production, in which highly skilled operators and automated robots can work safely and effectively side-by-side on the same manufacturing role to produce personalized and customized products. Such kinds of robots are known as collaborative robots and should be designed to completely accomplish different heavy-precise tasks with high consistency guarantee. Digital twins, recognized as virtual models of the process, product, or service enabling data analysis, system monitoring, and operation and performance assessment via simulations, are promising solutions to optimize business and manufacturer outcomes over managing the entire life cycle of a product~\cite{ramu2022federated}. Relying on the comprehensively predictive and descriptive capabilities of digital twins, customers can comprehend the experience of product functionalities along with operational optimisation fully, while manufacturers provide maintenance services to guarantee digital twins are manageable and profitable~\cite{Ali2021Industry}.  In short, 6G enables Industry 5.0 as 5G cannot provide massive and various personalised information for more accurate and fast production with a full-stack AI-based network. Similarly, the digital twin will go the extra mile due to 6G as it provides ultra-low latency that can interact with the real-life object in a much finer granularity and richer information which is far beyond image, audio, and video. In addition, AI integration in different segments (i.e., RAN, Edge, and Core) of 6G networks is needed to realize better digital twin applications than using AI only at the application level (i.e., 5G does not support network-level AI for the moment). To put it simply, the digital twin will completely characterize a virtual representation of the physical system along with remote sensing, computing, communication, security, and privacy technologies to enable 6G-based IoE services and applications with the capability of automatic planning, analysis, proactive monitoring, control, preventive maintenance, optimization, and business decision making.
In the meantime, the digital twin should exploit AI as a native part integrated with not only the smart application layer but also other ones, including the intelligent sensing layer, data mining layer, and intelligent control layer, thus allowing to deploy of different types of digital twins (e.g., from micro to macro level with component/part, asset, system/unit, and process) more accurately and consistently with the physical entity. 

\subsubsection{Requirements}
In the Industry 5.0 era, we expect to see an intensive upgrade from cyber-physical (i.e., using digital technologies to operate factories to reduce human participation) to human-cyber-physical. 
Interestingly, as with the foreseeable evolution of collaborative robots and digital twins, AI plays a vital role in processing raw data from sensors, analyzing high-level information, and proving decisions and recommendations automatically. 
At the centre stage of this new revolution, humans should work alongside collaborative robots with the support of digital twin systems, teach them to do tedious, repetitive, and dangerous tasks, and correct them when they make operation mistakes or conduct wrong decisions~\cite{Maddikunta2021Industry}. 
Besides the requirement for faster and smarter decision-making in such manufacturing tasks and processes, we desire collaborative robots and digital twins for Industry 5.0 to be more understandable and interpretable, which means, they are able to explain actions and decisions derived from AI/ML models. The complexity and sophistication of AI-powered automated manufacturing systems rapidly increase to argue that humans cannot understand the ambiguous mechanisms of AI systems, especially when they deliver unpredictable and unexpected decisions~\cite{Demir2019Industry}. 
For preparing the incoming wave of Industry 5.0 with mega-factory, collaborative robots, and digital twins, 6G should support offloading of real-time intensive computations tasks, hyper-fast data rate, extremely low latency communications, and highly flexible compatibility of massive IoT devices. In addition, as expected to become an essential requirement in the manufacturing environments, the trustworthiness should be improved with secure-by-design for trusted hardware/software and XAI for preventing adversarial ML.

\subsubsection{Existing AI solutions}
Nowadays, collaborative robots are being developed to support automatic inspection and corrective action in high-precision control systems~\cite{Brito2020Industry}, in which there is an AI-enabled intelligent module designed with DRL to effectively learn and adaptively act based on inspection results. The approach shows two primary advantages: first is learning the AI model continuously without shutting systems down, and second, its extensibility to different real-world scenarios.
In~\cite{Sun2020Industry}, a dual-input DL model was developed to improve the performance of human-robot collaboration and allow robots to learn from human demonstrations effectively.
This model synthesized the assembly contexts of multiple human demonstration processes and tasks to consequently accomplish suitable assistant actions. Compared with traditional feature-based approaches being more complex and time-consuming for labelling a huge amount of data, the proposed method can annotate data labels automatically by perceiving human demonstrations. 
In another work~\cite{Yu2020Industry}, RL with CNN architecture was leveraged to optimize the working sequence of human-robot collaborative assembly, which in turn increases the working performance of smart manufacturing systems. Some complicated learning use cases, such as robot random failure and human behaviour uncertainties, were further taken into consideration to satisfy real-world conditions.
For the promotion of fully smart manufacturing, cognitive digital twins~\cite{Ali2021Industry} were presented by incorporating different modern digital technologies, including industrial IoT, big data, ML, and virtual reality, which aimed to analyze and simulate operation modules, assets, systems, and processes. 
In~\cite{Gehrmann2020Industry}, a secure industrial automation system was built based on a digital twin replication model to identify and verify multi-level design-driving security requirements using sophisticated simulations and optimizations. The digital twin technology was also exploited in some heavy industry sectors~\cite{Wanasinghe2020Industry}, such as shipbuilding, steel production, and oil and gas, to enhance safety and productivity while reducing operational costs and minimizing health risks and accidents.

\begin{figure*}[!htb]
    \centering
    \includegraphics[width=0.75\textwidth]{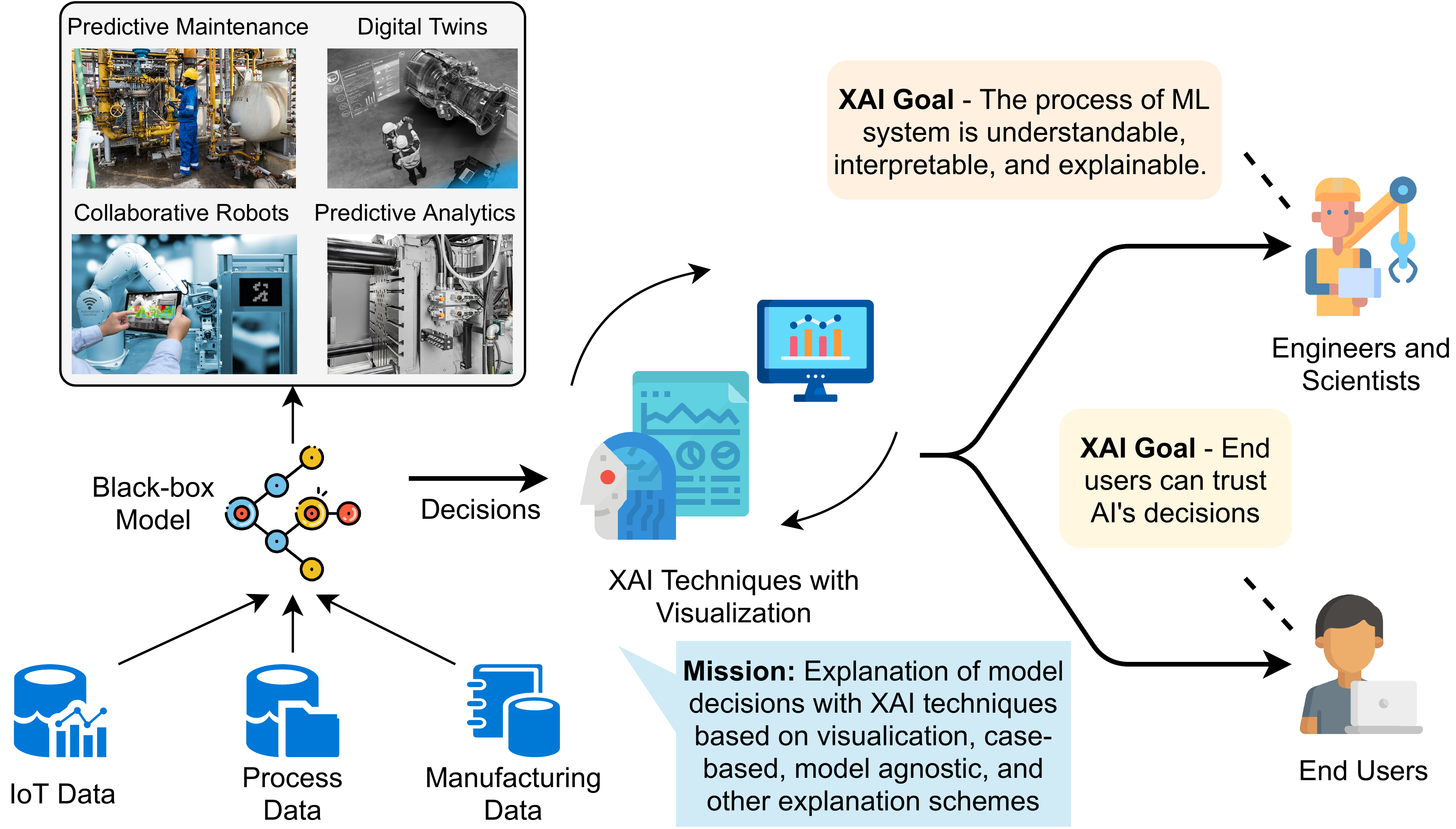} 
    \vspace{1mm}
    \caption{An illustration of XAI for Industry 5.0, where the goal of XAI for different subjects varies: XAI helps end users trust AU's decision while XAI makes engineers or scientists understand the process of ML systems completely.}
    \label{fig:xai_industry}
\end{figure*}

\subsubsection{How XAI can help}
XAI enables humans to understand certain aspects of AI-aided processes or systems, in which XAI answers such kinds of questions: why is the prediction reliable, what are the stable working conditions of an AI model, and when is it likely to crash; and provides more extra deep-analyzed information about system parameters, operation factors, manufacturing environments, and customer feedback to industry stakeholders like system engineers, operators, and factory managers.
Definitely, XAI becomes a sustainable solution for conventional and advanced AI models developed for applications and services in the manufacturing sector for trustworthiness improvement, transparency enhancement, and result interaction of AI-based predictive maintenance~\cite{Chen2020Industry}, ML-based manufacturing diagnosis~\cite{ srinivasan2021explainable}, and data-driven model-based human-robot collaborative assembly~\cite{Piroumian2021Industry}.

6G with massive URLLC and O-RAN can lead to remarkable advances in Industrial IoT (IIoT) and machine-to-machine (M2M) communications, wherein a huge number of automated connected edge devices and modules in smart factories requires a flexible interface, reliable and low-latency wireless communication, energy efficiency, and split computing with data synchronization, especially comprehensive XAI with the capability of self-detecting abnormality, self-monitoring operation, and predictive maintenance. From the viewpoint of system operators, XAI will improve the efficiency of automatic system diagnosis and maintenance based on the sensory data (e.g., sound, image, vibration signal, acceleration signal, etc.)  collected in the 6G intelligent sensing layer, thus reducing time and cost of periodic system maintenance.
In~\cite{Chen2020Industry}, a CNN-based bearing faults diagnosis method was proposed to classify vibration signals with the short-time Fourier transform, in which Grad-CAM was applied to generate the model’s awareness. By analyzing the awareness, visual explanations about the damage conditions of rolling element bearings can be carried out. 
In experiments, the explanations generated by Grad-CAM to articulate the decision of CNN were verified by neural networks, decision trees, and adaptive network-based fuzzy inference systems in terms of correctness and persuasiveness.
In~\cite{srinivasan2021explainable}, a LIME-based XAI framework for chiller fault detection and diagnosis systems, denoted XAI-FDD, was proposed to interpret data-driven classification and regression models like XGBoost and other traditional models. In association with AI models, LIME ranks the importance of sensors based on the failure detection rate, identifies the top of dominant handcrafted features for failure type classification, and estimates the correlation between configurable parameters (e.g., minimum of instance weight and maximum delta step) and learning convergence.
Moreover, XAI-FDD provides AI stakeholders with more meaningful diagnosis information, such as a set of frequently fault modules and components in a chiller system and a set of promisingly effective solutions to overcome failures, thus reducing the cost of system maintenance and avoid the system shutting down.
In industrial control systems, analysts should intervene and respond immediately when any sensor fault is automatically detected by AI models. However, many high-performance detection models have been limited by weak interpretability and explainability. In~\cite{hwang2021explainable}, a comprehensive XAI framework was deployed with Deep SHAP and Gradient SHAP to explain the prediction results derived from LSTM networks and to interpret the feature importance of predictions without compromising the accuracy of deep models. In experiments, it was concluded that Deep SHAP consumes much processing time to interpret complex deep networks (e.g., RNN and LSTM) and Gradient SHAP presents some advantages in handling multiple input models with a comfortable complexity.

Intelligent robots without explanatory capability may cause failures and dangerous actions unexpectedly, hence interpretability and explainability, including post-hoc rationalization and introspection, should be supposed to ensure that there is no conflict between supplementary explanations and regular requirements of applications. 
In the effort to reach human-like communication in human-robot communications, a novel XAI framework~\cite{Gao2020Industry} was developed for collaborative robots by building a human-learn hierarchical AI model with an And-Or graph-based explanation scheme. 
For hierarchical representation, a spatial-temporal-causal And-Or graph (STC-AoG) was designed to encode tasks and sub-tasks with the temporal relation, spatial relation, and causal relation between robots and an agent, which allows robots to understand human intentions. 
Based on the comparison between robot and human mental states that are updated and inferred by STC-AoG, it is possible to estimate the difference between expected human action and predicted robot response to consequently decide whether explanations are given. 
Besides graph-based techniques, reinforcement meta-learning XAI algorithms were studied in~\cite{daglarli2020computational} to interpret spatial-temporal attention and emotion of humanoid robots in collaborative tasks with agents, wherein the explanations are generated and synthesized by a deep belief network, rule-based fuzzy cognitive map, and genetic algorithm to justify robot’s self-awareness and decision-making. 
In the next industrial evolution, digital twins using explainable models and explanation interfaces can become a promising solution to enhance the trustworthiness of AI~\cite{Piroumian2021Industry}, where 6G will uplift digital twins in many aspects (e.g., effectively handle the massive volume of heterogeneous IoT data in real-time with ultra-low latency and ultra-high throughput, increase flexibility with IIoT deployment and configuration for digital twin refinement, improve analysis accuracy in manufacturing operation and reduce time and cost with predictive maintenance). Interestingly, as an exact virtual representation, or virtual clone/mirror, of a real-world asset, a digital twin enables XAI stakeholders to heighten the explanatory capability with augmented reality and virtual reality technologies, for example, the remaining life cycles of modules/components in a production line and the data transmission failure in a manufacturing network.
In summary, XAI helps to improve the efficiency of AI-based manufacturing operations sustainably with collaborative robots on a production line, wherein 6G enables nearly real-time multi-robot task assignment (i.e., factory throughput estimation, device-edge split computing, adaptive computation offloading, optimal energy allocation) in complicated multi-agent environments and support conventional AI, including ML and DL models, with more relevant information (e.g., useless features/characteristics of AI for removal without performance degradation and useful kernels/layers/modules of DL to improve the accuracy of robot's decision-making).
With the end-users, XAI plays the role of providing more detailed explanations and vindications on each action decision and more numerical analytics (simulation, experiment, and test of digital twins before implementing on the physical operation), thus enhancing the trust of customers, operators, and product line engineers.

\subsection{Connected Autonomous Vehicles, UAVs}

\subsubsection{Motivation}
Recently, the wealthy development of connected and automated vehicle technologies is expected to positively change the manner vehicles move and the approach travellers obtain mobility. Connected autonomous vehicle (CAV) is recognized as one of the important vertical industries in 6G with its various quality levels of on-demand services~\cite{He2021Vehicle}. CAV can mean autonomous vehicles which have the capabilities of connecting other vehicles/infrastructures over wireless communications and sensing the driving environment to achieve safe transportation with little or no human involvement. By incorporating advanced technologies, autonomous vehicles collaborate with each other directly over an intermediate infrastructure to improve the performance and efficiency of smart transportation systems if compared with individual autonomous vehicles without collaborative mechanisms. Indeed, the connected vehicle and automated vehicle technologies should be developed in parallel and closely cooperated to put forward completely smart transportation in the future~\cite{Chen2020Vehicle}. To this end, besides 6G-enabled wireless communications, AI plays one of the most important core technologies to process a massive amount of sensory data collected by multiple sensors, which helps autonomous vehicles to understand the surrounding environments and accordingly execute driving activities. In addition to CAVs, the emergence of flying platforms such as unmanned aerial vehicles (UAVs), a.k.a., drones, enables several key potential 6G applications and services in a broad spectrum of domains thanks to their mobility, flexibility, and adaptive altitude. 

\subsubsection{Requirements}
Besides some key connectivity requirements to achieve high-speed, real-time, and reliable data transmission with different communication scenarios, including vehicle-to-vehicle (V2V), vehicle-to-infrastructure (V2I), vehicle-to-cloud (V2C), vehicle-to-pedestrian (V2P), and vehicle-to-everything (V2X), CAVs should strictly demand about the QoS performance of autonomous driving systems. The requirements may vary according to the autonomous level of vehicles: no driving automation, driving assistance, partial automation, conditional automation, high automation, and full automation. To achieve the baseline autonomous requirement, a vehicle needs to be aware of its surroundings during the driving period by first perceiving information and subsequently acting with vehicle control. To fully understand the driving environment, a massive amount of data recorded by multimodal sensors should be processed automatically and accurately via a driving computer system with advanced AI/ML models. Due to the critical demand for safety, a self-driving AI-powered system is expected to operate flawlessly regardless of weather conditions, visibility, road surface quality, and other situational conditions. In this context, advanced ML algorithms and DL architectures have been recently exploited to cover all possible driving scenarios and surrounding environments~\cite{Dulaimi2020Vehicle}, however, the trustworthiness of AI systems embedded in CAVs is questionable. Besides providing useful driving recommendations and accurate driving activities immediately to drivers to minimize the probability of traffic accidents, a revolutionary AI system should be understandable and explainable to make sure that the driver feels confident about their decisions~\cite{Raats2020Vehicle}.

\subsubsection{Existing AI solutions}
For safe and efficient operation on roads, CAVs should understand the current state of nearby vehicles and surroundings to proactively predict future driving behaviours, which in turn allows AI systems to react automatically and immediately. In~\cite{He2021Vehicle}, a comprehensive survey on AI-aided driving behaviour prediction and potential risk analysis was presented, in which the advances of DL with different architectures, compared with conventional ML algorithms, were deliberated in terms of prediction performance. This survey showed that DL has been represented as a promising solution to effectively deal with different sensors (e.g., cameras, LiDAR, and radar) for complicated driving scenarios. The survey also categorized the state-of-the-art driving behaviour prediction approaches into three classes: input representation (track history of target and surrounding vehicles, bird’s eye view of the environment, and raw sensory data), output type (maneuver intention, unimodal trajectory, multimodal trajectory, and occupancy map), and prediction method (RNN, LSTM, and CNN). In~\cite{Basavaraju2020Vehicle}, an innovative road condition supervision system was developed by learning a deep network to reduce accidents caused by poor road quality, in which the sensory data of the accelerometer and the gyroscope were processed in coordination with GPS data. Suddenly changing lanes and braking of the leading vehicle, caused by driver’s distraction, misjudgment, and misoperation, increase the risk of accidents. An autonomous braking decision-making strategy was proposed with deep reinforcement learning in~\cite{Fu2020Vehicle} to facilitate low-level control of CAVs in emergency situations. Despite promising as the core of next-generation intelligent transportation systems, CAVs will face some hidden security problems relating to the complicated AI configurations/settings of autopilot systems~\cite{Sajjad2020Vehicle}, where end-users with limited AI experience and knowledge can be a nuisance and may cause accidents and risky dangers. ML and DL have been recently leveraged for various UAV-based applications in wireless communication systems (e.g., interference management, catching optimization, and resource allocation~\cite{Luong2021UAV }) and intelligent transportation systems (e.g., trajectory planning, traffic flow monitoring, and navigation~\cite{Samir2020UAV}). In cell-free and aerial-assisted vehicular networks~\cite{Wu2020UAV}, UAVs were used to assist CAVs to make decisions in driver assistance systems (for example, path planning and crash warning on highway), in which a supervised learning model was designed with CNN architectures and optimized with particle swarm optimization to make timely inference and facilitate online decisions in real-world scenarios.

\begin{figure}[!htb]
    \centering
    \includegraphics[width=0.5\textwidth]{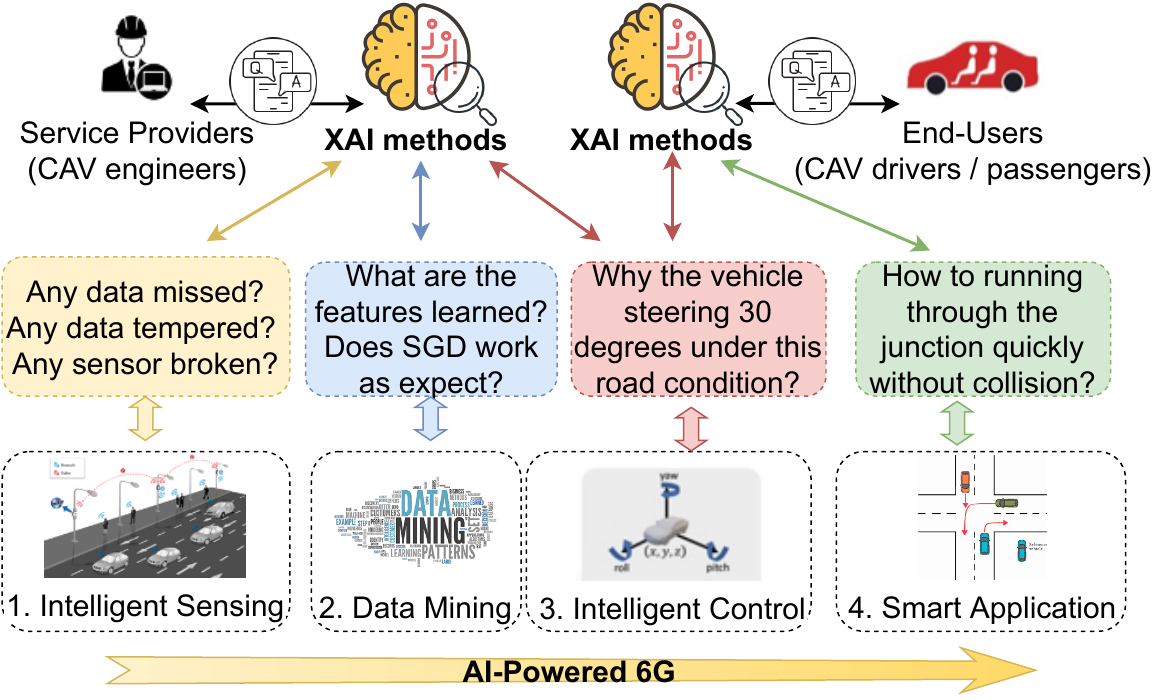} 
    \vspace{1mm}
    \caption{An illustration of typical questions that XAI can answer for various CAV stakeholders (e.g., engineers and passengers) across all AI-powered 6G layers.  Note that the three first questions (left-right) are related to \textit{Service Providers} and the two last questions are related to \textit{End-Users}.}
    \label{fig:xai_cav}
\end{figure}

\subsubsection{How XAI can help}

The important role of AI in driver assistance systems in CAVs is undeniable as we move forward to the next generation of intelligent transportation systems, however, there is an arguable issue of whether drivers completely feel confident and secure with AI-based decisions. 
Several advanced assistance systems developed by some big companies like Tesla are too complicated with numerous AI-based functionalities (e.g., automatic parking, adaptive cruise control, automotive navigation, collision avoidance system, driver drowsiness detection, and electronic stability control), which ask end-users to tune multiple settings to ensure that the systems will operate smoothly and properly. 
The transparency of system operation and the explainability of decision-making are very immature in the field of driverless cars. For example, a self-organizing neuro-fuzzy model coupled with a density-based feature selection technique was leveraged in~\cite{Soares2019Vehicle} to explain automated reacting decisions (e.g., braking, speeding up, changing lane to left), in which the outputs of an AI model were capable of being interpreted by human-understandable if-then rules. 
In~\cite{Keneni2019UAV}, the rule-based XAI was also investigated for different flying events of UAVs, in which the decision of changing the lying path can be explained regarding the weather conditions, surrounding environments, and relative enemy locations based on the representation of if-then rules derived from a fuzzy inference model.
In another work~\cite{Hamilton2020Vehicle}, XAI with a random forest algorithm was proposed to improve the self-confidence level of autonomous navigation in an autopilot system. The intermediate information extracted from trees was helpful in explaining navigation autonomy and AI-driven decisions.
Most of the existing works of XAI for CAVs have focused on explaining the decision of a single system instead of the decision derived from multiple interactive systems, where changeable environments and conditions can affect the outputs of prediction systems and consequently lead to some potential threats. Indeed, explaining why a vehicle changes a lane or hits a break is a more complicated task. 
For an early risk-aware system, a hierarchical AI model~\cite{Khonji2020Vehicle} was studied to predict uncertainty and collision risks under the constraint of account perception, intention recognition, and tracking error. 
At the highest level of automation (i.e., a vehicle is free from geofencing with the capability of reacting as an experienced human driver), explainability is one of the most prior requirements for not only system developers and manufacturers but also customers, passengers, and society.
Fig. \ref{fig:xai_cav} shows some exemplary questions CAV stakeholders may raise from each intelligent 6G layer that XAI can help to answer. Specifically, for the service providers, XAI will improve the quality of data that is collected in the 6G intelligent sensing layer across heterogeneous devices (e.g., vehicles, pedestrians, UAVs) and wireless network conditions (e.g., highway, congested/un-congested urban scenarios), ensure the trained AI models on the 6G data mining layer are robust and generalised to diverse scenarios (e.g., traffic predictions for major/minor roads), and improve the efficiency of network resource management in the 6G intelligent control layer in supporting all intelligent functions (e.g., more switches or routers allocated to congested urban scenarios.). Those (i.e., sensing layer, data mining layer, and intelligent control layer) are the main changes required in 6G when compared with existing 5G architecture. For the end-users, at the 6G smart application layer, XAI will provide more detail on each driving behaviour decision to enhance the trust of drivers/passengers in CAVs.

\subsection{Smart Grid 2.0}


\subsubsection{Motivation}
The core infrastructure in the smart grid is provided by IoT. The smart grid, characterized by automation, informatisation, and interaction, provides a diverse and quality power supply for customers efficiently and securely. To provide supervision of the assets reliably, information interaction in real-time, peer-to-peer trading of energy, load management, and other electrical services, the smart grid needs a communication infrastructure, which is flexible and also highly reliable \cite{liu2021smarter}. 
One of the challenges in making the smart grid more sustainable is to manage the remote communication among various systems that are connected by smart meters \cite{dragivcevic2019future}. Smart grid 2.0 is a futuristic evolution, where seamless connectivity of several power generation sources such as large-scale renewable energy sources is offered. In smart grid 2.0, machine-to-machine communication is facilitated through AI-based algorithms with no-third party intervention to automate the operations of the smart grid. Customers can choose economically viable local microgrids and at the same time pay attention to reducing the impact on the environment. Smart cities, next-generation transportation vehicles such as electric vehicles, unmanned aerial vehicles, and autonomous vehicles would be benefited from their capability of locating the nearest charging pile that offers the best electricity price. In smart grid 2.0, all the exchanges of energy transactions will be autonomously executed by smart autonomous contracts if they meet the pre-defined conditions. In the futuristic smart grid 2.0 the real-time measurements are acquired through IoT sensors, which would be processed in distributed edge devices and the data would be transmitted to energy management systems that use cloud computing. The fusion of big data analytics and deep learning based algorithms would help in realizing self-resilient smart grids \cite{yapa2021survey}.  6G can help in realizing intelligent applications of smart grid 2.0-like remote monitoring and remote controlling of distributed energy resources, automation of demand response, etc \cite{hui20205g}. A smart meter needs the deployment of a distributed network, and wide coverage is essential in preventing blackouts and also to ensure the smart grid's self-healing capabilities. 6G can also help in realizing some of the applications of the smart grid 2.0 that require high-speed connectivity, such as predictive maintenance, video surveillance in real-time or during natural calamities, recovering proactively during emergency times, etc \cite{tariq2020vulnerability,borenius2021smart}.

\subsubsection{Requirements}
To deal with the massive volume of data generated due to constant communication and connectivity in the smart grid, sophisticated techniques that can analyze the data and can assist in the decision-making process are required \cite{sundararajan2020adapting}. ML can solve the problems arising due to the large volumes of data generated from smart grids and can assist the smart grids in the collection of data, analyzing the patterns existing in the data, and also in making decisions to run the smart grid. An ML-enabled 6G network can benefit the smart grid to solve some of the issues in real-time such as automated detection of intruders in the network, forecasting the price of electricity consumption, electricity thefts, line maintenance, generation of power based on demand, optimal scheduling, detection of faults, demand response, prediction of the stability of a smart grid, etc \cite{alazab2020multidirectional}. 

\subsubsection{Existing AI Solutions}

To address the security of the smart grid, Babar~et al. \cite{babar2020secure} proposed a secured demand-side management engine for the smart grid using Naive Bayes, a machine learning algorithm, to efficiently preserve the energy utilization in the smart grid. Due to the increased data generated at a rapid pace in 5G and beyond, enabling the smart grid, data acquisition, and processing by a smart meter is vital. The redundant data present in the data acquired can be reduced by using event-driven sampling. To address this issue, Qaisar~et al. \cite{qaisar2021appliance} employed the SVM algorithm to identify the relevant features for analyzing the consumption patterns of appliances. Providing on-demand services for electric vehicles through vehicle-to-grid systems is very important because of the height manoeuvrability of electric vehicles. In this regard, Shen~et al. \cite{shen2020ev} proposed a hybrid architecture based on cloud and fog computing with applications in the 5G-enabled vehicle-to-grid networks. The proposed architecture allows the bi-directional flow of information and power between smart grids and schedulable electric vehicles to improve the cost-effectiveness and QoS of the energy service providers. For proper scheduling, the selection of suitable electric vehicles is very important. To improve scheduling efficiency, finding the categories of target electric vehicle users is required. To accurately identify target electric vehicles, the authors proposed an artificial intelligence method that is based on the electric vehicle's charging behaviour. In a similar work, Sun~et al. \cite{sun2020integrated} proposed a novel architecture for a 5G-enabled smart grid based on edge computing and network slicing for providing provide on-demand services for electric vehicles. This architecture collects the bidirectional information of traffic between the electric vehicles and the smart grids to decrease the cost of the energy providers and improve the charging experience of electric vehicles. To improve the scheduling efficiency of electric vehicles, the authors proposed LSTM-based electric charging behaviour prediction, KNN-based classification of electric vehicle charging and k-means-based clustering of electric vehicle charging. 
 
\subsubsection{How XAI can help} 

The use of XAI in 6G-based smart grid systems will enable the collection of the most significant data through smart meters and sensing devices located at the intelligent sensing layer. This data will be subjected to XAI algorithms to generate justifiable/interpretable patterns in the data mining layer (demand response, faulty line prediction, closest charging pile detection). Also, the use of XAI will ensure optimized resource management, efficient network automation using ZSM, and accurate information broadcasting through intelligent radio in the intelligent control layer of 6G architecture.  These changes involving XAI in the sensing layer, data mining layer, and intelligent control layer are required in the 6G architecture when compared to the existing 5G architecture. In the application layer of the 6G architecture, the use of XAI will enable more informed and justifiable decision-making capabilities for the end users to ensure enhanced trust and confidence in the smart grid applications. 

Kuzlu~et al. \cite{kuzlu2020gaining} proposed a forecasting approach based on an XAI methodology to predict the generation of PV power, to increase the trustability of AI models, and hence improve the acceptance of AI in smart grid applications. Zhang~et al.~\cite{zhang2021explainable} proposed a Shapley additive explanations (SHAPs) based backpropagation deep explainer, termed as Deep-SHAP method, that produces an interpretable model for emergency control applications in smart grids. 

\begin{figure*}[!htb]
    \centering
    \includegraphics[width=0.95 \textwidth]{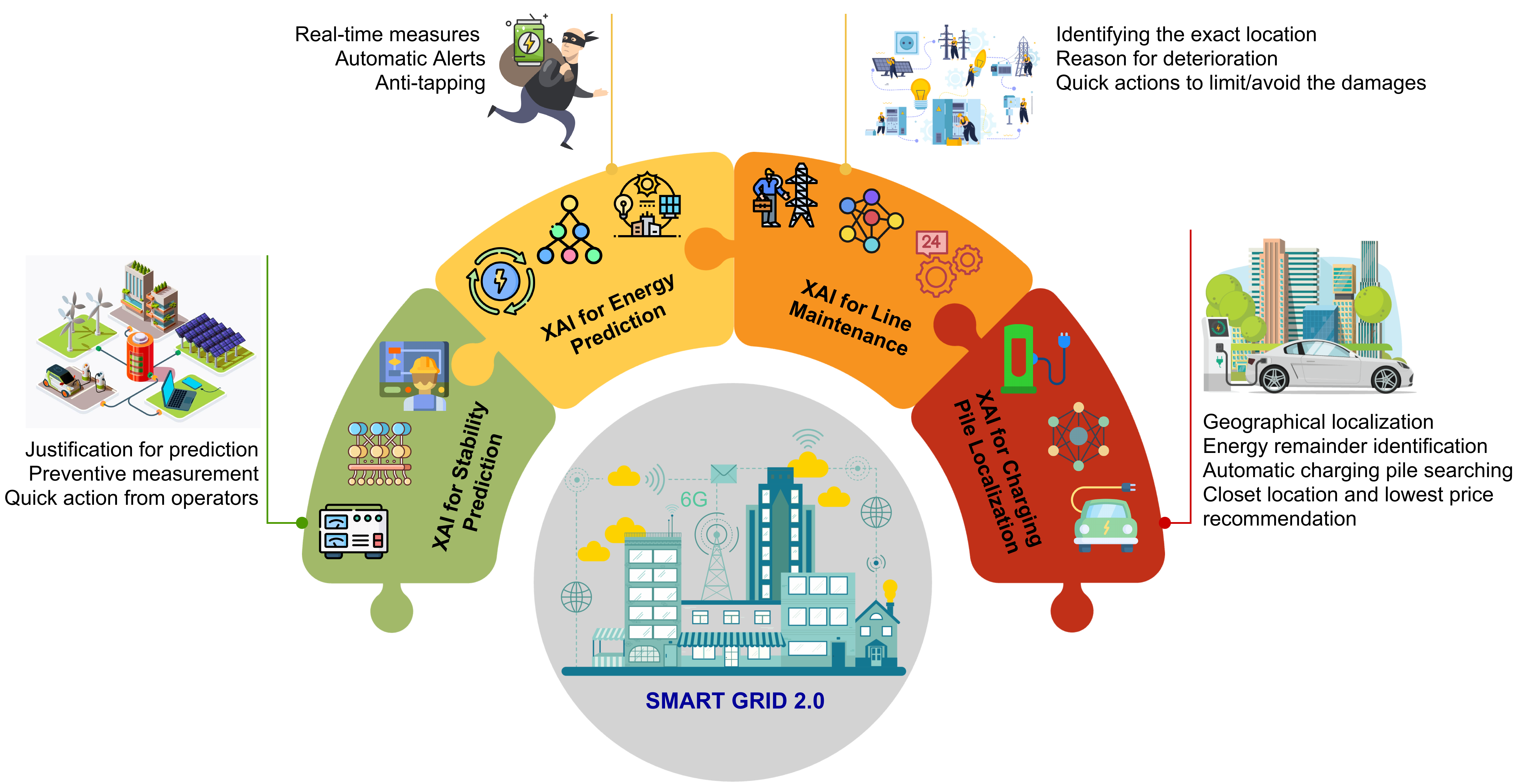}
    \vspace{1mm}
    \caption{An illustration of XAI for 6G-enabled smart grid 2.0: stability prediction, energy theft prediction, line maintenance, and charging pile localization.}
    \label{fig:xai_sg}
\end{figure*}

Some of the use cases of XAI-enabled 6G for the smart grid are discussed below and are summarized in Fig. \ref{fig:xai_sg}.

\textit{Use Case 1: Stability predictions of a smart grid} Maintaining the stability of a smart grid is of paramount importance. There are two criteria concerning the stability of a smart grid. The first criterion is to have a reserve of battery storage to meet the dynamic demand for electricity. The second criterion is providing enough capacity for the stability of the voltage at every location. The instability of a smart grid may lead to power outages and blackouts, which may lead to huge losses in revenue in several industries \cite{singh2014stability,england2020real}. The data from several sensors in the 6G-enabled smart grid can be analyzed by AI/ML algorithms to predict the stability of a smart grid. The AI/ML algorithms should be able to accurately predict the stability of a smart grid early so that the operators can take necessary actions \cite{bashir2021comparative}. Since traditional AI/ML algorithms could not justify the predictions, the operators of a smart grid might be reluctant to take preventive or corrective measures earlier to avoid losses due to instability. XAI can help smart grid operators to identify the causes of the under-voltage load shedding, the need for secondary control power, frequency stability, deterministic frequency deviations, post-fault transient stability issues, and transient stability issues that make the smart grid unstable. The insights provided by XAI may encourage them to take immediate action without further investigation.

\textit{Use Case 2: Detection of Energy Theft in Smart Grid} It is estimated that approximately 96 Billion dollars are being lost every year by the utilities due to energy thefts, which lead to increased prices of energy for the consumers \cite{syed2020detection}. The energy thieves make use of several methods such as tapping a line between a house and the transformer, hacking into meters of neighbours/their own meter, and tampering with the meters \cite{yan2021electricity}. To minimize electricity thefts, we have to identify the most likely cases of theft that can be investigated further. By training the ML models on the data from the smart meters, other external factors like geographic risk in a particular area, and weather in a 6G enabled smart grid, we can generate such a list in real-time that will enable the operators to take appropriate measures immediately \cite{ismail2020deep}. There is no fixed solution regarding the action taken by the investigators/operators of the smart grid. For example, a meter that is reversed may have to be disconnected, if a meter is intruded, the household has to be alerted, and also the meter that is altered has to be replaced, etc. XAI can be helpful in this scenario to explain what kind of theft may happen so that the operators of the smart grid can take relevant action to address the issue.  XAI can help smart grid operators can also identify energy theft caused by physical obstruction, electrostatic attacks on electronic metres, obfuscation of the energy metre, the introduction of foreign material into the energy metre, and direct line connection.

\textit{Use Case 3: Line maintenance in Smart Grid} The reliability of a smart grid depends on the proper maintenance of the infrastructure. The deterioration/ageing of a transformer or power lines due to several reasons such as weather conditions has to be detected at an early stage to prevent the failure of equipment that may cause power outages and blackouts \cite{hu2020adaptive}. The conventional AI/ML methods in a 6G-enabled smart grid can predict the ageing/deterioration of the lines based on several data generated from the sensors such as humidity, weather, and so on in real-time. XAI can be used efficiently in these scenarios to identify the exact location and reason for the deterioration of the power lines so that the maintenance crew can be sent to the exact location with the heads up on the reasons so that they can take appropriate actions to limit/avoid the damages.

\textit{Use Case 4: Location of Closest Charging Pile by Electric Vehicles}
In the futuristic Smart Grid 2.0, electric vehicles can choose the closest charging pile that offers the best price for charging electricity. To realize this, a 6G communication network, which offers ultra-reliable and low-latency continuous services, is essential to constantly monitor the geographical location of the electric vehicles and also to keep searching for charging stations nearby the location of electric vehicles. ML algorithms can help to identify the requirement of the amount the charging for the vehicles and also the apt charging station. However, the black-box nature of the ML algorithms makes it difficult for electric vehicles to choose the charging station among the available piles. XAI can come in handy in this situation. With justification provided by XAI, owners of electric vehicles can get a clear picture of which charging stations can offer electricity charging at optimal pricing considering many factors such as the distance of the charging station from the location of the vehicle, incentives offered, etc. This will help owners of electric vehicles select a charging station based on the analysis provided by XAI.  Hence, XAI-enabled 6G can play a vital role in realizing several futuristic and autonomous services of Smart Grid 2.0 for electric vehicles.

It is assumed that the XAI algorithms such as PIRL, SHAP, and Facets can be integrated with 5G and beyond cellular networks, so that the grid operators can benefit from the justifications/explanations provided by the XAI algorithms that can help them in making apt decisions in several smart grid 2.0 applications/scenarios. Even though XAI has a tremendous potential to improve the operations of smart grid 2.0 by bringing in more transparency and justification of the results obtained by AI algorithms, some of the challenges are to be addressed to reap the benefits of XAI-enabled 5G and beyond in smart grid 2.0 applications. One of the challenges is that due to the lack of metrics to measure the performance of the XAI algorithms, grid operators may face a severe dilemma while taking decisions in real-time.

\subsection{Multi-sensory XR Applications, Holographic Telepresence, Metaverse}


\subsubsection{Motivation}

Extended Reality (XR) is a combination of all immersive technologies such as virtual reality (VR), augmented reality (AR), mixed reality (MR), etc. These immersive technologies are used to extend the reality that is experienced by the creation of an experience that is fully immersive or by amalgamating the real and virtual worlds using multiple sensors \cite{alizadehsalehi2020bim,bilbow2021developing}. 
XR has many real-time and practical applications in many sectors where the travel cost and time of the customers can be saved, such as entertainment, retail, healthcare, real estate, marketing, remote working, disaster handling, etc \cite{andrade2019extended,yastrebova2018future}. 

Holographic Telepresence is a technology, where the systems can project real-time, full-motion, realistic 3D images of people located in distant places into a room, with real-time audio communication, which will make the users feel as if they are communicating with people in person. Unlike in AR or VR, users don't require any device, sensors or headsets to experience holographic telepresence. In holographic telepresence, the captured images of people at remote locations along with their surrounding objects will be compressed and then transmitted through a broadband network. These images will be decompressed at the users' end and then projected through laser beams. Holographic Telepresence has the potential to revolutionize traditional communication through mobile phones by giving immersive experiences to users. It has huge potential in many other applications of communications such as telemedicine, enhanced television and movie experiences, gaming, advertising, robot control, aerospace navigation, 3D mapping, and other simulations \cite{anjos2019adventures,clemm2020toward}.    

Metaverse – a trending term which is the convergence of two principal ideas: virtual reality and digital second life, has been very recently attracting much more interest from various academic communities and big tech companies \cite{gadekallu2022blockchain}. Different from AR which can deliver to users the experiences of video streams and holograms in the real physical world, VR is responsible for conveying immersive experiences in the virtual world. With a VR headset, users can experience numerous services and applications in the metaverse, and create their own hyperreal content. In this context, a variety of AI algorithms have been applied and developed in VR devices to improve the human-machine interactive experience based on modeling and learning visual information.
Some metaverse projects (for example, Decentraland and Sandbox) have successfully built virtual reality platforms, in which AI contributes in many aspects, such as image retrieval, image quality enhancement, and 3-D video rendering. 

\subsubsection{Requirements}


XR and holographic telepresence technologies require a communication network that has near-zero latency, and fast processing of the information from sensors. 6G, through its attributes like connection density, user-experienced data rate, scalability, mobility, reliability, and traffic volume density, can play an efficient role in realizing the true benefits of XR \cite{strinati20216g,giordani2020toward,imoize20216g,piran2019learning}. 

\subsubsection{Existing AI solutions}

AI can be effectively used for self-managing of the participating devices in 6G-enabled multisensory XR and holographic telepresence technologies. Some of the potential applications of AI in the end devices in these applications are to understand the environment by applying computer vision to analyze the multidimensional knowledge from the images captured by the devices, reduction of network volume by enabling AI-based applications in mobile devices, etc. Some of the potential applications of AI/ML in multi-sensory XR applications and Holographic Telepresence are \cite{ARVRAI}:
\begin{itemize}
    \item \textit{Estimation of the Position of Objects:} The object's position (such as fingers, hands) can be inferred for controlling the content of XR \cite{vzidek2019automated}.
    \item \textit{Labelling of Scenes and Images:} Triggering XR labels with image classification \cite{manni2021snap2cad}. 
    \item \textit{Semantic Segmentation and Occlusion:} Segmentation and occlusion of the specified objects\cite{can2021semantic}.
    \item \textit{Object Detection:} The object's extent and position in a scene can be estimated to form colliders and hitboxes to enable interactions between virtual and physical objects \cite{kim2018ar}.
    \item \textit{Recognition of Audio:} Triggering the effects of AR through recognition of keywords \cite{vretos2019exploiting}.
    \item \textit{Recognition and Translation of Text:} The application interfaces of XR can be used for overlaying the text detected from an image into the 3D world \cite{wu2019machine}.
    \item \textit{Content Generation:} Designing of environment,  characters, and other graphical objects \cite{sicat2018dxr}.
    \item \textit{Virtual Humans:} Training of animations so that they can respond in real-time \cite{schmid2018future}.
    \item \textit{Virtual Assistants for Dynamic Customer Experiences:} Training of virtual assistants that answer the queries of customers to provide a virtual experience on the latest trends \cite{mclean2019hey}. 
\end{itemize}

\subsubsection{How XAI can help}
XAI can provide valuable reasoning and justification on the predictions/classifications that can convince the providers of XR/Holographic Telepresence applications to make decisions based on the outcome of the AI/ML algorithms. For instance, accurate estimation of the position of objects such as fingers/hands of humans is very important in controlling the content of XR. AI/ML algorithms can be used to estimate the position of the objects.  Taking decisions based on the recommendations from AI/ML algorithms in real-time may sometimes lead to inaccurate content of the XR due to false positives. If the 6G is enabled with XAI, the application providers will understand the reasoning behind the predictions/classifications by the AI/ML algorithms that can help them in taking accurate decisions in real-time. Similarly, if 6G is enabled with XAI, the virtual assistants can provide accurate information to customers \cite{turner2022next}.

In holographic telepresence-based applications, the use of XAI will enable the generation of realistic 3D images (eg. human anatomy, clinicians, human holograph images) of individuals located in geographically distributed locations in the intelligent sensing layer. This data will be processed using XAI algorithms providing a realistic and immersive experience to the users without the use of any physical devices, or sensors for achieving optimum quality holographic presence. In the intelligent control layer, the use of XAI would enable seamless and faster processing of data through network automation and ZSM, optimized resource management ensuring near zero latency while streaming of the 3D images ensuring justifiable decision making by the stakeholders in the intelligent control layer in the 6G architecture. The aforementioned changes are essential for the integration of XAI in 5G architecture for realizing the true potential of holographic telepresence applications in 6G. The use of XAI will enable the realistic presence of the end users and the accurate positioning of objects in a holographic telepresence environment ensuring accurate decision-making at the 6G smart application layer.
The authors in \cite{hayes2020trustworthy} proposed guidelines for using XAI techniques and simulations using XR for secured human-robots interactions. The authors suggested that the proliferation of high-fidelity VR-based simulation environments will result in the reduction of barriers in cataloguing and performing postmortems in operations by robotics that may result in the characterization of more rigorous autonomous system behavior and promote the adaption of explainable techniques in their controllers.

When XAI is integrated with 5G and beyond cellular networks, it may lead to performance degradation of the AI algorithms to enable explainability.
In the metaverse, many VR-based services and applications leverage advanced AI models to enhance user experiences with interactive activities; however, they are usually presented as black boxes without interpretability and explainability. In the effort to completely explain AI decision-making processes, a variety of XAI algorithms and methods can be studied for many development tasks (e.g., object detection, semantic segmentation, image super-resolution, 3D video rendering, etc.) in the metaverse framework. For instance, with LIME and LRP, 3D designers and computer vision scientists who apply DL to build virtual worlds can understand and explain what is happening inside deep models and when they are likely to be broken down.

\begin{table*}[htbp]
\caption{The requirement analysis of XAI for typical B5G/6G use cases}
\centering
 \begin{threeparttable}

\label{tab:use_cases}
\renewcommand{\arraystretch}{1}
\begin{tabular}{|p{4cm}|p{6.5cm}|p{0.5cm}|p{0.5cm}|p{0.5cm}|p{0.5cm}|p{0.5cm}|p{0.5cm}|p{0.5cm}|}
        \hline
        \rowcolor{Gray}
         &  & \multicolumn{3}{c|}{\textbf{Key Stakeholders\tnote{a}}} & \multicolumn{4}{c|}{ \textbf{XAI Challenges\tnote{b}}} \\ \cline{3-9} 
         
        \rowcolor{Gray}
        \multirow{2}{*}{\textbf{6G Use Cases}} & \multirow{2}{6cm}{\textbf{Typical High-Stake AI Decisions}} & {\rotatebox[origin=c]{90}{ Service Providers}} & {\rotatebox[origin=c]{90}{Legal Auditors}} & {\rotatebox[origin=c]{90}{End Users}} & {\rotatebox[origin=c]{90}{Explainability Metrics}} & {\rotatebox[origin=c]{90}{ Exp. Perf. Trade-off }} & {\rotatebox[origin=c]{90}{ Legal Engagement }} & {\rotatebox[origin=c]{90}{Privacy Leakage}} \\ \hline
         
    Intelligent Health and Wearable, Body Area Networks & Medical Image Diagnosis, Health Risk Prediction, Treatment Planing and Supervision & \cellcolor{yellow!30}M & \cellcolor{green!30} H & \cellcolor{yellow!30}M & \cellcolor{yellow!30}M & \cellcolor{red!30} L & \cellcolor{green!30} H & \cellcolor{green!30} H \\ \hline
    
    Industry 5.0, Collaborative Robots, Digital Twin & Predictive Maintenance and Analytic, Quality Inspection, Machine Condition Monitoring and Supervision & \cellcolor{green!30} H & \cellcolor{yellow!30}M & \cellcolor{yellow!30}M & \cellcolor{green!30} H & \cellcolor{green!30} H & \cellcolor{red!30} L & \cellcolor{red!30} L \\ \hline
    
    CAVs and UAVs & Collision Avoidance & \cellcolor{green!30} H & \cellcolor{green!30} H & \cellcolor{green!30} H & \cellcolor{yellow!30}M & \cellcolor{green!30} H & \cellcolor{green!30} H & \cellcolor{yellow!30}M \\ \hline
    
    Smart Grid 2.0 & Predictive maintenance, Theft detection & \cellcolor{green!30} H & \cellcolor{yellow!30}M & \cellcolor{red!30} L & \cellcolor{yellow!30}M & \cellcolor{yellow!30}M & \cellcolor{yellow!30}M & \cellcolor{red!30} L  \\ \hline
    
    Multi-sensory XR Applications, Holographic Telepresence, Metaverse & Remote operation of machines & \cellcolor{green!30} H & \cellcolor{yellow!30}M & \cellcolor{green!30} H & \cellcolor{yellow!30}M & \cellcolor{red!30} L & \cellcolor{yellow!30}M & \cellcolor{green!30} H \\ \hline
    
    Smart Governance & Policy recommendation & \cellcolor{yellow!30}M & \cellcolor{green!30} H & \cellcolor{green!30} H & \cellcolor{green!30} H & \cellcolor{yellow!30}M & \cellcolor{green!30} H & \cellcolor{green!30} H \\ \hline
    
    
  \end{tabular}
  \begin{tablenotes}
            \item[a] Detailed discussions can be found in Table \ref{tab:stakeholders}.
            \item[b] Detailed discussions can be found in Section \ref{sec:xai_6g_lim}.
        \end{tablenotes}
        \end{threeparttable}

\begin{flushleft}
\begin{center}
\begin{tikzpicture}
\node (rect) at (1,2) [draw,thick,minimum width=1cm,minimum height=0.4cm, fill= red!30, label=0:Low Demand] {L};
\node (rect) at (6.3,2) [draw,thick,minimum width=1cm,minimum height=0.4cm, fill= yellow!30, label=0:Medium Demand] {M};
\node (rect) at (12,2) [draw,thick,minimum width=1cm,minimum height=0.4cm, fill= green!30, label=0:High Demand] {H};
\end{tikzpicture}
\end{center}
\end{flushleft}
\end{table*}

\subsection{Smart Governance}

\subsubsection{Motivation}
Smart governance is perceived as the intelligent use of ICT and innovation to facilitate and support enhanced decision-making, planning, and citizens’ role through collaborative decision-making \cite{pereira2018smart}. The motivation of smart governance is similar to the ones realised under the ideals of good governance \cite{abdellatif2003good} in modern-day democracies, with an additional focus on ICT to uphold the ideals, ensuring the development and welfare of the public and public resources \cite{mani2007democracy}. The fundamental challenge that remains relevant in the existing governance is that of corruption \cite{pillay2004corruption} and unfair policies, and methods to improve education, security, transport, resource management, and economic infrastructure, which is where smart governance is envisioned to offer better solutions. 

Presently, 5G is enabling hyper-connectivity, decreasing latency, increasing traffic capacity, and improving throughput compared to 4G and predecessor networks. Causing an evolution in smart governance applications due to unprecedented levels of real-time information from any device, anytime, and anywhere, while improving public infrastructure and experience \cite{rao2018impact}. The main beneficiaries of smart governance are the general public and the government. The demand for smart governance will grow with the growth of 6G to drive innovative applications enhancing the user experience of governance by collecting user data and rewarding with enriched information through the advancement in AI and XAI \cite{luckey2020artificial}, such as finding the quickest path to a destination, following election campaigns, law enforcement decisions or guidelines, or postal service tracking.

\subsubsection{Requirements}
To fully realise the vision of smart governance, we need ICT services with high bandwidth connectivity and more devices to connect and communicate with more focus on understanding decision-making. The understanding of decision-making means the internet supports XAI, empowering users with the decision and accompanying explanations to keep the user informed of it. This decision-making and explanation would require real-time operations, such as while driving in cooperative traffic management; therefore, high latency would be catastrophic. Complex traffic management would require big data operation for faster and safer commutes when many devices and sensors produce data simultaneously. Besides this, by the time 6G goes into effect, the advancement in XAI will be furthered, meaning smart governance applications would require support from infrastructure to consume XAI to its futuristic potential. It will empower users with transparency, meaning 6G should allow application support in terms of explanation instead of focusing on improving hardware and signal processing.

Overall, the idea of the internet of everything will be a crucial requirement for smart governance, requiring all sensors and devices spread across smart cities to achieve high-speed and real-time connectivity. Also, data transmission would demand highly reliable data exchange to ensure QoS performance and optimal service to citizens, the government, and other stakeholders. Additionally, to implement 6G smart governance, the key difference compared with the existing 5G network is that a much higher level of security and privacy is required, especially at the intelligent sensing layer of the 6G architecture shown in Figure \ref{fig:xai_6g}. It is crucial to ensure the trustworthiness of the massive data collected from diverse sensors by the network itself, as humans cannot inspect data on such a large scale. Moreover, as most of the data collected in smart governance scenarios may endanger the privacy concerns of personal data, the whole data collection process needs legal compliance, which XAI has great potential to ensure through transparency.




\subsubsection{How XAI can help}

The 5G drive has emphasised  improving  the network infrastructure while neglecting the application side of services. Hence, it restricted applications from fully exploiting XAI in the context of smart governance. Such as, the current smart monitoring services have limited use of transparency in decision-making \cite{brauneis2018algorithmic}, which is one of the reasons engineers don't fully trust automated decision-making, where human oversight plays a crucial role in the monitoring. 
 
For example, in the oil and gas sector, monitoring and maintenance of valves are critical to steady operations. Here the AI helps with decision-making by informing when a valve is degrading and needs to be replaced; however, the AI does not explain the decision, which makes it hard for hardware engineers to trust the decision. Things can go wrong when a valve is coming to the end of life, and the AI decides it is healthy. From a governance point of view, it will be highly catastrophic if oil gets leaked into the sea, causing significant environmental harm. Therefore, to avoid this circumstance, AI needs to explain the decision to bring transparency, empowering hardware engineers to make an informed judgement. 
To this end, recently
 some efforts have been made to explain decisions through graphs and reports based on semantics that allows engineers and decision-makers to understand automated decisions better \cite{thakker2020explainable, qureshi2020valve}. 
However, the efforts could be advanced using XAI approaches such as LRP, Bayesian RL, LIME, and SHAP  that can improve the quality of explanation by embedding explanation as a part of the system instead of as a mere add-on; this will be strengthened through 6G transparency and high-speed network. Also, here, explanations can indicate what has changed visually in the video feed or through sensor data points that may call human attention for monitoring purposes. 
Another area of application is public engagement in the system of governance. At present, governments are opting for limited use of dashboards when it comes to smart cities, and smart governance \cite{matheus2020data} to share insight and stats. In this regard, there were some applications  developed to bring insights during the election campaigns \cite{qureshi2016twittercracy} and movements of public concerns and social causes \cite{poghosyan2016topy} (e.g., austerity, Brexit, refugee crisis) with the use of AI.

Though these types of dashboards bring some transparency, such as which politician or political party has a specific stance concerning a domestic political issue on their social media, what is the public voice concerning austerity, and the voice of different media outlets. However, these dashboards don't explain their decisions to the degree where XAI is making advancements. With XAI's inherent strength of explaining the decision and 6G's interface that ensures high-quality real-time data ingestion, 
\definecolor{Gray}{gray}{0.9}
\definecolor{LightCyan}{rgb}{0.76,0.98,1}
\definecolor{LightGreen}{rgb}{0.90,1,0.76}

\newcolumntype{g}{>{\columncolor{LightGreen}}p{0.25cm}}
\newcolumntype{q}{>{\columncolor{LightCyan}}p{0.25cm}}

\afterpage{}
\clearpage
\thispagestyle{empty}
\begin{landscape}
\onecolumn
\setlength\LTleft{-180pt}            
\begin{longtable}{@{\extracolsep{\fill}}|c|p{3.5cm}|p{10.5cm}|g|g|g|g|g|g|q|q|q|q|q|q|q|@{}}
    \caption{Summary of key related works in XAI algorithms and goals for 6G use cases and technical aspects 
    }
    \label{tab:ReferenceSummery}\\
  \hline
    \multirow{2}{*}{Ref.} & \multirow{2}{*}{XAI Algorithm} & \multirow{2}{*}{XAI Goal} & \multicolumn{6}{c|}{\cellcolor{LightGreen}6G Use Cases} & \multicolumn{7}{c|}{\cellcolor{LightCyan}6G Technical Aspects} \\ \cline{4-16}
    & & & {\rotatebox[origin=c]{90}{~Intelligent Health ~}}
    &{\rotatebox[origin=c]{90}{~Industry 5.0~}}
    &{\rotatebox[origin=c]{90}{~CAVs and UAVs~}}
    &{\rotatebox[origin=c]{90}{~Smart Grid 2.0~}}
    &{\rotatebox[origin=c]{90}{~XR, Holographic Telepresence~}}
    &{\rotatebox[origin=c]{90}{~Smart Governance~}}

    &{\rotatebox[origin=c]{90}{~Intelligent Radio~}}
    &{\rotatebox[origin=c]{90}{~Trust and Security~}}
    &{\rotatebox[origin=c]{90}{~Privacy~}}
    &{\rotatebox[origin=c]{90}{~Resource Management~}}
    &{\rotatebox[origin=c]{90}{~Edge Network / AI~}}
    &{\rotatebox[origin=c]{90}{~Network Automation, ZSM~}}
    &{\rotatebox[origin=c]{90}{~Other technical aspects~}}\\ \cline{1-16}
    

    \cite{mahbooba2021explainable} & 6-layer Decision Tree & 
    Enhance the trust management for Intrusion Detection System.
    &   & \checkmark & & \checkmark & & & & \checkmark &  & & & \checkmark &  \\
    \hline

    \cite{morichetta2019explain} & LIME & 
    Analyzing the rationale behind the unsupervised machine learning model for encrypted video traffic clustering.
    &   &  & &  & \checkmark & & &  &  & & &  & \checkmark \\
    \hline

    \cite{ayoub2022application} & SHAP & 
    Analyzing how feature importance varies among different machine learning models when predicting the bit error rate of the lightpath.
    &   & \checkmark & &  &  & & &  &  & & &  & \checkmark \\
    \hline

    \cite{barnard2022resource} & SHAP & 
    Demystify the behaviour of complex AI models used in short-term resource reservation within sliced networks.
    &   &  & &  &  & & &  &  & \checkmark & & \checkmark & \\
    \hline

    \cite{terra2020explainability} & LIME, SHAP, XGBoost, PI, Eli5 & 
    Evaluate the performance of multiple XAI methods in identifying root-cause of SLA violation prediction.
    &   &  & &  &  & & & \checkmark &  &  & & \checkmark & \\
    \hline

    \cite{ayoub2022using} & LIME, SHAP & 
    Achieve a deeper understanding of ML algorithm behaviours in automated failure-cause identification in microwave networks.
    &   & \checkmark & &  &  & & \checkmark & \checkmark &  &  & &  & \\
    \hline
    
    \cite{gulum2021improved} & Grad-CAM & Improving DL explanations of prostate lesion localization while keeping high performance in terms of accuracy of lesion classification.
    & \checkmark  & & & & & & &  \checkmark &  & & & & \\
    
    \hline
    
        \cite{Chen2020Industry} & Gradient class activation mapping (Grad-CAM) &
    Analyzing the rationale behind the AI model CNN in bearing faults diagnosis by analysing the vibration signals.
    & & \checkmark & & & & & & & & & \checkmark & &  \\
    
    \hline
    
        \cite{Gao2020Industry} & Bayesian inference for ST-causal And-Or graph &
    Generated explanations significantly improve the performance and user perception of the human-robot collaboration.
    & & \checkmark & & & & & & \checkmark & & &  & &  \\
    
    \hline
    
    
    
    \cite{Keneni2019UAV} & Rule-based XAI &
    Explain the lying path decision of UAVs regarding weather conditions, surrounding environments, and enemy locations.
    &  & & \checkmark & & & & \checkmark &  & & & \checkmark & &  \\
    
    \hline
    
    

    
    \cite{kuzlu2020gaining} & LIME, SHAP, Eli5 &
    Gaining insights from AI models in predicting solar PV power generation for smart grid. 
    &  & & & \checkmark & & & & &  & &  & & \checkmark \\
    
    \hline

    

\cite{hayes2020trustworthy} & XAI Guideline &
     Guidelines for using XAI techniques and XR simulations for secured human-robot interactions. 
    &  & & &  &\checkmark & & & \checkmark & \checkmark & &  & &  \\
    
    \hline

\cite{nagisetty2020xai} & LIME &
     Using fewer data to achieve better performance of GAN-based privacy-preserving solution by XAI's explanations on the discriminator’s decision. 
    &  & & &  & & & & \checkmark & \checkmark & &  & &  \\
    
    \hline
    
    
    
    \cite{matin2021earthquake} & SHAP & 
    Using XAI explanation to improve the feature engineering for earthquake-induced building-damage mapping.
    &  & & &  & & \checkmark &  & & & &  & & \checkmark \\
    
    \hline

 \end{longtable}
\end{landscape}

\twocolumn
these dashboards will further advance and play essential roles during the elections, referendum and other political processes for all political stakeholders (public, politicians, and government). In these dashboards, XAI approaches will further open doors for explaining the insights, with some visualisations also depicting real-time data, such as during elections. These explainable approaches can be in the form of visual, textual or conversational nature, where the likes of surrogate, visual and textual XAI approaches will contribute to improving the overall user experience in explaining insights.


Different XAI algorithms shown in Table \ref{tab:xai_ai} can be applied in the context of smart governance, where visuals and reports based on semantics will be useful \cite{thakker2020explainable}. Model-agnostic approaches \cite{matin2021earthquake} can be beneficial for complex decision-making algorithms (such as DNN) where explanations may be demanded in a real-time setting (applications such as route planning and smart monitoring). However, model-specific approaches will be useful where the precision of the decision's explanation is significantly essential. These types of decision-making may not require real-time processing, such as when legal auditors are among stakeholders or when decision-making informs policy-makers, and other applications can be city infrastructure planning. Both model-agnostic and model-specific can also co-exist for several applications within smart governance. Here, model-agnostic approaches can provide a simpler and quicker explanation that may be useful for end-users. For a more detailed and accurate explanation, model-specific explanations can be provided. The applications here would include all areas of smart governance, such as smart monitoring, election campaigns, causes of public concern, law enforcement, and public order.

Despite the great potential of XAI in smart governance, a simple explanation through a simplified user interface of visuals or reports-based explanation might not inform the user of an accurate explanation. It may mislead individuals and become critical when applications belong to causes of public concern. Here, application providers must inform users of the potential risk of simplified explanations. Also, there is a danger of trying to explain election campaigns due to a lack of metrics of explanation, as the campaigns might be so dynamic and novel that it would be challenging even for a human expert to explain them due to subjectivity, let alone XAI to do the job. However, acknowledging the shortcomings of technology may help. On the contrary, a complex explanation may be more accurate but hard to comprehend. Also, real-time decision-making and explaining the decision may be critical for law and order applications requiring accurate information that may require more computational power available today, which may be achieved through technological advancements.
In particular, for stakeholders, XAI will offer facilities to citizens and help better govern cities and infrastructure, which is directly related to the data quality and reliability in high volume, which is ensured through 6G. Also, until IoE is fully realised, the promises of smart governance, cities, and offered smart infrastructure cannot be truly achieved. Especially most of the network support for IoE would be wireless, which is where promises of 6G are important. From a stakeholder's perspective, getting informed about a high volume of data from IoE means a correct decision that can be trusted through explanation. Overall, for smart governance, the efficiency of mobility and resource management are critical issues that would be potentially dealt with through the vision of 6G, similar to Open RAN real-time operations via xAPPS deployed in NR-RIC.

\subsection{Summary of the XAI impact on 6G applications and technical aspects}

This section reviews the attempts and potentials of XAI for a wide range of 6G applications. Different from 6G technical aspects, 6G applications involve more types of stakeholders than engineers only: such as end-users and legal auditors. Therefore, the requirements of XAI for 6G applications will have to be analysed case by case. In Table \ref{tab:use_cases}, for each 6G use case, we describe its typical high-stake AI-powered decisions that need the XAI most. Then, we identify the level of demand for XAI for each stakeholder, and for each XAI challenge that needs to be addressed in the future. For instance, collision avoidance of CAVs or UAVs is a typical high-stake AI decision as the incorrect decision can lead to the loss of life. Thus, all stakeholders would need the explainability for such decisions at the highest level. Moreover, collision avoidance requires both high model accuracy and explainability to give evidence for legal experts to judge responsibilities on various occasions. Another example from Table \ref{tab:use_cases} is the quality inspection of Industry 5.0. If a product is mistakenly qualified, it would be risky for service providers mainly as it affects their reputation. As most Industry 5.0 activities are within the factory, the demand for wider legal engagement and higher privacy protection is low.

In Table \ref{tab:ReferenceSummery}, we also summarised the key references that are discussed in both this section and the previous section about 6G applications and technical aspects. Lots of existing work focused on the security, privacy, and edge AI technical aspects as these contain the most high-stake decision-making process.  In addition, these technical aspects also have AI solutions deployed already in many existing systems. However, the lower-level intelligent radio and the backhaul ZSM are lacking attention, partly because these technologies are still being studied in the early stage. The existing research interests are roughly evenly distributed across all 6G applications discussed in this paper. CAVs and UAVs are the ones that have attracted the most interest so far due to their large existing research communities, while intelligent health still demands more work as it requires close interdisciplinary collaborations.

\section{Limitations and Challenges of XAI for 6G} \label{sec:xai_6g_lim}
This section summarises the major limitations of the recent research of XAI for 6G described in the earlier sections. The corresponding challenges and future research directions to move XAI for 6G forward are also elaborated.

\subsection{Limitations of XAI for 6G}

As previously mentioned, XAI can increase the trust and transparency of AI-powered ``human-centric” 6G networks for all stakeholders. However, several well-known limitations \cite{arrieta2020explainable} of existing XAI methods would delay the successful deployment of XAI in the future 6G infrastructure.
\begin{itemize}
    \item \textit{Lack of in-model XAI methods}: There is a widespread concern \cite{gunning2019darpa} that the performance of the AI models will go down with the growth of its corresponding level of explainability due to the ever-increasing level of AI model complexity. Recently, in-model XAI methods are likely to be satisfactory in both AI performance and its corresponding explainability. It is because the in-model XAI methods are designed to be self-explanatory, rather than an add-on to the XAI method after the AI decision is made (i.e., post-hoc XAI methods). However, most existing XAI methods are still post-hoc (e.g., LIME, SHAP), as they are more straightforward \cite{arrieta2020explainable} to be developed and pluggable with existing AI algorithms compared to in-model methods.
    \item \textit{Lack of quantifiable explainability metrics}: Visual and textual explanations are two commonly observed formats of XAI methods output. These explanations are intuitive for human beings but difficult to measure objectively using quantifiable metrics. Therefore, it is challenging for XAI system designers to achieve standard/unified systems that are simple to deploy and use for all stakeholders. 
    \item \textit{Lack of engagement of stakeholders and legal experts}: A strong motivation for introducing XAI technologies is to address the legal requirements. A typical example is the ``right to explanation” in the EU GDPR which requires machine algorithms to be capable of giving explanations for their outputs.  In the last few years of early research into XAI, many new technologies are being proposed and applied by computer scientists. However, two important points are missing. Firstly, a meaningful engagement of legal experts is required to ensure that XAI complies with legal requirements. Secondly, a deep engagement of stakeholders is needed to ensure the explanations provided by the new XAI methods make good sense to them.
\end{itemize}
Apart from the three limitations mentioned above, possible privacy leakage \cite{grant2020show} is also a significant concern that exists inherently without any assuring solutions. This possible privacy leakage refers to the fact that when XAI is applied, more information will inevitably be exposed externally concerning the AI decision-making process, which likely leads to the leakage of users' data. Anonymisation might be a possible solution to protect private information. However, if one can easily violate such protection, the risk of privacy leakage is still high.

\subsection{Challenges of XAI for 6G}

This subsection details future research challenges, according to the previously mentioned limitations of using XAI for 6G. These challenges mainly include: devising quantifiable metrics to measure the effectiveness of explainability, proposing new XAI methods that can achieve a better trade-off between interpretability and model performance, and improving societal and economic engagements through the application of XAI for 6G. 

To measure the effectiveness of explainability generated by XAI tools, the effort that researchers made in the early days when DARPR XAI program \cite{gunning2019darpa} was launched in 2017, was focused on proposing general assessment frameworks or metrics across different domains. Doshi-Velez and Kim \cite{doshi2017towards} proposed a taxonomy of XAI assessment methodology which contains three classes: application-grounded (i.e., measured for specific applications), human-grounded (i.e., measured for specific stakeholders), and functionality-grounded (i.e., measured for specific AI algorithms). Hoffman et al. \cite{hoffman2018metrics} discussed the evaluation of XAI in depth from both psychometric and AI perspectives. They proposed an evaluation process for measuring: the goodness of explanations, user satisfaction, user understanding of the AI system, user motivation for explanations, user trust and reliance on the AI, and the performance of the human-XAI work system. Inspired by the success of the system usability scale (SUS) that has been widely used for assessing the usability of the system-human interface for over three decades, Holzinger et al. \cite{holzinger2020measuring} proposed a system causability scale (SCS). The SCS has ten questions to measure if XAI-generated explanations quickly meet the users' intention.

XAI researchers soon realised that to make explainability measurement more effective, the quantifiable metrics have to be designed specifically for stakeholders and scenarios and cannot be domain-agnostic. For example, in the AI health domain, Kaur et al. \cite{ kaur2021trustworthy } proposed a metric called ``Trustworthy Explainability Acceptance" that measures the Euclidean distance between XAI explanations and domain experts’ reasonings in predicting Ductal Carcinoma in Situ (DCIS) recurrence using AI. In the computer network domain, Li et al. \cite{li2020trustworthy} proposed a metric called quality of trust (QoT) to quantify the level of trust when a particular XAI model is applied for 6G applications. The QoT contains a physical and emotional trust, representing the objective and subjective assessment of explainability, respectively. The key attempts of the research mentioned in this section are summarised in Table \ref{tab:metrics}.

Quality of trust (QoT) \cite{li2020trustworthy} is an exemplary initial work in proposing quantifiable XAI metrics for 6G. However, there are still many challenges to implementing a full set of XAI metrics for all key stakeholders in different 6G scenarios. For example, as AI will be used across all layers of the computer network infrastructure, the number of specific 6G AI scenarios that require explanations will be significantly higher than the existing 5G networks. Do we need to propose a set of metrics for each of these 6G scenarios? Can we reuse some metrics across various 6G AI scenarios? What are the metrics that we have to design specifically for particular use cases? Additionally, although SCS \cite{holzinger2020measuring} is a good strategy for measuring user satisfaction with the generated explanation, the low-latency 6G networks may need quick solutions to capture the users’ feedback. Moreover, for use cases where the stakeholders are engineers or scientists such as radio spectrum allocation, it is necessary to gain expert explanations in advance to ensure high effectiveness of XAI explanations for example   \cite{ kaur2021trustworthy }. However, the acquirement of explanations from domain experts needs to be carefully planned to avoid the huge potential cost in time and labour.

\begin{table}[h]
    \centering
    \caption{Key Recent attempts on measuring the explainability of the XAI outcomes.}
\label{tab:metrics}
\rowcolors{2}{gray!15}{white}
\begin{tabular}{p{0.4cm} p{0.4cm} p{2.7cm} p{0.6cm} p{1cm} p{1cm}}
\rowcolor{gray!50}
\toprule
Ref. & Year & Metrics Proposed & Quanti-fiable & Domain Specific & 6G Applicability\\
\midrule

\cite{doshi2017towards} & 2017 & application-grounded, human-grounded, functionality-grounded & No & General & Medium \\

\cite{hoffman2018metrics} & 2018 & the goodness of explanations, user’s motivation for explanations, the performance of human-XAI work system & No & General & Medium \\


\cite{holzinger2020measuring} & 2020 & SCS & Yes & General & Medium \\

\cite{li2020trustworthy} & 2020 & QoT & Yes & 6G & High \\

\cite{kaur2021trustworthy} & 2021 & Trustworthy Explainability Acceptance & Yes & Health & Low \\

\bottomrule
\end{tabular}
\end{table}

\subsubsection{Performance and Interpretability Trade-off in large-scale 6G Infrastructure}

During the last decades, researchers were mainly focused on the performance of AI, paying little or no attention to interpretability \cite{gacto2011interpretability}. However, the GDPR shifted the focus, a regulation recognising ``an explanation of the decision reached after assessment" for the users and holding automated algorithms accountable. The caused shift is now deriving from the trend towards the interpretability of the models.

In AI, the system's performance can be measured by two popular metrics, which capture the effectiveness of the predictions. For classification, the metric used is called accuracy, which represents the percentage of correct predictions over the total predictions. For regression, the metric used is called mean square error (MSE), which captures the average of the squares of the errors between the predicted and actual value. On the other hand, interpretability is the ability of how well the model's decision-making ability can be made transparent to the user, irrespective of whether the decision is right or wrong. It is important to note that interpretability can be subjective, which explains why it lacks a quantification metric and is still seen as a quality \cite{gacto2011interpretability}.

On the other hand, interpretable models are those in which it is possible to understand their inference capability due to the simplicity with which they were designed \cite{gunning2019xai}. For example, in a simple decision tree where the nodes represent the values of specific attributes, edges are rules, and the leaf nodes represent the class. With a simple decision tree, it is more manageable to understand the overall decision by following rules through edges and nodes, and eventually, the leaf node's final decision. In simple words, it will be like if-else rules concluding a decision, therefore, making it interpretable. However, these rules will grow with the depth of the decision tree but remain interpretable, although it is more laborious for human cognition when depth increases significantly. 

Even though simple decision trees are interpretable but unlikely to be best at performance, to overcome the performance limitation, a committee of multiple trees was proposed, such as the Random Forest, however with a compromise on interpretability. Random Forest calculates the final decision based on voting among multiple trees. The strategy of exploiting different ML models to predict a decision is called ensemble methods, and in doing so, undermines interpretability. Some other methods like non-linear Support Vector Machine are not interpretable because of the complexity due to the non-linear decision boundary \cite{gunning2019xai}. Examples of interpretable models are Bayesian classifiers, linear models, and rule-based models. These models can be defined by several rules that retain interpretability at the expense of performance \cite{rai2020explainable}.

\begin{figure}[!htb]
    \centering
    \includegraphics[width=0.45\textwidth]{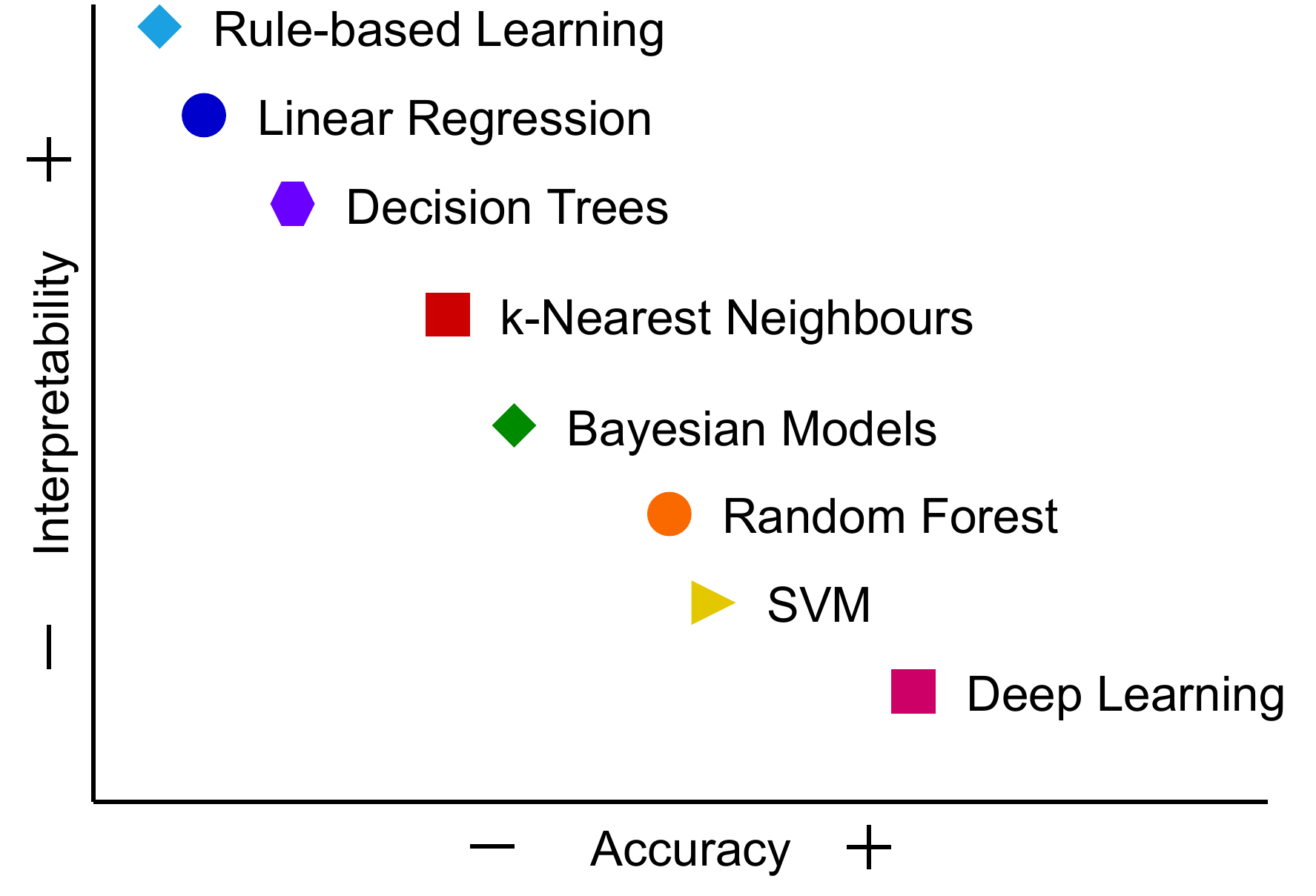} 
    \vspace{1mm}
    \caption{Representation of the Interpretability versus Accuracy according to \cite{gunning2019xai}. Due to the increasing model complexity (i.e., from linear to non-linear) and training data size (i.e., both in volume and number of features), it normally shows the trend that the model interpretability decreases with the increasing model accuracy.}
    \label{fig:trade-off}
\end{figure}

As seen in Figure \ref{fig:trade-off}, the least interpretable model is DL which is a clear example of this paradigm. Furthermore, the more interpretable models tend to be less accurate. DL is an evolution of the ANN, with an increase in the number of intermediate layers that allow it to learn complex relations from huge datasets. DL models could be composed of 30 layers with thousands of nodes where each node captures a distinct feature value. For example, the largest deep neural network until February 2020 is called Turing-NLG (from Natural Language Generation) and was released by Microsoft. Turing-NLG has around 17 billion parameters \cite{rasley2020deepspeed}. They are called black boxes because it is hard to see the inference behind the predictions emanating from the combination of all the nodes across the multiple layers.

The trade-off between the performance (accuracy) and the simplicity (interpretability) of a model has been studied many times in the literature \cite{gunning2019xai,bohanec1994trading,plate1999accuracy}. The more complex a machine learning model is (such as a higher number of nodes, more rules, branches, or layers), the less likely it is to be interpretable. Adding complexity to the model is likely to model complex decision boundaries, making the model prediction more accurate. The fundamental challenge is to ensure higher accuracy without compromising on interpretability. In some domains, interpretable methods can provide similar levels of accuracy, and therefore they are recommended \cite{rudin2019stop}. That is why when selecting the algorithms to build the large-scale 6G infrastructure, we need a clear knowledge of which of the two (accuracy or interpretability) is more important in each domain.

Some methods are more accurate, and others are more interpretable \cite{gunning2019xai}. In the same way, the simpler the model is, the faster it will be at prediction time. Selecting a suitable method depending on the domain and application is essential. In some fields, such as redirecting IP packages over a network, it is not crucial to understand the model's internal reasoning as long as it is very efficient. On the contrary, there are domains such as medicine or high-stack decision concerning an organisation in which understanding the model's internal reasoning is paramount. 

Explainability is different from interpretability because it can explain through additional strategies which may have no direct link with the inference of the true AI model (see Section \ref{subsec:explain_and_inter}). Explainability patches black-box models without making them fully interpretable, aiming to provide users with justified explanations. Suppose we want both high performances and justify why a decision has been taken by the model. In that case, we can use different strategies like model-agnostic approaches or posthoc explanations to translate the complex reasoning of black-box models into a human-understandable for a specific type of audience. It is always vital to consider the target audience's needs to which explanations are presented; something readily understandable by a specialised group such as what may work for cardiologists might not work for financial experts. One of the core challenges towards incorporating ML models into businesses is that business executives do not fully accept or trust ML models unless satisfactory explanations behind the decision are provided. Therefore, it is essential to develop more reliable and explainability-based techniques \cite{vilone2020explainable}.

Another position is to entirely avoid black-box models because they hide the actual inference process, which increases the chance of making blunders without knowing why. Similarly, DL methods that automate the feature selection prevent developers from identifying important features against redundant or superfluous features \cite{rudin2021interpretable,rudin2019stop}, which significantly hinders overcoming black-box challenges. There are two possible approaches to overcoming the black-box limitation. First, promote the usage of simple models with limited but acceptable performance \cite{rudin2021interpretable}. Second, develop better and more effective explainability techniques, which is the growing trend in the DL community \cite{vilone2020explainable}.
It is vital to decide which algorithmic design should be preferred in each of the modules that compose 6G. In some modules like redirecting IP packages efficiently or selecting the best antenna for a given user, we will be more interested in accuracy than in interpretability. Nevertheless, other modules will interact with humans in a more direct way, such as route suggestions for cars, in which we need to prioritise the interpretability of the models \cite{letaief2019roadmap}.


\subsubsection{Societal and Economical Engagement in Integrating XAI with 6G infrastructure}
The third challenge of adopting XAI for 6G is about societal and economic engagement. Specifically, we first examine trends concerning laws and ethics. Then, we discuss commercial challenges concerning technology producers and intellectual property. Finally, we present the need for new laws and regulations to evolve and support XAI for 6G.

\begin{itemize}
    \item \textit{Laws \& Ethics} With the growing demand and development of XAI and IoT, the challenges concerning ethics will continue to evolve \cite{atlam2020iot}, empowering the user with more choices and preferences on transparency in automated decision-making. Moreover, with the growth of IoT to IoE within 6G, more devices would be connected with decision-making capability, raising security and privacy concerns. Several devices will likely interact directly without requiring a human in the loop. However, the endpoint will serve human demands. Here, additional laws and an understanding of ethics would be in demand to ensure humans control all activities. Laws like GDPR are necessary to mention, which will lead and enable innovation with the optimal level of governance over automated decision-making. However, GDPR alone does not encompass a full spectrum of sets of rules all around the globe. In addition, the Chinese PIPL and other regional laws would collectively undertake the future challenges of ethical practices concerning automated decision algorithms and devices.
    \item \textit{Commercial Side} Constituting decision-making algorithms explainable through the law or practice can cause concern for the producer of technology. The producer would find making algorithms transparent a risky business to protect the intellectual property of the technology. This concern would further grow when 6G starts to converge on the application-centric approach to automated decision-making, with the vision of IoE. It is here, that the approach taken by the US is vital to consider which views data protection and data integrity as a  commercial asset, unlike GDPR. It is important to note that explaining decision-making can leak critical information to competitors, which can quickly compromise commercial assets and undermine freedom of ownership. The right approach would balance GDPR and US laws regarding data privacy and protection while safeguarding consumer rights and businesses.
    
    \item \textit{Compliance of New Laws} With the boom of 6G, continuous evolution would take place from IoT to IoE, focusing on applications that could explain automated decision-making. However, unlike hardware infrastructure, software applications get upgraded quickly, and many new types of applications keep popping up, unlike new hardware development. At this pace, future applications that would be required to explain decisions can be made restricted by strict laws that do not permit flexibility to confirm all types of growing needs. Here, the laws should tolerate flexibility to assert the law's spirit that encourages best practices. These laws should enable consumer rights protection as well as adequate protection of commercial assets, ensuring both personal freedom and freedom of ownership. Finally, future laws should promote globally accepted rules while guaranteeing regional and transnational freedoms that provide some adjustment and flexibility. Otherwise, international acceptance of law that confirms the right to explanation to exact detail internationally may be too ideal to agree upon before 2030. The common future data privacy laws should strike a balance that protects businesses and users from exploitation while respecting national policies globally.

\end{itemize}

A possible solution is to evolve current laws with a balanced approach that recognises consumer rights in terms of ethical considerations and ensures technology producers' intellectual property concerns. In addition, these laws should be flexible to allow rapid application development that matches the pace of 6G adoption. Since the upcoming 6G is a global phenomenon, future products would greatly benefit from common international law. This common international law would better connect the world, safeguarding consumers' rights and technology producers while ensuring rapid growth and adoption of 6G products internationally.

\subsection{Summary and our opinion}

At the beginning of this section, as XAI is still in its infancy, we have briefly discussed three main limitations of using XAI for 6G systems. Then, we described three promising research challenges to address these limitations accordingly. Firstly, more in-model XAI methods should be proposed to achieve a better trade-off between the interpretability and the performance of the model. Secondly, more quantifiable metrics are required to verify the XAI goals. Thirdly, societal and economic engagement should be boosted to meet XAI requirements that are set in local regulations.

Even when XAI becomes mature, XAI is still a double-edged sword, the same as every other technology. When applying XAI for 6G, XAI's explanations help build trust for stakeholders. However, these explanations also expose extra information about the AI decision-making process. Such a risk of privacy leakage could be alleviated by closely working with stakeholders to perform privacy violation checks when developing XAI solutions.


\section{Lessons learned and Future Research Directions} \label{sec:future}





This section briefly summarises the lessons learned from the topics discussed in the previous sections, which include the background of AI and XAI, the impact of XAI on typical 6G technical aspects and use cases, limitations, and challenges when developing XAI for 6G, and major research projects and standardisation on XAI and 6G. The corresponding future research directions for each of these topics are also discussed.

\subsection{XAI Technique}



\subsubsection{Lessons learned}

Before 2010, although some AI scientists made great efforts in designing explainable ML models, the focus was mainly placed on improving the models’ accuracy. Therefore, the complexity of the models sharply increased (from simple rule-based models like decision trees to DL now). In contrast, the non-technical impact of complicated AI models was overlooked. After 2010, when ML models were applied to areas that affected humans like diagnosing cancer or approving a bank credit, the focus shifted to concerns on transparency of AI decisions, such as making sure decisions were not made based on attributes of the person such as the ethnic background or race of people. 
These concerns involve more questions like why the system decides in a particular way. In the imminent 6G age, where various model devices can talk to each other, more AI decisions will be made at a much faster pace. The transparency and trustworthiness of responsible AI have to be considered formally so that experts from academia and industry can improve the overall technology ecosystem for the future user experience.

XAI is a promising set of technologies that provide a level of transparency in the decision-making process behind the AI black box. Though XAI is still in its early stages, the current important lesson learned so far is the trade-off between interpretability and performance. When stakeholders need a higher degree of explainability, AI system designers may have to compromise the quality of the prediction/classification results. XAI has also a very important role in validating the model. Sometimes models can be biased and make predictions based on non-related factors to the output.

Another important lesson to be aware of is that ML models are heavily dependent on the quality of the training dataset \cite{stoica2017berkeley}. In many cases, incorrect AI predictions or classification will lead to massive losses for 6G stakeholders in both economical and non-economical (e.g., health and life) aspects. It means that the robustness of AI systems is of particular importance. Most of the existing solutions \cite{stoica2017berkeley} focus on improving the AI robustness, which is to carefully design the data pipeline, including data collection, pre-processing, augmentation, and dimensionality reduction. This helps with reducing the error rates.

\subsubsection{Key Research Problems}
Explainability is the cornerstone for professionals adopting AI models and validating the logic of the models. Some of the important questions that arise in this area are:

\begin{itemize}
  \item Is it possible to improve the explainability techniques by creating an extra layer to translate into layman's terms the logic behind ML models so we can provide models with both accuracy and suitable explanations?
  
  \item Would the next generation of ML models fill the demands of current legislations such as GDPR which highlight the importance of the “right to explanation” so that people are entitled to explanations of the outputs' algorithms that affect an individual?
  
  \item To what extent ML models will expand their use to other domains by providing better explanations to increase the professionals' trustfulness?
    
\end{itemize}

\subsubsection{Preliminary Solutions}
One of the most prominent lessons learned during several decades of research, development, and commercialisation of AI is that the performance of AI, especially ML systems, in certain domains is not as important as the explainability/interpretability of the model. And that an important strategy to involve professionals in implementing ML models is to increase their truthfulness in the model by providing explanations of the decisions models make \cite{ribeiro2016should}. These explanations can be in the shape of text, graphs, or by providing an interpretable model \cite{arrieta2020explainable}. Currently, they are many technologies used to generate explanations such as LIME and SHAP. However, they have a few limitations like that they do not well with all the models (Lime does not work well with XGBoost) or that SHAP is terribly slow in some methods such as k-NN. LIME is in general faster than SHAP, but LIME’s explanations based on linear models do not guarantee the same levels of consistency as SHAP.

\subsubsection{Future Directions}
The wide and deep convergence of XAI to the existing AI systems is foreseen to be increasingly important. Some promising future research directions for this deployment include are: how to measure the level of explainability, how to satisfy the explainability demands from multiple stakeholders, how to ensure high performance while still being able to provide a high level of explainability, and how to work collaboratively in a multi-disciplinary team (i.e., typically, ICT researchers with legal experts). Generating clear and consistent explanations for accurate models in an easy way remains a future challenge for AI with great potential to make AI more transparent and accessible to humans.

\subsection{Standardization and Research Projects}

\subsubsection{Lessons learned} XAI is becoming an exciting research area under 6G.  For the moment, IEEE is leading the standardisation activities related to XAI. Especially, the IEEE Computer Society / Artiﬁcial Intelligence Standards Committee (C/AISC) and IEEE Intelligence Society - Standards Committee (CIS/SC) are leading these tasks. However, the current 6G SDOs, such as ETSI, and 3GPP, have not focused entirely on the XAI domain. Legal frameworks for XAI have already been developed at the global level, including in the EU and USA. 

Several reputable research projects for 6G using XAI have already started. Mainly, the EU H2020 funding program, EU Christ-era funding program, EU MSCA program, and US DARPA program have funded many projects in the XAI domain. Many of these projects are not directly related to 6G. However, most of these projects focus on technologies and applications associated with B5G and 6G networks.

\subsubsection{Key Challenges}

The SDOs and funding organizations have to solve the following key challenges to support XAI integration for the 6G domain:

\begin{itemize}
  \item How to establish collaborations between telecommunication SDOs, AI organizations, and funding agencies to define basic XAI requirements for 6G service deployment?
  \item Encourage the development of open-source XAI projects and encourage funding organizations to invest in XAI enabled telecom projects.
  \item Provide training and education programs for SDOs and funding agencies to increase their XAI expertise.
  \item How to foster partnerships between SDOs and funding organizations with leading XAI experts and companies to facilitate the integration of XAI into 6G services.
\end{itemize}

By addressing these challenges, SDOs and funding organizations can support the successful integration of XAI in the 6G domain,  leading to  secure, accountable and responsible AI-enabled telecom services.

\subsubsection{Preliminary Solutions}

Standardization of Explainable AI (XAI) can be integrated with current standardization efforts for Zero-Touch Service and Network Management (ZSM). The focus of ZSM standardization is AI-powered service management in B5G networks. Leading telecom SDOs, such as ETSI, NGMN, 3GPP, and ITU-T, should consider incorporating XAI into their 6G standardization plans.

\subsubsection{Future Direction}
Since AI will play a significant role in 6G networks, it is necessary to consider the necessity of XAI for 6G applications. Initially, research projects can build new knowledge on utilising XAI for B5G and 6G networks. EU funding programs such as Horizon Europe and Eureka programs can be ideal venues for funding research related to XAI and 6G integration. In addition, global level 6G programs such as Japan 6G/B5G Promotion Strategy and South Korea MSIT 6G research program can also be possible venues to obtain research funding for XAI and 6G integration. 6G standardisation is essential to deﬁne the technological requirements of 6G networks and select suitable technologies to deploy 6G networks. XAI will be considered one of the critical technologies to utilise in 6G networks. 
\color{black}
\subsection{XAI for 6G Technical Aspects}



\subsubsection{Lessons learned}
Conventional AI/ML algorithms and innovative DL architectures have been applied for different tasks in 6G networks when considering technical aspects. The objective includes accuracy improvement in intelligent radio and edge networks, reliability enhancement in network security and data privacy, and optimisation in resource management. While the system performance and automatic decision of communication networks mostly depend on AI models, it does not usually provide descriptions and explanations about results, especially from the how-why-when perspective. XAI, owning to three principal features, including explainability, interpretability, and accountability, represents a promising technique to help not only end-users but also AI stakeholders understand how an AI model processes data and conducts outcomes automatically, which in turn allows end-users to be confident with its decision as well as engineers to comprehend their systems. Some ML models present good interpretability, but their performance in terms of accuracy is unacceptable. Therefore, the balance between interpretability and accuracy should be considered in XAI-based system design. 
For example, some XAI approaches have exploited the interpretability of AI models like a rule-based model and linear regression to generate explanations. However, their accuracy can not satisfy the baseline QoS in 6G. On the other hand, although DL showed high performance in dealing with many fundamental tasks such as detection, classification, and recognition, they offered little or no interpretability.
Furthermore, depending on the input data type, storage infrastructure, computing platform, and communication infrastructure, XAI for explanation generation should be appropriately chosen to deal with a specific technical problem while ensuring a reasonable performance in terms of accuracy and complexity. Besides, the explanation should be simple for end-users with less domain knowledge and advanced for AI stakeholders with high expertise, which can be numerical results, text, graphs, images, and simulations. It can contain details on how a statistical AI model causes a prediction from a feature set, a decision path from a decision tree, a rule from a simple model, and a visual operation graph of information flow. Lastly, explanations brought by XAI may bring extra information leakage unintentionally to potential privacy violators. XAI can help strengthen AI accuracy and efficiency, but it can also tell others how existing AI models work, which in some cases should be confidential. Thus, extra attention should be put when applying XAI to privacy-based 6G solutions.

\subsubsection{Key Research Problems}

Although XAI can help enhance the technical aspects of 6G networks related to intelligent radio, trust and security, privacy, resource management, edge AI, network automation and ZSM by improving the explainability, interpretability, accountability, and justification of decisions made by AI models. However, to integrate XAI with 6G, a few challenges have to be addressed which are discussed below.

\begin{itemize}
  \item  In the aspect of intelligent radio, with its interpretability and explainability, XAI has the potential to revolutionise intelligent radio in 6G, but it also increases the system's complexity. This creates a research problem on how to enhance explainability and interpretability without increasing system complexity.
  \item In the aspect of trust and security, thought XAI will provide administrators and stakeholders with significantly insightful, comprehensive security and trust information. The data may also be altered, which will influence the XAI model's decisions. This raises a research problem regarding the development of tamper-resistant storage and sharing of data. 
  \item  In the aspect of privacy, accountability will be improved by XAI, but the privacy of the shared data will be a problem since the data might be collected by a third party without the consent of the legitimate user.
  \item In the aspect of resource management, XAI can help with resource management in 6G, the algorithms used for resource management cannot be explained at a high level because results vary from user to user. Context-dependency is a research problem which is needed to be addressed.
  \item In the aspect of edge AI, XAI can improve the performance and explainability, but the issues, including the decreased performance of the AI algorithms and the absence of metrics to measure the performance of the XAI algorithms, must be addressed in order for 6G to be enabled by XAI with edge AI.
  \item In the aspect of network automation and ZSM, XAI can improve the interpretability and justification of the decisions made by AI. However, performance deterioration of AI algorithms due to integration of XAI is a concern that has to be addressed since the judgements made by ZSM may be mission-critical and may influence the network bandwidth and resource allocation.
  \end{itemize}
By addressing these research problems, XAI can further enhance 6G networks.
\subsubsection{Preliminary Solutions}
A critical component for reducing system complexity is a model governance environment. A good model of governance reduces the risk of a compliance audit and establishes the platform for transparent, ethical AI that eliminates bias in 6G networks. The use of blockchain for data storage, which is later fed to the XAI for decision making in 6G networks, will enhance trust and security. The blockchain with its consensus, cryptography, and decentralisation principles helps XAI to train on data which is tamper resistant  and trustworthy. By eliminating the need for data from local models in the creation of the global model, Federated Learning will improve the privacy of XAI in 6G networks. The use of incentives based on XAI will enhance resource sharing in 6G networks and help ZSM respond faster in mission-critical situations.

\subsubsection{Future Direction}
Despite certain benefits of interpretability and explainability, the utilisation and development of XAI for different technical aspects in 6G networks are still limited. In this context, future work can focus on incorporating DL with several explanation techniques for XAI at multiple levels, from processing units to operation modules and systems. For example, the visual explanation technique can be applied to capture and then visualise the feature activation maps of a trained CNN, which are helpful to explain the labels outputted by a DL-based automatic modulation classifier in intelligent radio. Besides visual explanation techniques (e.g., class activation mapping, peak response map, and class-enhanced attentive response), some other textual explanations (e.g., question-answering and semantic information retrieval) and numerical explanations (e.g., concept activation vectors and local interpretable model agnostic) approaches should be leveraged for a broad spectrum of 6G technical aspects. The combination of different explanation techniques (such as numerical and visual explanations) can be helpful for a sequence of interactive decisions produced by hierarchical AI models in complex systems. Many existing XAI works have concentrated on explainability for ML at different stages in developing AI systems to address several technical tasks in 6G. However, there remain some gaps in data explanation methods, which are helpful in selecting and exploring a better-suited model later. Additionally, XAI methods should be designed to incorporate domain knowledge to explain useful inferences under clear and uncertain circumstances. One last promising future research direction is to identify to what certain level of explainability provided by XAI could be potentially harmful to preserving the privacy of XAI stakeholders.

\subsection{XAI for 6G use cases}


\subsubsection{Lessons learned}
Existing AI methodologies can provide prediction/classification for future 6G-based applications such as healthcare, Industry 5.0, CAV, smart grid, multi-sensory XR applications, and smart governance to help make decisions in real-time. Decision-taking in mission-critical applications such as healthcare, smart grid, and smart governance should be done very carefully as it may result in the loss of properties and lives and cause significant danger. However, the black-box nature of AI-based algorithms makes it very difficult for decision-makers to trust the results of these algorithms as they lack justification/explanation. The explanation should be technologically aware and thoroughly address the ethical, legal, and societal questions. 
To address these issues, XAI will be essential in future 6G-based applications (especially healthcare, autonomous driving, and smart governance) to trust, understand, and improve the accountability of the decisions made by AI-based algorithms. It will help instil confidence in end-users as they can understand the decision-making process of these algorithms. However, several key challenges and open issues need to be addressed to realise the full potential of XAI in the 6G-based applications which are discussed below.

The improved interpretability may result in reduced performance of AI algorithms in terms of real-time decision-making and prediction accuracy, which is unacceptable in mission-critical applications such as smart healthcare, autonomous vehicles, and smart grids. Hence, the trade-off between interpretability and performance is an open issue that needs addressing. Another important challenge is addressing the issue of the high dimensionality of the data generated from the applications based on IoT in real-time due to the high bandwidth and reduced latency of the 6G network infrastructure. Furthermore, the generation of labels for the data in real-time in the big data era makes it suitable for classification which is a tedious and demanding task. In the case of 6G-enabled applications that use heterogeneous networks, providing explainable and customised decisions is another open issue that needs to be addressed in future research. Another critical issue is the privacy preservation of sensitive data generated from applications such as healthcare, connected and autonomous vehicles, and smart grid. The malicious users or attackers can gain access to the private and sensitive data generated from these 6G applications through several means such as poisonous attacks.

\subsubsection{Key Research Problems}
Intelligent health and wearable, body area networks, industry 5.0, collaborative robots, digital twin, connected autonomous vehicles, UAVs, smart grid 2.0, multi-sensory XR applications, holographic telepresence, the Metaverse and smart governance are some of the use cases which benefit from the integration of XAI with 6G networks. However, there are still some challenges which are needed to be addressed.

\begin{itemize}
  \item  In the case of intelligent health and wearables, and body area networks, healthcare stakeholders may benefit from explanations and assistance from XAI in interpreting AI models' decisions. However, the information used to feed XAI models may potentially come from unreliable sources, and the information from these sources can yield inaccurate results. Therefore, the decisions taken by stakeholders can have grave consequences. Identifying these unreliable sources is a research problem to be addressed.
  \item  In the case of industry 5.0, collaborative robots, and digital twins, the stakeholders will be benefited by the  trustworthiness improvement, transparency enhancement, and result interaction provided by XAI. However, in the process of decision-making requires data from multiple sensors, any fault in these sensors can lead to erroneous decisions, identifying the fault sensor in real time is an issue to be addressed.
 \item In the case of connected autonomous vehicles and UAVs, the drivers will benefit from the suggestions provided by XAI on issues related to collision alerts, driving alerts, and navigation assistance. However, most drivers lack the knowledge necessary to comprehend the justification and evaluate the decisions provided by XAI. Because of this issue, the aid provided by XAI is rendered ineffective in some circumstances. 
 \item In the case of smart grid 2.0, the accountability provided by XAI will help identify the theft, the reason for electric outrage, and also help the experts respond in an emergency. However, the complexity of the system will increase as it becomes necessary to get data from the whole chain of operations from multiple sources in order to produce decisions, which is an issue to be addressed.
 \item In the case of multi-sensory XR applications, holographic telepresence, and the metaverse, XAI can improve the quality of service and experience in these use-cases. However, the issue related to system complexity, security, and privacy is still an issue to be addressed. 
 \item In the case of smart governance, XAI can provide accountability and transparency for decisions made. However, due to the politicization, there is a chance of possible conflict and the chance of having faulty outcomes, which is an issue that needs to be addressed. There is also a need for standards and guidelines for the integration of XAI with 6G. 

  \end{itemize}

\subsubsection{Preliminary Solutions}
There are many factors that influence the decision-making of XAI models in 6G networks. The security and privacy of data are crucial factors that can be improved by combining blockchain and federated learning with XAI, which will also increase the trust in XAI decisions. In the use cases relating to 6G networks, a robust governance model can minimise bias, promote transparency, and decrease the chance of erroneous outcomes.
\subsubsection{Future Direction}
Some of the potential research directions that can address the aforementioned challenges and open issues are as follows. Researchers should focus on developing XAI algorithms that maintain the balance between explainability and the performance of the AI/ML algorithms by using technologies such as techno-economic analysis \cite{righi2020ai,de2019ai}. Several soft computing techniques such as meta-heuristic algorithms, principal component analysis, and fuzzy systems can be considered to address the challenge of high dimensionality through dimensionality reduction \cite{reddy2020analysis}. Unsupervised learning algorithms such as clustering that do not require labels for prediction/classification can be used to address the issue of generation of labels in real-time for 6G-based applications \cite{kaur2021machine}. Federated learning (FL), which is a recent development of ML, can be adopted in XAI-enabled 6G applications to provide customised decisions to heterogeneous networks \cite{gadekallu2021federated}. Furthermore, FL can be integrated with XAI-enabled 6G applications to address the issue of privacy preservation \cite{liu2020privacy,song2022eppda}.

\subsection{Limitations and challenges of XAI for 6G}



\subsubsection{Lessons learned}
The recent studies of XAI methods in 6G have three limitations. Firstly, there are not enough in-model XAI methods proposed so far. Most of the existing XAI can only explain black boxes after the 6G AI decision-making results are given. It prevents the achievement of a higher level trade-off between the interpretability and model performance in 6G. Secondly, although many research studies have emphasised the importance of XAI measurements, there are no widely recognised quantifiable metrics for explainability in typical AI applications in 6G. Thirdly, there is a lack of multidisciplinary collaborations between experts in AI and the legal community.

\subsubsection{Key Research Problems}
Explainable AI can be one of the engines to propulse the development of 6G in the coming era. However, there are still a few challenges that need to be solved.

\begin{itemize}
  \item Is it possible to develop better in-model XAI technologies for 6G to achieve a higher level of explainability and high decision-making performance?
  
  \item Will researchers be able to apply well-recognized metrics to evaluate explainability in 6G for end-users and stakeholders?
  
  \item How can XAI methods in 6G by engaging legal experts and experts from other disciplines fulfill the demands of current legislation and user satisfaction in all major stages (from design to evaluation)?

\end{itemize}
  
\subsubsection{Preliminary Solutions}
According to recent literature \cite{guo2020explainable}, most of the core systems that compose the wireless communications (signal detection, antenna detection, channel estimation, power allocation, etc.) which are essential in 6G have low, very low, or none explainability. This means that there is a long way to provide 6G with high levels of explainability to increase the trust of the users, especially in critical domains such as autonomous driving or remote surgery. There are currently some frameworks proposed for integrating XAI in 6G and future wireless networks to help the understanding of the system by users and engineers in charge of the network infrastructure. These frameworks also consider malicious attacks from external threats \cite{guo2020explainable}.

The lack of explainability was addressed in 2017, by the Defense Advanced Research Projects Agency (DARPA) in the US. DARPA launched an initiative to promote XAI techniques to explain to humans the decisions taken by ML models \cite{dw2019darpa}. This new challenge involved different areas such as designing explainable interfaces, understanding how the human mind understands concepts, and learning new explainable models. This was a four-year project and there were two different teams. One team is the XAI developers that worked on creating new effective techniques to provide useful explanations based on Human-computer interaction. The other team created an evaluation framework based on psychology to test the quality of the explanations. 6G can be used for very critical services such as remote surgery or autonomous driving \cite{guo2020explainable}.

\subsubsection{Future Direction}
In the upcoming years, we expect to see more applications of XAI in filling the gaps within the existing AI-driven 6G use cases and technical aspects. 

The upcoming 6G brings several new challenges to some of the emerging technologies such as quantum computing and blockchain 3.0, which may need XAI intensively in the future. Quantum computing has evolved quite quickly in recent years, after a long-term silence since it was theoretically discussed in the early 1980s \cite{feynman1982simulating}. Quantum computers have a great potential to solve a particular type of computing task that used to be considered infeasible for classical computers. As one type of quantum computing, the power of adiabatic quantum computation was validated in a 6G smart transportation pilot project for assigning optimal bus routes in the city of Lisbon, Portugal \cite{yarkoni2020quantum}. Such quantum computing-powered AI decision-making will be exponentially faster in the 6G ages due to more input data every second. It requires XAI technology to explain high-stake decisions under strict performance pressure where data flows are extremely high in volume. Blockchain 3.0 \cite{maesa2020blockchain} generally refers to all non-cryptocurrencies blockchain applications such as electronic voting and supply chain management. In the 6G era, many heterogeneous blockchain systems need to be connected, which poses a great challenge to balancing network performance and system security and privacy demands. Moreover, the high heterogeneity of blockchain 3.0 also implies more diverse stakeholders from different organisations involved in 6G AI-assisted decision-making. Ensuring all types of stakeholders are satisfied with the outcomes of XAI and making XAI methods compliant with local regulations would remain vital.

\section{Conclusion}\label{conclusion}

This paper comprehensively reviews and analyses the potential of applying XAI methods to make a future AI-based 6G system more transparent and trustworthy. The existing ideas for designing 6G networks are overviewed at the beginning of this paper, along with the exhaustive survey of the state-of-the-art AI and XAI methods. Later in this paper, several representative 6G technical aspects and use cases are carefully analysed in their existing AI-based solutions and the trend of applying XAI to enhance the trustworthiness of 6G network systems. At the end of this paper, the lessons learned about the limitations of existing work are summarised to remind researchers and practitioners that XAI can not solve all problems. Accordingly, research challenges are also highlighted that are promising to overcome or alleviate the potential limitations of XAI. We hope this survey can guide the future 6G developments in a more sustainable direction.

\section*{Acknowledgement}
This work is partly supported by the European Commission in SPATIAL (Grant no: 101021808)  Academy of Finland in 6Genesis (Grant no. 318927), Science Foundation Ireland under CONNECT phase 2 (Grant no. 13/RC/2077\_P2) projects, ADAPT Centre Phase 2 (Grant No., 13/RC/2106\_P2), and Industry Fellowship (Grant No. 21/IRDIF/9839).


 
\bibliographystyle{IEEEtran}
\bibliography{references.bib}

\begin{IEEEbiography}[{\includegraphics[width=1in,height=1.25in,clip,keepaspectratio]{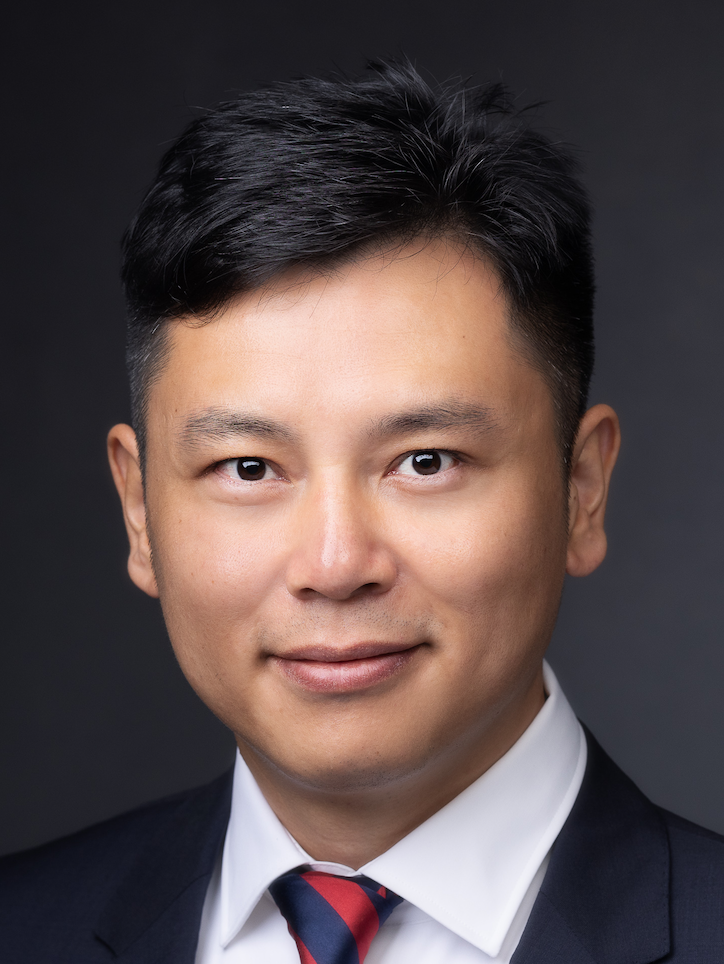}}]{Shen Wang}(Member, IEEE)
is currently an Assistant Professor at the School of Computer Science, University College Dublin, Ireland. He received an M.Eng. degree from Wuhan University, China, and a Ph.D. degree from Dublin City University, Ireland. Dr. Wang is a member of the IEEE and has been involved with several EU projects as a co-PI, WP, and Task leader in big trajectory data streaming for air traffic control and trustworthy AI for intelligent cybersecurity systems. Some key industry partners of his applied research are IBM Research Brazil, Boeing Research and Technology Europe, and Huawei Ireland Research Centre. He is the recipient of the IEEE Intelligent Transportation Systems Society Young Professionals Travelling Fellowship 2022. His research interests include connected autonomous vehicles, explainable artificial intelligence, and security and privacy for mobile networks.
\end{IEEEbiography}

\begin{IEEEbiography}[{\includegraphics[trim=25 0 25 15 ,width=1in,height=1.25in,clip,keepaspectratio]{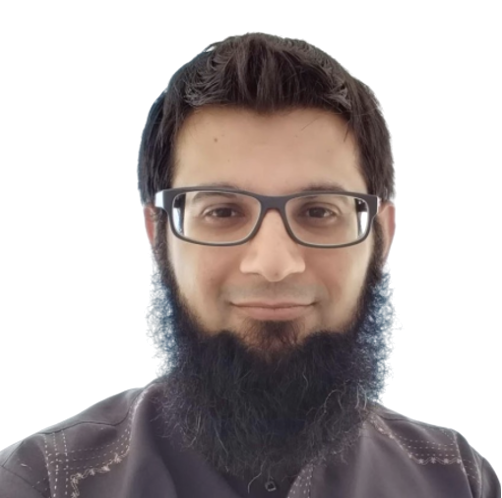}}]{M. Atif Qureshi} is a Lecturer/Assistant Professor at Technological University Dublin, Ireland. 
He received a joint PhD in Computer Science from the National University of Ireland Galway (Ireland) and the University of Milano-Bicocca (Italy), an MS degree in Computer Science from Korea Advanced Institute of Science and Technology (South Korea), and a BS in Computer Science from the University of Karachi (Pakistan). His research interests include natural language processing, machine learning, disinformation space, and explainable artificial intelligence. Atif has contributed to various projects funded by Science Foundation Ireland and the Irish Research Council as principal investigator and technical lead for those funded by Enterprise Ireland and the EU calls and licensed an outcome to a leading media organisation of Ireland in the space of social media analytics. Atif is passionate about applied research focusing on the tight coupling of business needs and analytics to create value and impact.
\end{IEEEbiography}

\begin{IEEEbiography}[{\includegraphics[width=1in,height=1.25in,clip,keepaspectratio]{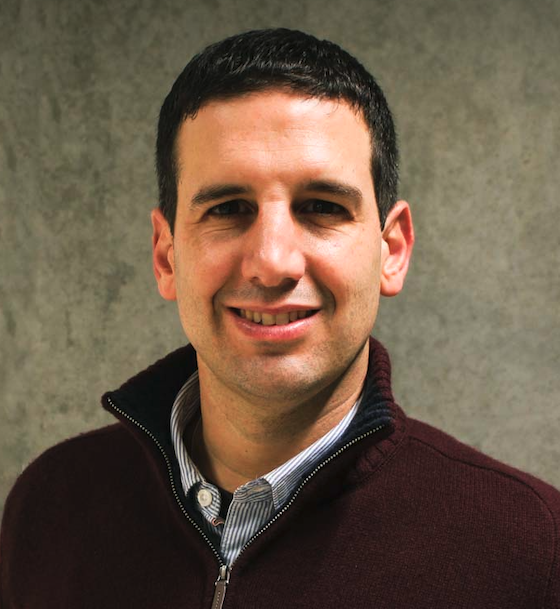}}]{Luis Miralles-Pechuán} is currently an Assistant Lecturer at Technological University Dublin. He obtained his Ph.D. and Bachelor in Computer Science at the University of Murcia (Spain). He worked as a full-time researcher/lecturer at University Panamericana in Mexico for more than three years. He decided to start a Ph.D. in 2012 on creating new approaches within the Online Advertising world. During his Ph.D., he got familiar with ML and he published a good number of papers on topics related to how to apply ML to online advertising. After finishing his Ph.D., he worked in postdoc levels I and II in CeADAR, University College Dublin, and there, he won the prize for the best student paper at the Digital Forensic conference. Currently, his favourite topic is how to apply Reinforcement Learning to fight the COVID-19 pandemic and to plan the containing levels considering both public health and the economy. Lastly, he has expertise in human activity recognition and generalised zero-shot learning (GZSL) and applying machine learning to improve the accessibility of websites. 
\end{IEEEbiography}

\begin{IEEEbiography}[{\includegraphics[width=1in,height=1.25in,clip,keepaspectratio]{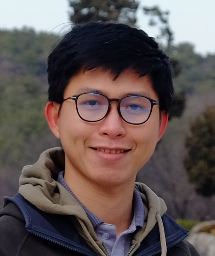}}]{Thien Huynh-The} (Member, IEEE) received the B.S. degree in Electronics and Telecommunication Engineering and the M.Sc. degree in Electronics Engineering from Ho Chi Minh City University of Technology and Education, Vietnam, in 2011 and 2013, respectively, and the Ph.D. degree in Computer Science and Engineering from Kyung Hee University (KHU), South Korea, in 2018. He was a recipient of the Superior Thesis Prize awarded by KHU. From March 2018 to August 2018, he was a Postdoctoral Researcher with Ubiquitous Computing Laboratory, KHU. From September 2018 to May 2022, he was a Postdoctoral Researcher with the ICT Convergence Research Center, Kumoh National Institute of Technology, South Korea. He is currently a Lecturer in Department of Computer and Communication Engineering, Ho Chi Minh City University of Technology and Education (HCMUTE), Vietnam. He was a recipient of the Golden Globe Award 2020 for Vietnamese Young Scientist by Central Ho Chi Minh Communist Youth Union associated with the Ministry of Science and Technology. His current research interests include digital image processing, radio signal processing, computer vision, wireless communications, IoT applications, machine learning, and deep learning. 
\end{IEEEbiography}

\begin{IEEEbiography}[{\includegraphics[width=1in,height=1.25in,clip,keepaspectratio]{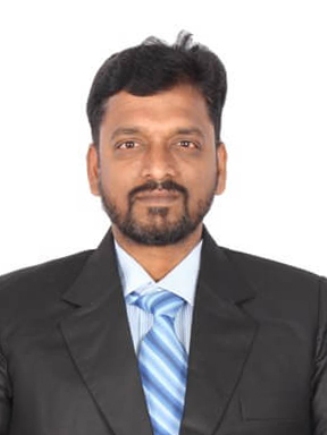}}]{Thippa Reddy Gadekallu} (Senior Member, IEEE) is currently working as Associate Professor in School of Information Technology and Engineering, VIT, Vellore, Tamil Nadu, India. He obtained his Bachelor's in Computer Science and Engineering in the year 2003 from Nagarjuna University, India, Master in Computer Science and Engineering from Anna University, Chennai, Tamil Nadu, India, and completed his Ph.D. in the domain of Machine Learning from Vellore Institute of Technology, Vellore, Tamil Nadu, India in the year 2017. He has more than 14 years of experience in teaching. He has published more than 80 international/national publications. Currently, his areas of research include Machine Learning, Internet of Things, Deep Neural Networks, Blockchain, and Computer Vision.
\end{IEEEbiography}

\begin{IEEEbiography}[{\includegraphics[width=1in,height=1.25in,clip,keepaspectratio]{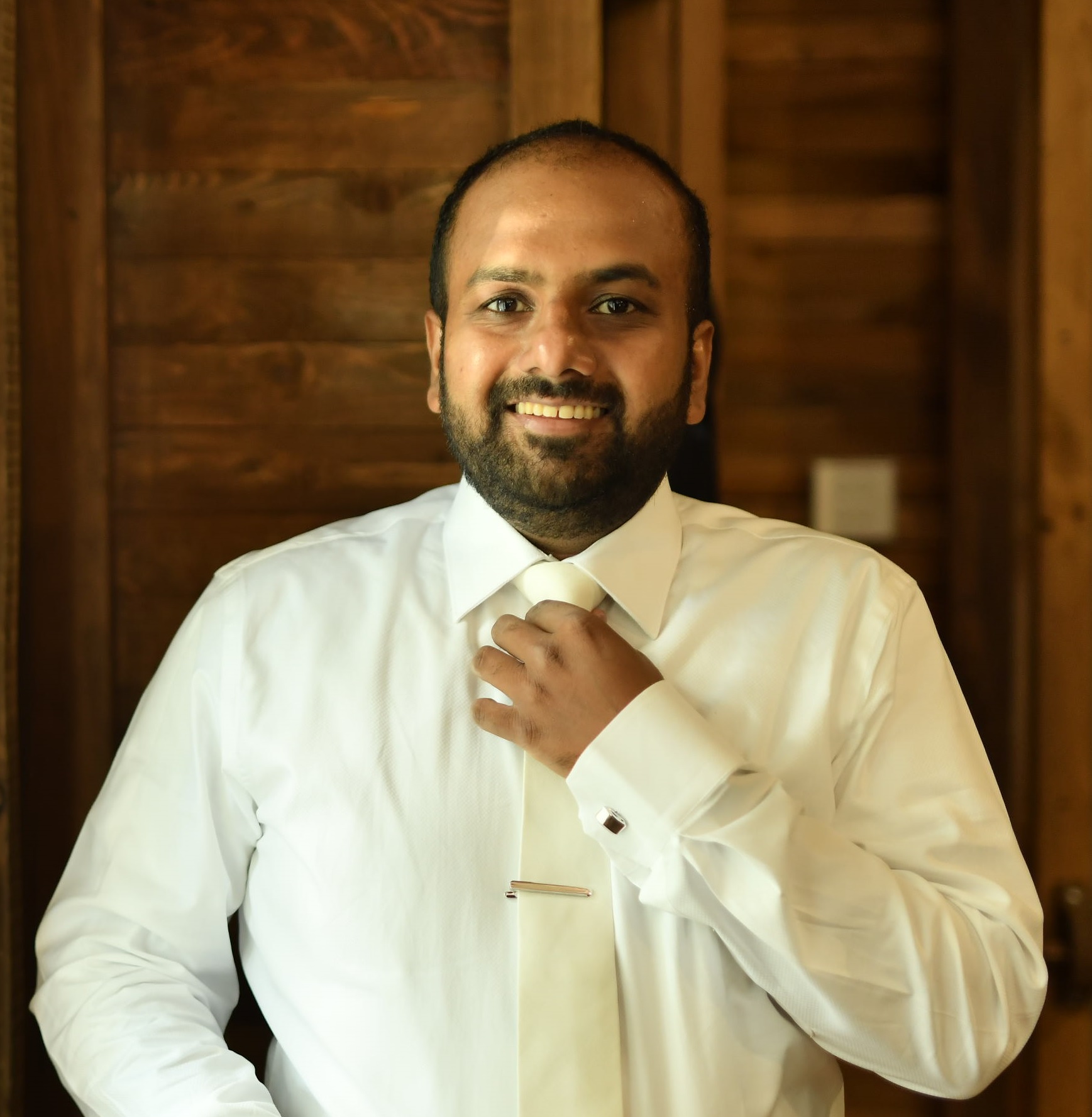}}]{Madhusanka Liyanage} (Senior Member, IEEE) received his the Doctor of Technology degree in communication engineering from the University of Oulu, Oulu, Finland, in 2016. From 2011 to 2012, he worked as a Research Scientist at the I3S Laboratory and Inria, Sophia Antipolis, France. He is currently an assistant professor/Ad Astra Fellow and the Director of Graduate Research at the School of Computer Science, University College Dublin, Ireland. He is also acting as an adjunct Processor at the Center for Wireless Communications, University of Oulu, Finland and Department of Electrical and Information Engineering - University of Ruhuna, Sri Lanka. He was also a recipient of the prestigious Marie Skłodowska-Curie Actions Individual Fellowship during 2018-2020. During 2015-2018, he has been a Visiting Research Fellow at the CSIRO, Australia, the Infolabs21, Lancaster University, U.K., Computer Science and Engineering, The University of New South Wales, Australia, School of IT, University of Sydney, Australia, LIP6, Sorbonne University, France and Computer Science and Engineering, The University of Oxford, U.K. He is also a senior member of IEEE. In 2020, he received the ``2020 IEEE ComSoc Outstanding Young Researcher" award from IEEE ComSoc EMEA. Dr. Liyanage is an expert consultant at European Union Agency for Cybersecurity (ENISA). In 2021, Liyanage was elevated as a Funded Investigator of Science Foundation Ireland CONNECT Research Centre, Ireland. He was ranked among the World's Top 2\% Scientists (2020) in the List prepared by Elsevier BV, Stanford University, USA. Also, he was awarded an Irish Research Council (IRC) Research Ally Prize as part of the IRC Researcher of the Year 2021 awards for the positive impact he has made as a supervisor.
Moreover, he is an expert reviewer at different funding agencies in France, Qatar, UAE, Sri Lanka, and Kazakhstan. Dr. Liyanage's research interests are 5G/6G, SDN, IoT, Blockchain, MEC, mobile, and virtual network security.  More info: \url{www.madhusanka.com}
\end{IEEEbiography}

\vfill

\end{document}